\documentclass[lettersize,journal]{IEEEtran}
\usepackage{amsmath,amsfonts}
\usepackage{amssymb}
\usepackage{physics,amsmath}
\usepackage{algorithm}
\usepackage{algpseudocode}
\usepackage{array}
\usepackage{bm}
\usepackage{color}

\usepackage{soul}
\sethlcolor{yellow}
\usepackage{booktabs}
\usepackage{subcaption}
\usepackage{mathtools}
\usepackage{xcolor}
\usepackage{tabularx}
\usepackage{tikz}
\usetikzlibrary{chains,positioning,arrows.meta,calc}
\usepackage{makecell}
\usepackage{textcomp}
\usepackage{stfloats}
\usepackage{lipsum}  
\usepackage{bm}
\usepackage{url}
\usepackage{verbatim}
\usepackage{graphicx}
\usepackage{cite}
\usepackage{pdfpages}
\usepackage{tikz-3dplot}
\usepackage{relsize}
\usetikzlibrary{positioning, arrows.meta}
\usepackage{xcolor}
\usepackage[hidelinks]{hyperref}
\usepackage{siunitx} % Required for proper alignment and unit formatting
\usepackage{array} % Required for defining new column types

\usepackage{tikz}
\usetikzlibrary{3d}
\usetikzlibrary{arrows.meta} % for more arrow tip kinds

\definecolor{mycustomcolor1}{rgb}{0.6627, 0.1412, 0.1255} 
\definecolor{mycustomcolor2}{rgb}{0.8196, 0.5686, 0.2431}
\definecolor{mycustomcolor3}{rgb}{0.0549, 0.1922, 0.3765}
\definecolor{mycustomcolor4}{rgb}{0.0745, 0.3255, 0.5647}
\definecolor{mycustomcolor5}{rgb}{0.7882, 0.3490, 0.3725} 
\definecolor{mycustomcolor6}{rgb}{0.2157, 0.1843, 0.1843}
\definecolor{mycustomcolor7}{rgb}{0.9843, 0.8078, 0.2275}
\definecolor{mycustomcolor8}{rgb}{0.8784, 0.8157, 0.7216}

\newcolumntype{C}[1]{>{\centering\arraybackslash}m{#1}}
\newcolumntype{L}[1]{>{\centering\arraybackslash}m{#1}}

\usepackage{courier} % Load the courier package for typewriter font
\usepackage{algorithm} % For algorithms
\usepackage{algpseudocode} % For pseudocode

\algrenewcommand\alglinenumber[1]{\small\ttfamily\textcolor{black}{#1}}
\algrenewcommand\algorithmicrequire{\textbf{\small\ttfamily Input:}}
\algrenewcommand\algorithmicensure{\textbf{\small\ttfamily Output:}}
\algrenewcommand\algorithmiccomment[1]{\hfill\#\ \eqparbox{COMMENT}{\small\ttfamily #1}}

\begin{document}

\newcommand{\orcidiconAbk}{\href{https://orcid.org/0009-0006-1187-7782}{\includegraphics[scale=0.1]{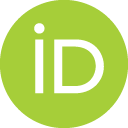}}}

\newcommand{\orcidiconfatih}{\href{https://orcid.org/0009-0009-5597-6717}{\includegraphics[scale=0.1]{figures/orcidID128.png}}}

\newcommand{\orcidiconOba}{\href{https://orcid.org/0000-0003-2523-3858}{\includegraphics[scale=0.1]{figures/orcidID128.png}}}

\title{Information and Communication Theoretical Foundations of Plant Communication: Mechanisms, Models, and Empirical Approaches}

\author{Ahmet B. Kilic\orcidiconAbk,~\IEEEmembership{Student Member,~IEEE}, Fatih Merdan\orcidiconfatih, 
        and Ozgur B. Akan\orcidiconOba,~\IEEEmembership{Fellow,~IEEE}                
        \thanks{The authors are with the Center for neXt-generation Communications (CXC), Department of Electrical and Electronics Engineering, Koç University, Istanbul, Turkey (e-mail: \{ahmetkilic20, fmerdan25, akan\}@ku.edu.tr).}
        \thanks{Ozgur B. Akan is also with the Centre for neXt Communications (CXC), Department of Engineering, University of Cambridge, CB3 0FA Cambridge, U.K (e-mail: oba21@cam.ac.uk).}
	    \thanks{This work was supported in part by the AXA Research Fund (AXA Chair for Internet of Everything at Ko\c{c} University).}
}% <-this % stops a space

% The paper headers
%\markboth{Journal of \LaTeX\ Class Files,~Vol.~14, No.~8, August~2021}%
%{Shell \MakeLowercase{\textit{et al.}}: A Sample Article Using IEEEtran.cls for IEEE Journals}

%\IEEEpubid{0000--0000/00\$00.00~\copyright~2021 IEEE}
% Remember, if you use this you must call \IEEEpubidadjcol in the second
% column for its text to clear the IEEEpubid mark.
	
\maketitle

\begin{abstract}
Plants exchange information through chemical, electrical, and mechanical signaling, engaging not only with neighboring plants but also with insects, animals, and microorganisms. These interactions support a broad range of ecological functions, from resource coordination and reproductive signaling to collective responses to environmental change. Despite extensive biological documentation of these processes, their systematic analysis from an Information and Communication Technology (ICT) perspective remains limited. This article addresses that gap as a tutorial with survey elements. It first establishes the biological foundations of plant communication across chemical, mechanical, and electrical modalities, covering the full range of interacting agents and signal functions. Building on this foundation, it reformulates plant signaling within ICT frameworks by defining transmitter, channel, and receiver models for each modality and introducing the relevant mathematical descriptions. Empirical studies and state-of-the-art sensing approaches are then reviewed to ground the models in experimental reality and identify where theory and measurement diverge. The paper concludes with a lessons learned section that distills cross-cutting insights on the fidelity of biology-to-ICT mappings, the distinctive channel properties of plant communication, multimodal coupling, receiver design principles, and observability constraints. By integrating biological depth with communication-theoretic rigor, this article provides a comprehensive foundation for the communications research community and identifies the open problems that will define interdisciplinary progress in this field.
\end{abstract}


\section{Introduction}
 
\IEEEPARstart{P}LANTS engage in sophisticated communication across a wide range of partners and contexts. Through chemical, electrical, and mechanical signals, they exchange information with neighboring plants, insects, animals, fungi, and soil microorganisms \cite{karban2021plant, ninkovic2021plant, Volkov2018, Gagliano2012}. These interactions are not limited to responses to threat: they also mediate pollinator attraction, seed dispersal, kin recognition, nutrient sharing, and coordinated responses to environmental change \cite{detectthyfamily, Veits2019, Simard2012}. Plant communication is therefore a multimodal, multispecies signaling system whose functions span the full ecological life of the plant.

Plants communicate through three principal modalities. Chemical communication is the most extensively studied, occurring aboveground via volatile organic compounds (VOCs) dispersed through air \cite{midzi2022stress, ninkovic2021plant} and belowground through root exudates that diffuse through soil pores \cite{bais2004underground, delory2016root}. Chemical signals convey information about physiological state, identity, and environmental conditions to receivers ranging from neighboring plants to insects and soil microbes. Mechanical communication encompasses acoustic vibrations and direct physical contact, both of which elicit physiological responses in receiver organisms \cite{Veits2019, Khait2023, deWit2012}. Electrical communication enables rapid signaling within and between plants: action potentials, variation potentials, and system potentials propagate through plant tissue and, under appropriate conditions, through the soil between root systems \cite{Volkov2018, Szechynska-Hebda2022}. Mycorrhizal fungal networks provide a shared physical infrastructure through which chemical and electrical signals can travel between connected root systems, extending the reach of both modalities \cite{Simard2012, babikova2013underground}.

The biological richness of these signaling systems makes plant communication an analytically demanding subject for an ICT researcher. Each modality has a transmitter, a channel, and a receiver with quantifiable parameters and well-defined physical governing equations. Each exhibits modulation, noise, interference, and constraints on information transfer that have direct analogs in communication theory. The aboveground chemical channel is dispersive and memory-laden; the soil chemical channel couples transport, sorption, and microbial degradation; the acoustic channel through air and substrate supports multiple propagation regimes. Approaching these systems through an ICT lens does not merely redescribe known biology in engineering language. It enables new questions: what is the effective channel capacity of VOC-mediated signaling under realistic wind conditions? What receiver architecture best matches the nonlinear, saturation-bounded uptake kinetics of root chemical reception? How does multimodal coupling between chemical and electrical channels affect the degrees of freedom available for information transfer? These are questions that biology alone is not structured to answer, and they are questions whose answers would in turn enrich biological understanding by quantifying what plant communication systems can and cannot achieve.

\begin{table*}[t]
    \centering
    \renewcommand{\arraystretch}{1.2}
    \caption{Summary of related survey and tutorial papers.}
    \scriptsize
    \begin{tabular}{p{3cm}p{3cm}p{4cm}p{4cm}}
        \toprule
        \textbf{Paper} & \textbf{Scope} & \textbf{Distinct Contributions} & \textbf{Comparison to This Paper} \\
        \midrule
        \textbf{Karban (2021) \cite{karban2021plant}} & Overview of plant communication & Comprehensive review of plant communication & Explores plant communication from a biological perspective; this paper adopts an ICT perspective \\
        \textbf{Midzi et al. (2022) \cite{midzi2022stress}} & Overview of chemical communication between plants & In-depth review of chemical communication & Reviews chemical communication from a biological perspective; this paper provides ICT-based transmitter, channel, and receiver models \\
        \textbf{Ninkovic et al. (2021) \cite{ninkovic2021plant}} & Role of plant volatiles as mediators of plant interactions & Focus on VOCs in chemical communication & Focuses mainly on VOCs; this paper takes a holistic view across modalities and provides mathematical models for each step \\
        \textbf{Bais et al. (2004) \cite{bais2004underground}} & Root exudates in belowground communication & Covers different communication paradigms belowground & Discusses belowground communication broadly; this paper provides ICT-based models for belowground chemical communication \\
        \textbf{Niinemets (2010) \cite{niinemets2010mild}} & Quantifying VOC emissions under varying conditions & Comprehensive review of quantitative VOC emission models & Focuses on VOC emission quantification; this paper provides quantitative models across all communication modalities \\
        \textbf{Ahmed et al. (2022) \cite{ahmed2022molecular}} & Molecular communication network in crop sciences & Explores plant signaling through molecular communication & Focuses on molecular communication in plant signaling; this paper examines all communication modalities and includes sensing methods \\
        \textbf{Faiola et al. (2020) \cite{faiola2020impact}} & Impact of insect herbivory on plant volatiles & In-depth review of insect-induced chemical communication & Focuses exclusively on insect-induced VOC changes; this paper covers a broader range of signal types and interacting agents \\
        \textbf{Galieni et al. (2021) \cite{Galieni2021}} & Technologies for plant signal detection & Covers technologies for sensing chemical communication & Reviews sensing technologies; this paper also covers sensing across mechanical and electrical modalities \\
        \textbf{Wang et al. (2021) \cite{wang2021root}} & Root exudates in plant signaling & Focuses on belowground chemical communication & Reviews belowground chemical communication with focus on VOCs; this paper adopts a holistic view and provides mathematical models \\
        \textbf{Delory et al. (2016) \cite{delory2016root}} & Belowground VOCs in plant signaling & Focus on VOCs in belowground chemical communication & Examines belowground chemical communication; this paper takes a comprehensive approach with mathematical models for every step \\
        \textbf{Yan et al. (2009) \cite{Yan2009}} & Electrical signals in higher plants & Reviews measurement techniques for electrical signaling & Reviews electrical signal measurement; this paper provides transmitter, channel, and receiver models for electrical communication \\
        \textbf{Simard et al. (2012) \cite{Simard2012}} & Mycorrhizal networks: characteristics and functions & Models for characterization of mycorrhizal networks & Evaluates mycorrhizal networks broadly; this paper treats CMNs as infrastructure relaying chemical and electrical signals \\
        \textbf{Simard et al. (2004) \cite{simard2004mycorrhizal}} & Mycorrhizal networks and nutrient transfer & Reviews studies on nutrient transfer & Reviews nutrient transfer through mycorrhizal networks; this paper instead considers CMNs as a signal-relay pathway within other modalities \\
        \textbf{Oelmuller (2019) \cite{oelmuller2019interplant}} & Mycorrhizal networks in plant communication & Models for plant communication in mycorrhizal networks & Provides models for mycorrhizal network communication; this paper treats CMNs as shared infrastructure, without dedicated CMN models or validation  \\
        \textbf{Gagliano (2013) \cite{gagliano2013green}} & Plant communication through sound & Covers emission and detection of sound in plants & Reviews plant acoustic communication; this paper also covers sensing methods and mechanical communication more broadly \\
        \textbf{Demey et al. (2023) \cite{Demey2023}} & Plant responses to sound stimuli & In-depth review of sound perception in plants & Focuses on sound perception; this paper also covers sound emission and touch-mediated signaling \\
        \bottomrule
    \end{tabular}
    \label{tab:related_work}
\end{table*}

\begin{figure*}[t]
\centering
\begin{tikzpicture}[
    >=stealth,
    hdr/.style={rectangle, draw, fill=gray!25, very thick, align=center,
                font=\scriptsize\bfseries, text width=2.4cm,
                minimum height=1.0cm, inner sep=3pt},
    mod/.style={rectangle, draw, align=center,
                font=\scriptsize, text width=2.4cm,
                minimum height=0.9cm, inner sep=3pt, rounded corners=2pt},
    node distance=0.5cm and 0.7cm
]

%% ---- Column headers ----
\node[hdr] (h2) {\shortstack{Section II\\Biological\\Mechanisms}};
\node[hdr, right=0.7cm of h2] (h3) {\shortstack{Section III\\ICT Models\\(TX/CH/RX)}};
\node[hdr, right=0.7cm of h3] (h4) {\shortstack{Section IV\\Empirical \&\\Sensing}};
\node[hdr, right=1.0cm of h4] (h5) {\shortstack{Section V\\Lessons\\Learned}};
\node[hdr, right=0.7cm of h5] (h6) {\shortstack{Section VI\\Conclusion}};

%% ---- Modality rows ----
%% Chemical
\node[mod, below=0.6cm of h2]  (c2) {\shortstack{Chemical\\Communication}};
\node[mod, right=0.7cm of c2]  (c3) {\shortstack{Chemical\\TX/CH/RX\\Models}};
\node[mod, right=0.7cm of c3]  (c4) {\shortstack{Chemical\\Sensing \&\\Experiments}};

%% Mechanical
\node[mod, below=0.5cm of c2]  (m2) {\shortstack{Mechanical\\Communication\\(Acoustic \& Touch)}};
\node[mod, right=0.7cm of m2]  (m3) {\shortstack{Mechanical\\TX/CH/RX\\Models}};
\node[mod, right=0.7cm of m3]  (m4) {\shortstack{Mechanical\\Sensing \&\\Experiments}};

%% Electrical
\node[mod, below=0.5cm of m2]  (e2) {\shortstack{Electrical\\Communication\\(AP/VP/SP)}};
\node[mod, right=0.7cm of e2]  (e3) {\shortstack{Electrical\\TX/CH/RX\\Models}};
\node[mod, right=0.7cm of e3]  (e4) {\shortstack{Electrical\\Sensing \&\\Experiments}};

%% ---- Header dashed guides ----
\draw[->, dashed, gray] (h2) -- (h3);
\draw[->, dashed, gray] (h3) -- (h4);
\draw[->, dashed, gray] (h4) -- (h5);
\draw[->] (h5) -- (h6);

%% ---- Header II → modality rows ----
%% Leave from west (left-middle) of h2, run a vertical trunk to the left, branch to each row
\coordinate (trunk) at ($(h2.west) + (-0.4cm, 0)$);
\draw (h2.west) -- (trunk);
\draw[->] (trunk) -- ($(trunk |- c2.west)$) -- (c2.west);
\draw[->] (trunk) -- ($(trunk |- m2.west)$) -- (m2.west);
\draw[->] (trunk) -- ($(trunk |- e2.west)$) -- (e2.west);

%% ---- Horizontal modality paths ----
\draw[->] (c2) -- (c3); \draw[->] (c3) -- (c4);
\draw[->] (m2) -- (m3); \draw[->] (m3) -- (m4);
\draw[->] (e2) -- (e3); \draw[->] (e3) -- (e4);

%% ---- Convergence into Sec V header ----
\draw[->] (c4.east) -- ($(c4.east) + (0.3cm,0)$) |- (h5.west);
\draw[->] (m4.east) -- ($(m4.east) + (0.3cm,0)$) |- (h5.west);
\draw[->] (e4.east) -- ($(e4.east) + (0.3cm,0)$) |- (h5.west);

\end{tikzpicture}
\caption{Visual overview of the paper's organization. Three modality paths (chemical, mechanical, electrical) run horizontally from biological mechanisms (Section~II) through ICT models (Section~III) and empirical evidence (Section~IV), then converge into the cross-cutting lessons learned (Section~V) and conclusion (Section~VI).}
\label{fig:roadmap}
\end{figure*}

\subsection{Motivation and Objective}
 
Various studies have investigated plant communication from biological perspectives, as summarized in Table~\ref{tab:related_work}. These works have established a rich empirical foundation, but they leave a significant gap: the communication processes they describe have not been systematically reformulated in terms that allow quantitative ICT analysis. The complexity of biological mechanisms and the disciplinary distance between plant biology and communications engineering have historically constrained collaboration. Closing this gap requires translating biological mechanisms into ICT models with sufficient fidelity to support rigorous analysis. Such a translation, in turn, opens a path in the reverse direction: ICT-derived analyses can generate quantitative, testable hypotheses about plant communication that complement purely biological investigation.
 
Several interdisciplinary efforts have begun this work, particularly within molecular communication, where ICT-based models of chemical signaling have been developed for aboveground VOC transmission \cite{unluturk2016end, kilic2024endtoendmathematicalmodelingstress} and agricultural applications \cite{vakilipoor2025mcagricultureframeworknatureinspired, gulec2025decodingengineeringphytobiomecommunication}. However, these contributions address individual modalities in isolation and do not cover the full breadth of plant communication — its mechanical and electrical channels, its multispecies scope, or the coupling between modalities. A unified treatment that spans all three modalities, grounds models in empirical evidence, and draws cross-cutting lessons for ICT researchers does not yet exist.
 
This paper addresses that gap. Serving as a tutorial with survey elements, it provides the biological foundation needed by ICT researchers unfamiliar with plant biology, develops transmitter-channel-receiver models for chemical, mechanical, and electrical communication, and reviews the sensing methods and empirical studies that validate or constrain these models. The scope is deliberately broad: plant communication is treated as a multispecies signaling system involving plants, insects, animals, fungi, and soil microorganisms, and the range of functions covered extends beyond defense and stress response to include constitutive signaling, reproductive coordination, and kin-based resource allocation. This breadth is not incidental; it is necessary for an accurate characterization of the signal space, the range of receiver types, and the diversity of channel conditions that any complete ICT model of plant communication must accommodate.
 
\subsection{Contributions and Related Surveys}
 
This paper makes the following contributions to the ICT analysis of plant communication.
 
\begin{itemize}
    \item A comprehensive biological review of plant communication across chemical, mechanical, and electrical modalities, covering the full range of interacting agents and signal functions.
    \item An ICT-based modeling framework that defines transmitter, channel, and receiver components for each modality, synthesizes existing mathematical models, and critically evaluates their assumptions and conditions of validity.
    \item A systematic review of empirical studies and state-of-the-art sensing methods across all modalities, organized to connect experimental evidence directly to model parameters and channel characterization.
    \item A lessons learned section that distills cross-cutting insights on biology-to-ICT mapping fidelity, distinctive channel properties, multimodal coupling, receiver design principles, and observability constraints, directly addressing the open problems identified by the empirical review.
\end{itemize}
 
\subsection{Organization}
 
The remainder of this paper is organized as follows. Section~II provides the biological foundations of plant communication across chemical, mechanical, and electrical modalities. Section~III presents ICT-based transmitter, channel, and receiver models for each modality, including relevant mathematical formulations. Section~IV reviews empirical studies and sensing methods that ground the models in experimental reality. Section~V distills the cross-cutting lessons learned from the survey. Section~VI concludes the paper. Fig.~\ref{fig:roadmap} illustrates this organization and the flow of information across sections.

For convenience, Table~\ref{tab:acronyms} summarizes the acronyms that recur throughout the paper. Other acronyms are defined locally at their first occurrence.

\begin{table}[b]
\centering
\caption{Frequently Used Acronyms.}
\label{tab:acronyms}
\begin{tabular}{ll}
\hline
\textbf{Acronym} & \textbf{Definition} \\
\hline
ICT & Information and Communication Technology \\
VOC(s) & Volatile Organic Compound(s) \\
AP & Action Potential \\
VP & Variation Potential \\
SP & System Potential \\
CSK & Concentration Shift Keying \\
RSK & Ratio Shift Keying \\
ROS & Reactive Oxygen Species \\
ISI & Inter-Symbol Interference \\
CMN & Common Mycorrhizal Network \\
\hline
\end{tabular}
\end{table}

\section{Fundamental Mechanisms of Plant Communication}

\begin{figure}[t]
    \centering
    \includegraphics[width=\linewidth]{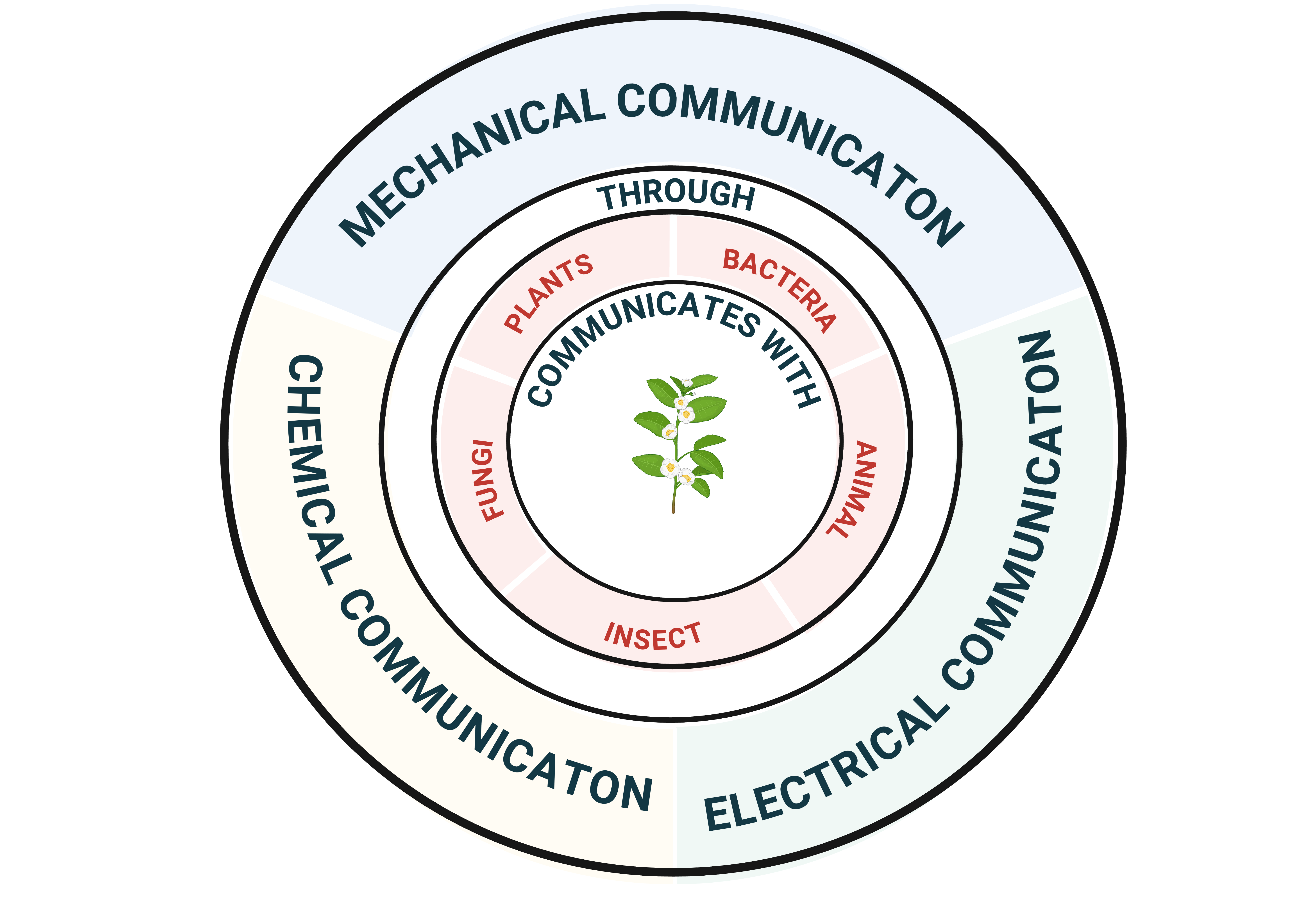}
    \caption{Overview of plant communication, illustrating the four primary signaling modalities (mechanical, electrical, chemical, and acoustic) and the inter-kingdom partners (bacteria, animals, insects, and fungi) with which plants exchange information. Created with BioRender.com.}
    \label{fig:plntcommovrw}
\end{figure}

Plants communicate through a diverse set of mechanisms, including
chemical signaling via volatile organic compounds (VOCs), electrical
signals, and mechanical stimuli such as acoustic vibrations and
direct physical contact \cite{ninkovic2021plant,boyno2022plant,
Volkov2018,Gagliano2012}. Mycorrhizal fungal networks further extend
these channels by providing a shared physical infrastructure through
which chemical and electrical signals can travel between connected
root systems. An overview of the three communication modalities and the inter-kingdom partners is shown in Fig.~\ref{fig:plntcommovrw}. Together they support a broad range of
biological functions, from resource sharing and reproductive
coordination to ecological interaction and collective defense.
Understanding the underlying mechanisms is essential before analyzing
plant communication from an ICT perspective, as it clarifies the
information sources, signal types, and interacting agents involved.

This section organizes plant communication into three modalities:
chemical, mechanical, and electrical. Chemical communication
encompasses both aboveground and belowground VOC-based signaling and
includes interactions among plants, animals, insects, and microbes.
Mechanical communication covers signaling mediated by physical
forces, including acoustic waves and touch-mediated contact signals.
Electrical communication addresses bioelectrical signal propagation
within and across plant individuals. Mycorrhizal networks are not
treated as a separate modality; instead, their role as a shared
channel infrastructure is discussed within the chemical and
electrical subsections where the evidence is most directly
established.
% -----------------------------------------------------------
\subsection{Chemical Communication}
% -----------------------------------------------------------

Plants continuously emit VOCs as chemical signals in both aerial and soil environments \cite{ninkovic2021plant}. These signals support a wide range of interactions: aboveground, VOCs mediate communication among neighboring plants, attract pollinators, repel herbivores, and recruit predators of herbivores; belowground, root-emitted compounds influence neighboring plants and interact with soil microbes \cite{bais2004underground,aktas2023odorbasedmolecularcommunicationsstateoftheart}. VOC emissions fall into two broad categories. Constitutive VOCs are released continuously under normal physiological conditions and serve functions such as pollinator attraction and herbivore deterrence \cite{preston2001methyl,midzi2022stress}. Induced VOCs arise when emission rates and blend compositions shift in response to biotic interactions (e.g., herbivory, pathogen attack) or abiotic conditions (e.g., drought, salinity, temperature extremes), producing signals that encode the nature and severity of the triggering event \cite{midzi2022stress,ninkovic2021plant}. From an ICT perspective, constitutive emissions resemble a continuous background carrier, while induced emissions correspond to modulated information-bearing signals, with information encoded in concentration levels analogous to concentration shift keying (CSK), or in blend composition analogous to ratio shift keying (RSK) \cite{kilic2024endtoendmathematicalmodelingstress,Kilic2024MRSK}.

% ------- Aboveground ----------------------------------------
\subsubsection{Aboveground Chemical Communication}
\label{sec:bioabochem}

In aboveground chemical communication, VOCs released from leaves, fruits, and flowers diffuse into the air and mediate interactions both within and beyond the plant kingdom \cite{ninkovic2021plant,delory2016root}. The emitting plant acts as the transmitter, while neighboring plants, insects, animals, or microbes that detect and respond to these signals act as receivers \cite{kilic2024endtoendmathematicalmodelingstress}.

VOC emission arises either from the release of compounds stored in specialized or nonspecific compartments, or from de novo synthesis \cite{niinemets2004physiological,grote2019new}. VOCs exit the plant through stomata or the cuticle: cuticular release proceeds by passive diffusion, while stomatal release is actively regulated by guard cells \cite{niinemets2004physiological,midzi2022stress}. Herbivory-induced wounding can also initiate VOC release \cite{midzi2022stress}. The major chemical families include terpenoids, fatty acid derivatives, benzenoid and phenylpropanoid compounds, and amino acid derivatives \cite{delory2016root}. Information is encoded either in the concentration of individual VOCs or in the composition and ratio of multi-compound blends \cite{ninkovic2021plant,ueda2012plant}, with each compound class capable of eliciting distinct physiological responses in receivers \cite{midzi2022stress}.

Once released, VOCs disperse through diffusion and wind-driven advection \cite{kilic2024endtoendmathematicalmodelingstress}. For single-compound signals, successful reception requires concentrations above background \cite{copolovici2011volatile}; for blends, the receiver must resolve the compositional ratio \cite{ninkovic2021plant}. Transmission fidelity depends on distance, wind speed, chemical degradation, and secondary atmospheric reactions. Blend integrity is particularly vulnerable to chemical transformations that may alter encoded information \cite{ninkovic2021plant}. Empirical studies confirm successful interplant VOC reception at distances up to 50~cm \cite{delory2016root}.

Aboveground chemical communication encompasses a rich array of interspecies interactions beyond pairwise plant signaling. Certain plant species produce VOC blends that mimic the alarm pheromones of honeybees, manipulating pollinator behavior or attracting predators of herbivores \cite{insects10100366}. Fleshy fruits exhibit a characteristic VOC shift during ripening—from defensive aldehydes to attractive esters and lactones—functioning as signals that recruit vertebrate seed dispersers such as bats and primates, while simultaneously attracting parasitoids of fruit-damaging insects and establishing a multitrophic chemical communication network \cite{plantxcom}. VOC signaling also encodes kinship information: plants distinguish genetically related neighbors from strangers through volatile cues, leading to coordinated responses and more efficient resource allocation at the group level \cite{detectthyfamily}.

Reception in neighboring plants occurs through stomatal and cuticular uptake \cite{trapp2007fruit}. Once absorbed, VOCs prime and activate defense-related pathways in receiver plants \cite{midzi2022stress}.

% ------- Belowground ----------------------------------------
\subsubsection{Belowground Chemical Communication}

Chemical communication also occurs belowground, where plant roots release VOCs and other compounds into the soil, enabling interactions with neighboring plants, soil bacteria, and fungi \cite{bais2004underground,delory2016root}. Compared with aboveground signaling, belowground communication is more complex and less studied \cite{bais2004underground,delory2016root}.

Root-emitted VOC blends vary in composition and quantity depending on genotype and physiological state \cite{ninkovic2021plant}. These blends typically include terpenoids, fatty acid derivatives, and sulfur-containing compounds \cite{delory2016root}, present both within root tissues and on outer surfaces \cite{lin2007volatile}. Emission occurs either by passive diffusion or through active transport proteins in root membranes \cite{wang2021root}.

Once released, VOCs diffuse through soil pores from high to low concentration \cite{horn1994,soildiff}. Transport is influenced by soil moisture \cite{eden2012}, soil type and pore size \cite{soilpore}, and degradation by soil microbes and enzymes \cite{thomsen1999}. The physicochemical properties of individual VOCs further determine achievable transport distances, which remain poorly quantified compared to aboveground communication \cite{delory2016root}.

Belowground chemical signals carry diverse information beyond immediate physiological state. Plants sense the specific chemical composition of root exudates—such as the nitrogen-rich compound allantoin—to enable kin discrimination, leading to reduced root competition and more cooperative resource allocation among related individuals \cite{detectthyfamily}. Mycorrhizal fungi that colonize plant roots and form common mycorrhizal networks (CMNs) also participate in belowground chemical signaling. These networks can carry specific signaling compounds between connected plants, potentially modulating host gene expression and enhancing phosphorus uptake and pathogen suppression \cite{detectthyfamily,Farag2013}. Chemical changes within fungal networks can further alter the surrounding rhizosphere, promoting beneficial bacteria that help connected plants resist soil-borne pathogens. From an ICT perspective, CMNs function as a shared broadcast medium: a chemical signal injected by one plant may reach multiple receivers through the hyphal network, resembling a multicast channel with shared infrastructure.

The final step in belowground communication is VOC absorption by the roots of receiver plants, which occurs primarily through passive diffusion along the concentration gradient \cite{nye1966}.

% -----------------------------------------------------------
\subsection{Mechanical Communication}
% -----------------------------------------------------------

Mechanical communication in plants encompasses signaling mediated by physical forces and the propagation of energy through media in contact with plant tissues. Two distinct submodalities are covered here. Acoustic communication involves the emission and reception of sound vibrations, including ultrasonic emissions and responses to externally generated sounds. Touch communication addresses direct physical contact between plant tissues or between plants and other organisms, and the systemic signals that propagate from such contact. Together, these submodalities constitute a physically mediated communication layer that complements the chemical and electrical signaling discussed in adjacent subsections.

% ------- Acoustic -------------------------------------------
\subsubsection{Acoustic Communication}
\label{sec:acocomm}

Plants both emit and respond to acoustic signals, and these interactions extend across species boundaries to include insects, pollinators, and other organisms. Plants emit sounds under a range of conditions \cite{Khait2023}, with emitted airborne vibrations capable of conveying both species identity and physiological state. Whether emissions are purely involuntary or also serve a deliberate signaling function remains unclear \cite{Demey2023}. Cavitation—the formation and collapse of air bubbles in the xylem—is a known source of clicking or popping sounds during water transport \cite{Demey2023}, though other mechanisms may also contribute.

Acoustic signals propagate through multiple media. Ultrasonic waves travel through air and are detectable at a distance \cite{Khait2023}. Cavitation sounds propagate internally via the vascular system \cite{Khait2023}, and soil has been proposed as an additional transmission medium \cite{Mishra2016}.

Plant responses to acoustic stimuli are well documented. \textit{Oenothera drummondii} flowers exposed to the sound of bee flight or synthetic frequencies at similar spectral content exhibited rapid increases in nectar sugar concentration within three minutes \cite{Veits2019}. This response was frequency-specific, with flowers reacting to pollinator sounds but not to higher-frequency tones, suggesting that floral structures may function as acoustic sensory organs. Plants also respond to vibrations produced by herbivores: feeding vibrations provide reliable cues for distinguishing herbivore identity \cite{Kollasch2020}, triggering targeted defensive responses. More broadly, sound vibrations have been shown to induce epigenetic changes and activate immune responses in \textit{Arabidopsis thaliana} against bacterial pathogens \cite{Jung2020}.

Sound perception in plants is mediated at multiple scales. Evidence supports acoustic sensing in flowers \cite{Veits2019}, leaves \cite{Appel2014}, and roots \cite{RodrigoMoreno2017,Gagliano2017}. Trichomes have been proposed to contribute through mechanosensing, though they may alternatively function as passive antennas \cite{Demey2023}. At the cellular level, mechanosensitive (MS) ion channels and receptor proteins sensitive to extracellular matrix deformation have been identified as candidate transduction mechanisms \cite{Hamant2017,Demey2023}. Genes regulated by touch are also upregulated by sound treatment, suggesting overlapping molecular pathways between tactile and acoustic stimuli \cite{Ghosh2016}.

While direct plant-to-plant acoustic communication has been demonstrated in a landmark study \cite{Gagliano2012}, the field is actively growing, with a rapidly increasing body of work targeting acoustic emission, reception, and response mechanisms in plants. Given the pace of research activity and the clear applicability of ICT models to this modality, we include acoustic communication as a dedicated subsection within mechanical communication.

% ------- Touch -------------------------------------------
\subsubsection{Touch Communication}
\label{sec:touchcomm}

Touch represents a direct mechanical interface through which plants exchange information with neighboring plants, animals, and insects. Unlike acoustic signals, which propagate through a medium, touch requires physical contact and triggers local and systemic responses through a well-characterized molecular cascade. Mechanosensitive (MS) ion channels—including the MscS-Like (MSL), Mid1-Complementing Activity (MCA), and Two-Pore Domain Potassium (TPK) families—serve as the primary transducers, converting membrane tension into ion flux upon contact \cite{Basu2017,Hamilton2015}. This initial ion flux, dominated by \(\mathrm{Ca}^{2+}\) influx, initiates a coupled wave of reactive oxygen species (ROS) and bioelectric potentials that propagates systemically through the plant \cite{Hansen2025}.

The behavioral outputs of touch signaling span two timescales. Rapid nastic responses—thigmonasty—are mediated by fast turgor changes in specialized motor organs. In \textit{Mimosa pudica}, contact triggers \(\mathrm{K}^{+}\) and \(\mathrm{Cl}^{-}\) efflux from extensor cells in the pulvinus, causing leaflet collapse within seconds \cite{Pouliquen2012}. The Venus flytrap (\textit{Dionaea muscipula}) employs a two-stimulus mechanism: two separate activations of trigger hairs within approximately 20~s are required before trap closure, effectively filtering out non-prey contact events \cite{Pouliquen2012}. This constitutes a form of signal integration analogous to a detection threshold or matched filter in communication systems. Over longer timescales, repeated mechanical stimulation drives thigmomorphogenesis—shorter stature, increased stem diameter, and denser wood—mediated by touch-responsive (TCH) genes encoding calmodulin-like proteins that regulate cell wall lignification \cite{Hansen2025}.

Touch communication extends beyond individual plant responses to mediate genuine plant-to-plant information transfer. When leaf tips of neighboring \textit{Arabidopsis thaliana} plants make contact, a mechanostimulation-induced \(\mathrm{Ca}^{2+}\) wave propagates from the trichomes at the leaf tip toward the petiole, triggering upward leaf movement (hyponasty) that positions the plant away from future competitors \cite{deWit2012}. This touch-induced response precedes phytochrome-based neighbor detection, establishing touch as an earlier and distinct signaling channel in competitive scenarios \cite{deWit2012}. Recent evidence further demonstrates that plants in physical contact with neighbors develop markedly greater resilience to excess light stress \cite{PelaezVico2025}. Systemic electric and \(\mathrm{H}_{2}\mathrm{O}_{2}\) signals propagate through a chain of physically touching plants, activating over 2,000 stress-response genes in receiver plants within one hour—including genes involved in responses to light, cold, wounding, and salinity \cite{PelaezVico2025}. These findings suggest that aboveground physical contact can establish community-wide signaling networks that coordinate collective responses.

Touch signaling also mediates plant–animal interactions of ecological significance. Crown shyness—the formation of canopy gaps between adjacent tree crowns—arises partly from mechanical stimulation caused by wind-driven branch contact \cite{vanderZee2021}. In pollinators, the touch of a stigma triggers rapid closure responses in some species, facilitating pollen deposition and cross-pollination \cite{Pouliquen2012}.

Whether touch-induced signals constitute true communication—where the sender encodes information specifically for a receiver—or represent eavesdropping on unavoidable mechanical byproducts of physical contact remains an active theoretical question. Current evidence, particularly the community-wide resilience effect \cite{PelaezVico2025} and the adaptive value of hyponasty \cite{deWit2012}, supports a functional communication role. From an ICT perspective, touch-mediated signaling maps to a wired, contact-dependent channel where the transmitter is the touching plant, the channel is the shared physical interface and the resulting ion and ROS cascade, and the receiver is the plant detecting the systemic propagating signal.

% -----------------------------------------------------------
\subsection{Electrical Communication}
% -----------------------------------------------------------
\label{sec:elccom}

\begin{table*}[t]
    \centering
    \scriptsize
    \caption{Comparison of electrical signals in plants (adapted from \cite{Zimmermann2009}).}
    \renewcommand{\arraystretch}{1}
    \setlength{\tabcolsep}{4pt}
    \resizebox{15cm}{!}{
    \begin{tabular}{p{4cm}p{3.5cm}p{3.5cm}p{3.5cm}}
        \toprule
        \textbf{Property} & \textbf{APs (Fast Signals)} & \textbf{VPs (Pressure-Based)} & \textbf{SPs (Slow Signals)} \\
        \midrule
        Trigger Mechanism & Voltage threshold & Turgor pressure & External stimulus \\
        Signal Type & Binary (All-or-none) & Analog (Graded) & Variable response \\
        Transmission Mode & Self-propagating & Localized & Self-propagating \\
        Speed (cm/min) & 20--400 & 6--60 & 5--10 \\
        Ion Involvement & Ca$^{2+}$, Cl$^-$, K$^+$ & H$^+$ pump deactivation & H$^+$ pump activation \\
        Membrane Effect & Depolarization & Depolarization & Hyperpolarization \\
        Duration & Short ($<$20 s) & Moderate (10 s--minutes) & Long (8--12 min) \\
        \bottomrule
    \end{tabular}
    }
    \label{tab:plant_signals}
\end{table*}

Electrical signaling enables rapid information transfer compared to chemical modalities, and can induce plant movement, ion channel activation, gene expression changes, enzymatic activity, wound healing, and growth responses \cite{Volkov2018}. Communication occurs belowground via soil conductivity between root systems \cite{Volkov2017Aloe,Volkov2019,Volkov2018} and aboveground through physical contact or conductive pathways between plant tissues \cite{Szechynska-Hebda2022}. Both intraspecific and interspecific electrical interactions have been observed \cite{Szechynska-Hebda2022,Volkov2019}.

Plant electrical signaling relies on three distinct signal types: action potentials (APs), variation potentials (VPs), and system potentials (SPs), which differ in their generation mechanism, amplitude characteristics, and propagation behavior \cite{Vodeneev2015,Zimmermann2009}. A detailed comparison is provided in Table~\ref{tab:plant_signals}.

APs are triggered by non-damaging stimuli such as cooling or gentle touch. They arise from \(\mathrm{Ca}^{2+}\), \(\mathrm{Cl}^{-}\), and \(\mathrm{K}^{+}\) fluxes and changes in plasma membrane \(\mathrm{H}^{+}\)-ATPase activity \cite{Vodeneev2015,Vodeneev2016}. Following an all-or-nothing principle, their amplitude remains constant once a voltage threshold is exceeded, regardless of stimulus intensity \cite{Zimmermann2009}. VPs result from harmful stimuli such as heating or crushing. They are generated by transient \(\mathrm{H}^{+}\)-ATPase inactivation combined with passive ion fluxes, and their amplitude scales with stimulus strength \cite{Vodeneev2016}. VPs can traverse non-living tissues via passive convective diffusion and turbulent water flow through xylem vessels, where both wound signals and hydraulic waves contribute to propagation \cite{Vodeneev2016}. SPs are induced by mild chemical or physical perturbations, are associated with \(\mathrm{H}^{+}\)-ATPase activation, and their amplitude depends on both stimulus intensity and type \cite{Zimmermann2009,Vodeneev2015}. APs and SPs propagate through active ion transport, while VPs rely on passive mechanisms \cite{Zimmermann2009}. From an ICT perspective, APs resemble binary digital signals with a fixed amplitude and threshold-based triggering, whereas VPs and SPs carry graded, analog information whose amplitude encodes stimulus properties. Despite their differences, all three signal types contribute to increasing plant resistance to environmental perturbations \cite{Vodeneev2015}.

Beyond plant-to-plant signaling, electrical communication also mediates plant–animal interactions. Flowers of many species maintain a negative surface charge arising from the atmospheric potential gradient, while foraging bees carry a positive charge. Bees detect floral electric fields through mechanosensory hairs, allowing pre-contact assessment of flower profitability \cite{Clarke2017}. When a bee lands on a flower, it transiently alters the flower's charge distribution, leaving an electrostatic signal detectable by subsequent foragers that may communicate nectar depletion status \cite{Clarke2017}. This electrostatic signaling represents a distinct channel through which plants communicate with animal partners beyond chemical cues.

Mycorrhizal networks also serve as a medium for electrical signal propagation. Electropotential waves—including action, variation, and system potentials—have been proposed as mechanisms for rapid signal transmission along fungal hyphae across tissue boundaries \cite{oelmuller2019interplant}. Root-generated electrical signals may propagate through the hyphal network to connected plants, enabling coordinated responses across individuals linked by the same fungal symbiont. This pathway complements the chemical signaling role of mycorrhizal networks described in the preceding subsection, with the two mechanisms potentially operating in parallel to convey different types of information across shared fungal infrastructure.

In receiver plants, electrical signal uptake proceeds through propagation of bioelectrical potentials across tissue interfaces. The propagation mechanisms of APs and SPs involve active ion transport, while VPs depend on hydraulic and passive mechanisms, as described above. Together, these signals enable rapid, long-distance information transfer within and between plants that chemical signals alone cannot provide.

\section{ICT Modeling and Analysis of Plant Communication Systems}

Plants can be regarded as communicating nodes within a multimodal network, where communication occurs through distinct physical channels for signaling. In this section, plant communication is examined through three major modalities: chemical, mechanical, and electrical communication. These modalities are governed by different physical principles, yet together they contribute to a broader system of information exchange. This section consolidates existing knowledge and develops a theory-oriented framework for modeling and analysis across these channels.

Approaching plant communication from the perspective of information and communication technology enables the abstraction of the process into transmitter, channel, and receiver output blocks. In this formulation, biological signaling can be directly connected to established theories of information transfer, while attention is drawn to the specific constraints of plant systems such as slow propagation, nonlinear responses, and strong dependence on environmental conditions. Within this framework, the modalities are examined through their underlying processes, mathematical formulations, and representative channel models reported in the literature, together with the assumptions embedded in these models and the conditions under which they remain valid. In general, the discussion emphasizes both simple models that clarify the dominant mechanisms and more recent models that incorporate biological constraints in a more detailed manner.

\begin{figure*}[t]
    \centering
    \includegraphics[width=0.85\linewidth, height=0.55\linewidth]{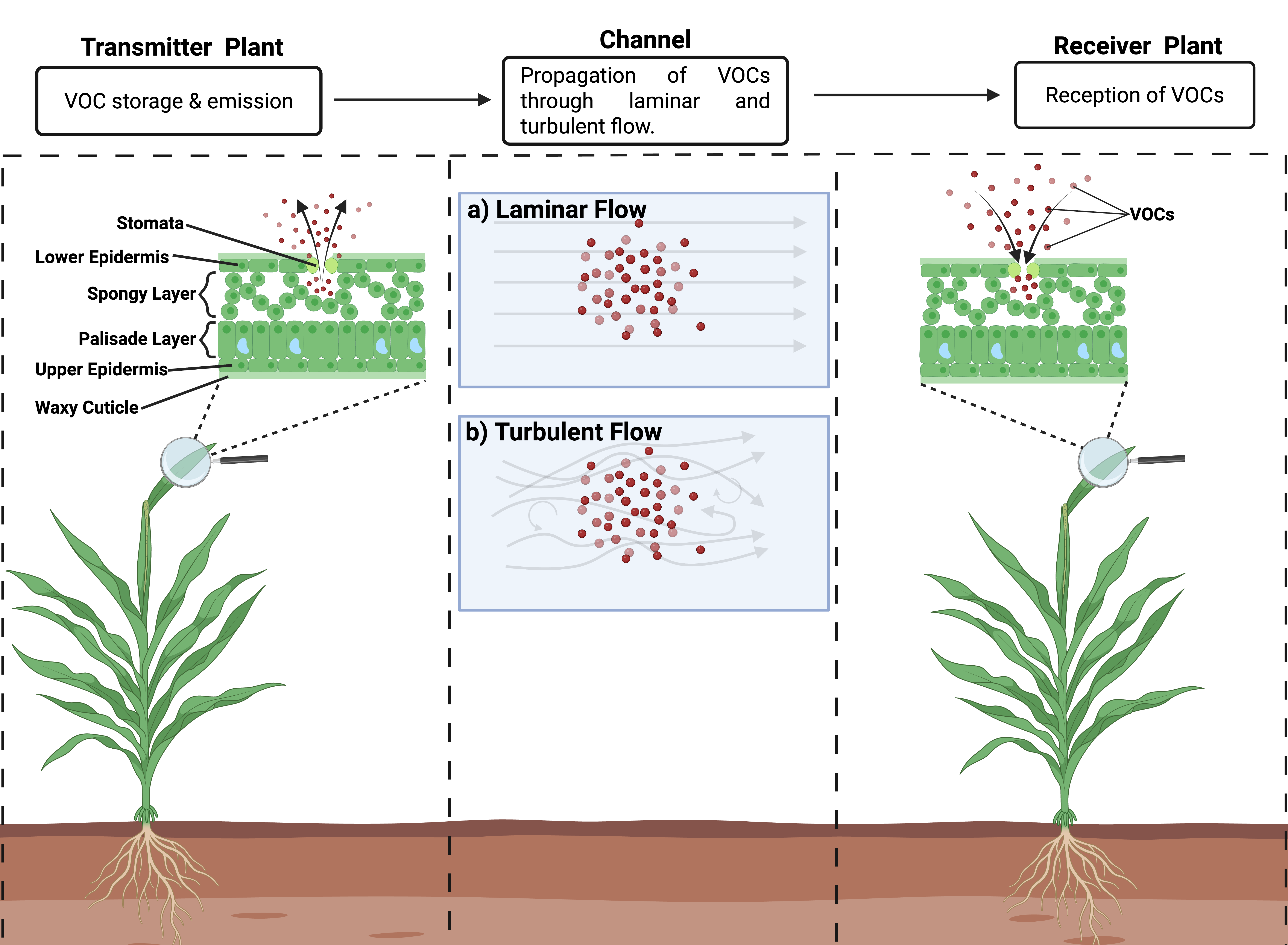}
    \caption{Illustration of aboveground chemical communication. The process involves VOC storage and emission in the transmitter plant, VOC transport through laminar or turbulent flow, and reception and uptake in the receiver plant that triggers a physiological response. Created with BioRender.com.}
    \label{fig:abvchemcom}
\end{figure*}

\subsection{Aboveground Chemical Communication}
\label{sec:abvmodl}
From an ICT perspective, aboveground chemical communication is a complete link composed of transmitter encoding, channel propagation, and receiver decoding. Each biological process maps to a communication-theoretic block and supports the analysis of plant-to-plant VOC signaling using established tools. An illustration of this process is shown in Fig. \ref{fig:abvchemcom}. Recently, a growing number of studies have begun considering end-to-end models of emission, propagation, and absorption  \cite{unluturk2016end, kilic2024endtoendmathematicalmodelingstress, maitra2026modelinganalysisvocbasedinterplant, merdan2026odorcommunicationgreenleaf}. This section presents transmitter, channel, and receiver models, along with end-to-end formulations, and highlights the implications for information transfer, interference, and noise.

\subsubsection{Transmitter Models}

To model aboveground chemical communication, the foundational elements of the transmitter plant should first be understood through theoretical and practical models proposed for odor-based molecular communication systems. This framework specifies three transmitter components: a storage unit, a delivery mechanism, and a controller \cite{kilic2024endtoendmathematicalmodelingstress}. In plants, VOCs reside in specific and nonspecific storage compartments and can be present in aqueous, lipid, and gas phases \cite{niinemets2004physiological}. Delivery to the air occurs through cuticles or stomata. Herbivory-induced wounding can also release VOCs \cite{midzi2022stress}, but this pathway is typically excluded when modeling controllable emissions. Delivery can occur by diffusion, thermal facilitation, or active pumping \cite{aktas2023odorbasedmolecularcommunicationsstateoftheart}; cuticular release is driven by concentration gradients, whereas stomatal release reflects guard cell control. Controller behavior is realized through cellular regulation. Stress inputs such as herbivory, drought, heat, and salinity act as information sources \cite{kilic2024endtoendmathematicalmodelingstress} that modulate gene expression and metabolic pathways \cite{dudareva2013, midzi2022stress}. Stress $s(t)$ is mapped to production and emission of VOCs by using nonlinear transcriptional regulation. For this purpose, VOC production rate $I(t)$ is written as follows
\begin{equation}
\label{eq:stressprod}
I(t) = \frac{\nu_{\max}}{1 + \exp(-w s(t) + c)} - k_d\, g(t),
\end{equation}
where $\nu_{\max}$ is the maximum transcription rate, $w$ is a regulation constant, $c$ captures transcriptional delay, $k_d$ is the degradation rate, and $g(t)$ represents degradation dynamics \cite{vu2007nonlinear}. The emitted message that encodes stress severity in amplitude and duration in symbol length can be written as
\begin{equation}
m(t) = \big(H(t-\tau_b) - H(t-\tau_e)\big) \int_{\tau_b}^{t} I(t') \, dt',
\end{equation}
where $H(\cdot)$ is the Heaviside step function, $\tau_b$ and $\tau_e$ mark onset and termination of emission, and $I(t')$ is defined in \eqref{eq:stressprod}. This is analogous to amplitude shift keying or pulse duration modulation \cite{kilic2024endtoendmathematicalmodelingstress}.

Another strategy employs a compartmental transmitter that partitions VOCs into aqueous, lipid, and gas pools, expressed as
\begin{align}
\frac{dS_a}{dt} &= \eta\, f(t) - k_a S_a, \\
\frac{dS_l}{dt} &= \big(1-\eta\big) f(t) - k_l S_l, \\
\frac{dS_g}{dt} &= k_a S_a + k_l S_l - k_g S_g,
\end{align}
where the emitted flux is given by
\begin{equation}
q(t) = k_g S_g(t),
\end{equation}
where $S_a$, $S_l$, and $S_g$ are aqueous, lipid, and gas pool concentrations, $f(t)$ is the production rate, $\eta$ is the aqueous partition fraction, $k_a$, $k_l$, and $k_g$ are transfer constants, and $q(t)$ is the emitted flux \cite{unluturk2016end, harley2013roles}. This structure acts like a filter bank: some pathways yield fast release, others produce delayed output from storage. The first model provides a simpler formulation, whereas the second model is more biologically realistic. The choice between them depends on whether the aim is to preserve analytical tractability or to represent biological details more accurately. Between these two levels of model complexity, additional stress emission models are  \cite{harley2013roles, grote2013leaf, grote2019new, guenther1995global, guenther2012model, lerdau1997plant}.

In transmitter plants, information can be encoded either in the concentration of a single VOC or in the relative ratios of compounds within a blend. These correspond to Concentration Shift Keying (CSK) and Ratio Shift Keying (RSK) in molecular communication \cite{ninkovic2021plant, kilic2024endtoendmathematicalmodelingstress, cskmsk, Kilic2024MRSK}. The choice of modulation is shaped by transmitter dynamics and environmental variability and, in turn, determines the symbol alphabet and detection strategy. Also, emission can be instantaneous or continuous. Instantaneous sources simplify theory \cite{Silva2013}, whereas continuous sources better reflect stress emissions that can persist for hours or days \cite{kilic2024endtoendmathematicalmodelingstress}.

\subsubsection{Channel Models}
After emission, VOCs propagate in the air via wind advection and molecular diffusion. This system is governed by a diffusion-advection equation written as
\begin{equation}
\label{eq:pde_c}
\begin{split}
\frac{\partial c(\mathbf{x}, t)}{\partial t}
&= D \nabla^{2} c(\mathbf{x}, t)
   - \mathbf{v}(t) \cdot \nabla c(\mathbf{x}, t)
   + S(\mathbf{x}, t),
\end{split}
\end{equation}
\noindent where $c(\mathbf{x},t)$ denotes the concentration of the particle at position \(\mathbf{x}\) and time \(t\), $D$ is the diffusion coefficient, $\mathbf{v}(t)$ is the wind velocity vector, $\nabla c(\mathbf{x},t)$ and $\nabla^{2} c(\mathbf{x},t)$ represent the spatial gradient and Laplacian of the concentration, respectively, and $S(\mathbf{x},t)$ denotes the external source term. There are many different approaches to solve this equation \cite{Guenneau2015, unluturk2016end, zoppouspatial, kumar2010analytical, sun2022approximate, sanskrityayn2021generalized}. In plant communication, the advection is due to the wind. Theoretically, the deterministic Navier-Stokes equations govern the flow behaviour. However,  the wind is generally taken as a random process for modeling purposes due to its chaotic behaviour. Therefore, using time-varying random advection models is the most appropriate for the air medium \cite{merdan2026airborneparticlecommunicationtimevarying}.

The solution for \eqref{eq:pde_c} yields the channel impulse response of the communication system. Each parameter in the advection–diffusion model has a direct communication-theoretic interpretation. Physically, the diffusion coefficient $D$ and the variance of the flow govern how molecules spread over time, broadening the plume. In ICT terms, this corresponds to channel dispersion, where an impulse response is smeared, as in multipath wireless or dispersive optical fibers. The mean wind vector shifts the entire plume downwind, displacing the peak concentration. From a communications perspective, this resembles modulation of one signal into another for better transport. Here, particles are modulated into the wind, which increases the achievable communication distance.

\subsubsection{Receiver Models}
Leaves serve as receivers that take up VOCs through stomata and cuticles. Several modeling approaches can be used for reception at leaves \cite{rein2011new, paterson1994model, hung1997uptake, trapp2007fruit, trapp2022generic, paterson1991fugacity}. The appropriate choice depends on study objectives, exposure duration, and environmental complexity. The simplest model for leaf uptake is the Robin (mixed) boundary condition, expressed as
\begin{equation}
- D \frac{\partial c}{\partial n} = k_a \big(c - c_{\text{int}}\big),
\end{equation}
where $\partial/\partial n$ is the outward normal derivative, $c$ is the ambient concentration, $c_{\text{int}}$ is the internal concentration, and $k_a$ is an effective mass transfer coefficient \cite{nye1977solute, barber1984bioavailability}. The limits $k_a \to \infty$ and $k_a = 0$ correspond to a perfect sink and a reflective boundary, respectively, with intermediate behavior summarized by the Sherwood number $\mathrm{Sh} = k_a L / D$, where $L$ is a characteristic leaf length and $D$ is the diffusion coefficient. From an ICT perspective, $k_a$ acts as a tunable receiver sensitivity that governs coupling between the signal and the receiver. Large $k_a$ corresponds to strong coupling and efficient capture, while small $k_a$ resembles a mismatched antenna with poor coupling. The Robin condition parallels impedance matching at the interface, and $\mathrm{Sh}$ provides a dimensionless measure of capture efficiency analogous to aperture efficiency in radio frequency (RF) systems. Because uptake both acquires signal and reduces the local concentration, absorptive receivers also shape the near-surface field, a dual role comparable to receiver filtering.

Recently, a more comprehensive absorption model has been introduced in \cite{An_Absorption_Model_of_Volatile_Organic_Compound_by_Plant_Leaf_The_Most_Influential_Site_in_the_Absorption_Pathway}. Considering air-liquid partitioning, partitioning into the plasma membrane, and metabolic conversion of the VOC in plant cells, the VOC absorption is modeled as
\begin{equation}
\label{eq:VOC_absorption}
\begin{split}
(\frac{r_{ias}}{F} + \frac{1}{\left(\frac{1}{r^G} + \frac{E}{2}\right)} + \frac{10^{3} \, r_{liq} }{H P_{a}}) A =
\frac{\left(\frac{1}{r^G} - \frac{E}{2}\right)}{\left(\frac{1}{r^G} + \frac{E}{2}\right)} C_a - \frac{10^{3}}{H P_{a}} C_{ct},
\end{split}
\end{equation}
\noindent where \(r_{ias} = \frac{\Delta L_{ias} \tau}{D_A f_{ias}}\) is the intercellular airspace resistance, \(\Delta L_{ias}\) is the effective diffusion path length in intercellular air spaces, \(\tau\) is the diffusion path tortuosity, \(D_A\) is the diffusion coefficient of VOC in gas phase and \(f_{ias}\) is the fraction of the intercellular air space volume in the total leaf volume. $F = \frac{273.15}{T_k \times 22.4e-3}$ is a scaling constant where $T_k$ is the temperature in $kelvin$. \(r_G = r_s + r_b\) is the total resistance from outside air to intercellular air spaces. \(r_b = r_b^{w} (\frac{D_w}{D_A})^\frac{2}{3}\) is the resistance of the leaf boundary layer, where \(r_b^{w}\) is the resistance of the leaf boundary layer for water vapor and \(D_w\) is the diffusion coefficient of water in the gas phase. Similarly, \(r_s = r_s^{w} (\frac{D_w}{D_A})\) is the stomatal resistance and \(r_s^{w}\) is the stomatal resistance for water vapor. $E$ is the VOC specific transpiration rate, $H$ is the VOC specific Henry's constant. $P_a$ is the pressure in $bar$ and \(r_{liq}\) is the sum of the resistances of the cell wall, plasma membrane, and cytosol. $A$ is the odor molecule absorption rate in \(mol \, m^{-2} s^{-1}\), \(C_a\) is the odor concentration in the atmosphere in \(ppb\) and \(C_{ct}\) is the odor concentration in cytosol in \(mol \, m^{-3}\).

\subsubsection{ISI and Channel Memory}

The diffusion-advection channel can be classified as diffusion or advection dominated depending on its P\'eclet number \cite{merdan2026airborneparticlecommunicationtimevarying}. For diffusion-dominated channels, a VOC pulse produces a sharp rise followed by a long decay. This slow clearance means that the channel has a very long memory, and even long symbol durations cannot eliminate overlap between emissions. In communication terms, this can cause severe intersymbol interference (ISI). On the other hand, for the advection-dominated systems, the channel memory shortens. In this case, the strong advection washes out the remaining particles further away from the receiver. In general, the wind can become both diffusion or advection dominated. Therefore, channel memory is a dynamic property, and any realistic analysis of plant signaling must account for this behavior.

\subsubsection{Noise and Multi-User Interference}

In aboveground chemical communication, randomness arises from thermal motion, turbulence that induces colored fluctuations, and emissions from multiple plants that create overlapping plumes. Laboratory sensors may register behavior that closely approximates additive white Gaussian noise (AWGN) after filtering, whereas natural environments are typically dominated by noise that is non-white and non-stationary. From an ICT perspective, the received signal combines diffusion noise, turbulence, and background VOCs. This parallels fading wireless channels with co-channel interference: noise is environment-dependent, and multiple transmitters compete in the same band. Hence, robust reception requires strategies such as diversity techniques, multi-user detection, or interference cancellation. In studies addressing aboveground chemical communication, it should be kept in mind that realistic analyses must incorporate both the stochastic and multi-user nature of the environment. Ignoring these aspects risks oversimplifying the channel and overlooking challenges that are central to biological interpretation as well as ICT system design.

\begin{figure*}[t]
    \centering
    \includegraphics[width=0.8\linewidth, height=0.4\linewidth]{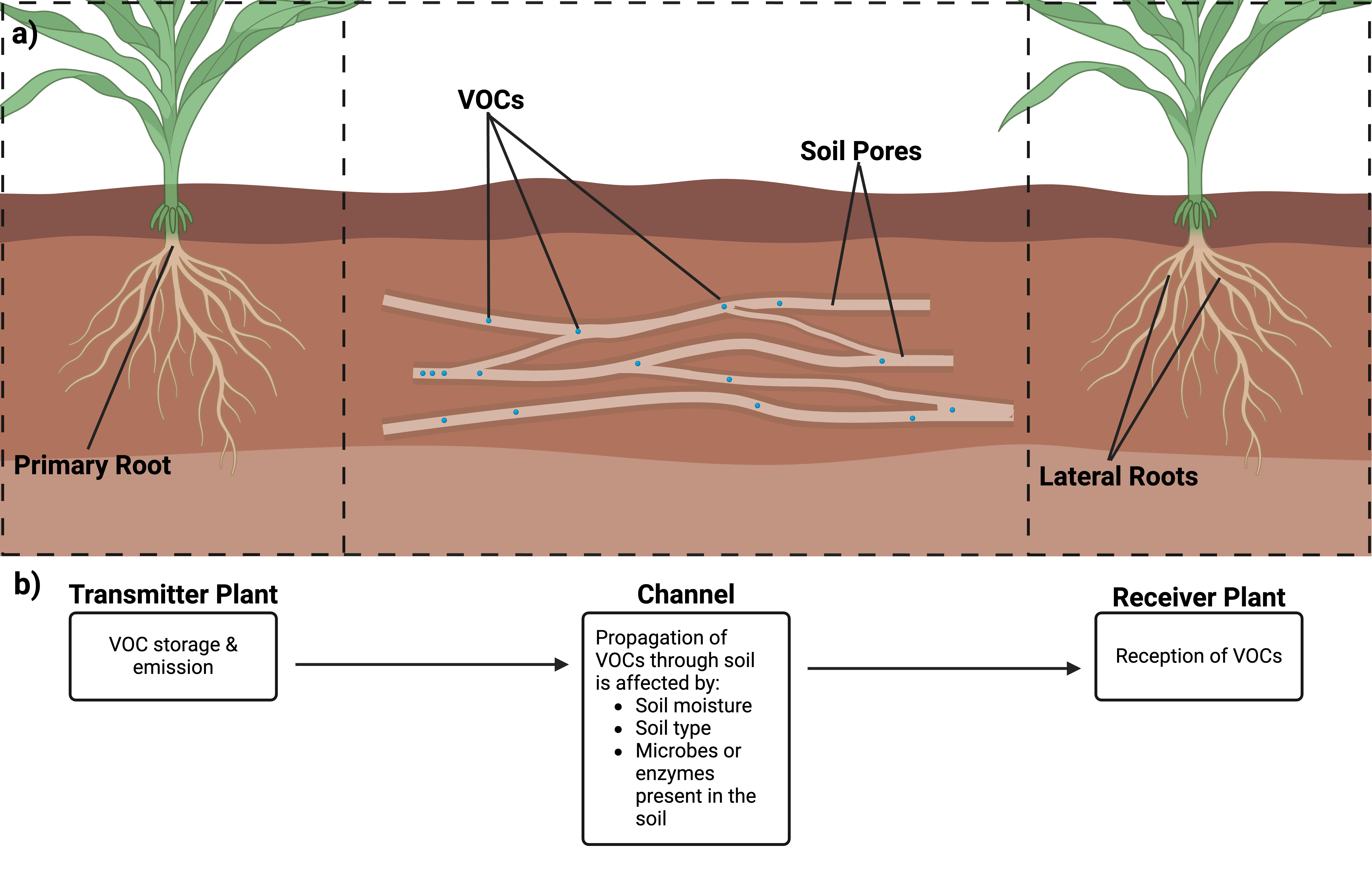}
    \caption{Modeling of belowground chemical communication: a) overview of the system, b) ICT-based modeling. The link comprises root emission (transmitter), dual-phase transport through porous soil (channel), and root uptake (receiver). Created with BioRender.com.}
    \label{fig:blwchemmdl}
\end{figure*}

\subsection{Belowground Chemical Communication}
From an ICT perspective, soil VOC and root exudate communication is a complete underground chemical link composed of transmitter emission at roots, propagation through porous media, and receiver uptake by neighboring fungi, roots, or bacteria. Dual-phase transport is central: compounds partition between soil air and soil water and move through both pathways at different rates while also interacting with soil particles. Each biological process maps onto communication-theoretic blocks that support analysis of belowground chemical signaling with established ICT tools. An overview of the system, along with ICT-based modeling, can be found in Fig. \ref{fig:blwchemmdl}. While many studies model individual components, comprehensive end-to-end formulations are rare for belowground communication compared to aboveground systems.

\subsubsection{Transmitter Models}

As in the aboveground case, the transmitter framework of storage unit, delivery mechanism, and controller \cite{kilic2024endtoendmathematicalmodelingstress} provides a useful structure for interpreting root emission, though the underlying biology and operating constraints differ substantially. Definitive storage compartments in plant roots have not been fully characterized; VOCs are believed to reside both within root tissue and at the outer cell layer \cite{lin2007volatile}. Delivery can occur through passive diffusion driven by concentration gradients, or through specialized transport proteins in root cell membranes as an active pumping mechanism \cite{wang2021root}. Controller behavior is realized through biochemical regulation: plants regulate gene expression in response to a range of internal and external cues, including nutrient status, developmental state, and biotic interactions, which triggers metabolite production through internal signal transduction pathways involving terpenoid, fatty acid derivative, and sulfur-compound biosynthesis \cite{dudareva2013, delory2016root, midzi2022stress}. Unlike aboveground emission into free air, root exudates are released into a dense, resistive soil medium, limiting communication range and primarily targeting rhizosphere organisms such as bacteria, fungi, and neighboring root systems \cite{THE_ROLE_OF_ROOT_EXUDATES_IN_RHIZOSPHERE_INTERACTIONS_WITH_PLANTS_AND_OTHER_ORGANISMS, Root_Exudates_Mediate_Kin_Recognition_in_Plants}.

Root emission can be represented by adapting the Neumann boundary condition used in soil science nutrient uptake models \cite{nye1977solute, barber1984bioavailability}. Applied here to the emission direction, the root surface $\Gamma_r$ acts as a source boundary in the soil domain, with the outward flux $q_{\mathrm{tx}}(t)$ serving as the channel input:
\begin{equation}
- D_w \frac{\partial c_w}{\partial n}\Big|_{\Gamma_r} = q_{\mathrm{tx}}(t),
\end{equation}
where $c_w$ is the solute concentration in soil water, $D_w$ is the aqueous diffusivity, and $\partial/\partial n$ is the outward normal derivative. The emission flux $q_{\mathrm{tx}}(t)$ can be expressed using a Michaelis–Menten type relation, adapted from root surface kinetics \cite{nye1977solute, barber1984bioavailability, darrah1991models}, where the internal physiological state $u(t)$ serves as the control variable:
\begin{equation}
q_{\mathrm{tx}}(t) = q_0 + \frac{q_{\max}\, u(t)}{K_s + u(t)},
\end{equation}
where $q_0$ is the basal secretion rate, $q_{\max}$ is the maximum inducible flux, $K_s$ is the half-activation constant, and $u(t)$ represents the internal physiological state driving secretion. In the original literature, this kinetic form describes concentration-dependent uptake at the root surface; here it is adapted to model emission driven by an internal state. The nonlinear form ensures that secretion saturates at high $u(t)$, analogous to a power-limited transmitter with bounded dynamic range.

For triggered emissions, such as those induced by biotic interactions, nutrient signals, or rapid physiological changes, the release profile is often approximated by an exponentially decaying pulse, consistent with pulse inputs in soil transport models \cite{jury2004soil} and transmitter formulations in molecular communication \cite{farsad2016}:
\begin{equation}
q_{\mathrm{pulse}}(t) = q_0 + A\, H(t-\tau_b)\, e^{-(t-\tau_b)/\tau_{\mathrm{rel}}},
\end{equation}
where $A$ is the pulse amplitude, $H(\cdot)$ is the Heaviside step function enforcing onset at $\tau_b$, and $\tau_{\mathrm{rel}}$ is the characteristic release time constant. In ICT terms, this profile resembles a pulse-amplitude modulated symbol whose duration is governed by the plant's biochemical response dynamics.

A third approach models secretion through a surface exudate pool $E(t)$ that accumulates internally and is gradually released, consistent with carbon pool dynamics described in rhizosphere models \cite{darrah1991models, roose2016rhizosphere}:
\begin{align}
q_{\mathrm{pool}}(t) &= k_{\mathrm{rel}}\,E(t), \\
\frac{dE}{dt} &= P(u(t)) - k_{\mathrm{rel}}\,E(t) - k_{\mathrm{met}}\,E(t),
\end{align}
where $P(u)$ is the production rate driven by the internal physiological state $u(t)$, $k_{\mathrm{rel}}$ is the release constant, and $k_{\mathrm{met}}$ is the metabolic consumption rate. The pool $E(t)$ acts as an internal buffer that shapes the emission profile before it enters the soil, analogous to a low-pass transmit filter that limits signal bandwidth.

Together, these three models span the operating range of a belowground chemical transmitter. The Michaelis–Menten formulation captures saturating, state-proportional secretion, analogous to a transmitter amplifier with limited gain. The exponential pulse model describes triggered emission bursts that decay in time, similar to a symbol generator producing short pulses. The surface pool model represents buffered, filtered release from an internal store. In all cases, the resulting flux serves as the channel input to the soil medium. Regarding signal encoding, root emissions consist of blends of compounds rather than single molecules \cite{ninkovic2021plant, delory2016root}, so information can be encoded in both absolute concentrations and relative blend ratios, analogous to CSK and RSK in MC \cite{Kilic2024MRSK}.

\subsubsection{Channel Models}
To model belowground chemical communication, the propagation of VOCs through the soil must be considered. VOCs are released from the roots of the transmitter plants and travel to the receiver through small openings in the soil called pores \cite{horn1994}. This movement occurs due to the diffusion of VOCs from areas of high concentration to areas of low concentration \cite{soildiff}. The physical characteristics of the soil structure influence the dispersion of VOCs \cite{soildiff}, including factors such as soil moisture \cite{eden2012}, soil type \cite{soilpore}, and the presence of microbes or enzymes in the soil \cite{thomsen1999}.

The literature provides various models that relate VOC diffusivity to the physical characteristics of the soil. For example, the effect of soil type on VOC diffusivity is modeled in \cite{hamamoto2012organic}. The influence of root characteristics on soil structure and gas transport in the subsoil is explored in \cite{uteau2013root}. VOC diffusivity in undisturbed soil is modeled in \cite{moldrup2000predicting,moldrup2004three}, while the impact of soil moisture on VOC diffusivity is examined in \cite{moldrup2007predictive}. The effect of microbes or enzymes in the soil on VOC diffusivity is discussed in \cite{soilmicrobialdiff}. In addition to VOC diffusivity, other models describe the diffusion of VOCs through soil \cite{soilarands,Wang2024,mendozadiffmodel,sleepdiffmodel}. Beyond the dependence of VOC diffusivity on the physical characteristics of soil, a general porous-media balance for a volatile signal can be expressed in the form of a diffusion-advection-reaction equation as 
\begin{equation}
\theta_a \frac{\partial c_a}{\partial t} + \theta_w \frac{\partial c_w}{\partial t} = D_{\mathrm{eff}} \nabla^2 c_a - \mathbf{v}\cdot\nabla c_a - R(c_a,c_w),
\end{equation}
where $c_a(\mathbf{r},t)$ and $c_w(\mathbf{r},t)$ are the concentrations in air and water phases, $\theta_a$ and $\theta_w$ are volumetric fractions, $\mathbf{v}$ is bulk soil-gas velocity, $D_{\mathrm{eff}}$ is the effective diffusivity in the pore network, and $R(c_a,c_w)$ represents the loss rate \cite{jury2004soil}. A single effective equation is often used under the assumption of instantaneous partitioning, with $D_{\mathrm{eff}}$ chosen to represent soil structure and phase exchange. Compared with the aboveground chemical communication, here the channel memory is much stronger, and extreme fading is observed with increasing distance. Therefore, the range of communication is very limited in this topology. Plants mostly utilize belowground chemical signaling for communicating with the surrounding bacteria and fungi networks \cite{THE_ROLE_OF_ROOT_EXUDATES_IN_RHIZOSPHERE_INTERACTIONS_WITH_PLANTS_AND_OTHER_ORGANISMS}. Moreover, such signaling has been implicated in kin recognition \cite{Root_Exudates_Mediate_Kin_Recognition_in_Plants}. From an ICT perspective, this model highlights that soil acts as a channel with severe limitations compared to free-air propagation. The dual-phase pathway resembles a parallel channel with two sub-links: one slow but persistent (liquid phase) and one faster but constrained by pore structure (gas phase). Partitioning and adsorption introduce long memory, analogous to channels with heavy delay spread. Finally, $D_{\mathrm{eff}}$ and $R(c_a,c_w)$ act as effective channel parameters, but unlike engineered systems, they vary with soil moisture, porosity, and microbial activity, making the soil channel strongly time-varying. In ICT terms, this is equivalent to a channel that combines high attenuation, long delay spread, and slow fading.

A sudden root emission, such as a pulse of a semi-volatile allelochemical, produces a dispersed and delayed concentration response at some distance, a pattern well known in soil science as a breakthrough curve \cite{leij1991breakthrough}. When bulk flow is negligible, the impulse response is similar to the free-air Green’s function but with the free-air diffusivity replaced by the much smaller soil effective diffusivity, and with further spreading caused by phase partitioning. Sorption and desorption at soil particle surfaces create an additional long tail, often summarized by a retardation factor that shifts the peak later and prolongs its persistence \cite{hillel1998environmental}. In practice, underground pulses can persist for many hours or even days. For example, measurements of methyl ethyl ketone diffusion through dry kaolinitic soils indicate that such semi-volatile compounds may spread only a few centimeters while still remaining detectable over long time scales \cite{Itakura2003}. From an ICT perspective, the soil impulse response has an extremely long memory. Dual-phase diffusion and sorption act like long delay lines that smear each symbol into the next, creating severe ISI. Unlike wireless channels, where ISI can often be cleared with a guard interval, in soil, the residual never fully disappears because the heavy tail continues for very long times. This makes conventional symbol-by-symbol signaling impractical. Instead, communication in soil is closer to a quasi-static channel where signals are interpreted as persistent changes in baseline concentration, or to an event-driven system where the timing of rare large emissions carries the information. In other words, the soil medium transforms fast inputs into slow, lingering outputs, forcing plant communication to operate with very low symbol rates or to exploit alternative coding strategies.

\subsubsection{Receiver Models}
To model belowground chemical communication, VOC uptake by the receiver’s roots must be included. Both dynamic \cite{rein2011new,fantke2011plant,fantke2013dynamics} and steady-state \cite{trapp2003fruit,trapp2007fruit,trapp2022generic,paterson1991fugacity} models have been proposed for modeling VOC uptake from the roots. When choosing between steady-state and dynamic models, it is important to consider the specific goals of the study, the time frame of interest, and the complexity of environmental interactions. While steady-state models simplify analysis by focusing on equilibrium conditions, dynamic models are essential for capturing transient effects and time-dependent behaviors. A standard representation, adapted from soil physics and root uptake models \cite{barber1984bioavailability}, is a linear uptake boundary condition expressed as
\begin{equation}
- D_w \frac{\partial c_w}{\partial n} = k_{a,\mathrm{root}} \big(c_w - c_{\mathrm{int}}\big),
\end{equation}
where $D_w$ is the aqueous diffusivity, $k_{a,\mathrm{root}}$ is the effective root uptake coefficient, $c_{\mathrm{int}}$ is the internal concentration inside the root, and $\partial/\partial n$ denotes the outward normal derivative. This is mathematically identical to the Robin-type boundary condition used earlier to describe VOC uptake at leaf surfaces, underscoring that both above- and below-ground receivers can be modeled within the same framework of partial absorption at the boundary. At higher concentrations, uptake becomes transporter-limited and follows Michaelis--Menten kinetics \cite{nye1977solute,barber1984bioavailability,darrah1991models}, which can be written as
\begin{equation}
J = \frac{J_{\max} \, c_w}{K_m + c_w},
\end{equation}
where $J$ is the uptake flux, $J_{\max}$ is the maximum flux at saturation, and $K_m$ is the half-saturation constant. At low $c_w$, uptake is approximately linear ($J \approx k_c c_w$), whereas at high $c_w$ the flux saturates.

From an ICT perspective, the root operates as a nonlinear front end. At low concentrations, the response is linear, similar to an ideal matched receiver that scales with input amplitude. At higher concentrations, the response saturates, analogous to a receiver with a limited dynamic range or an amplifier that clips at strong input signals. This nonlinearity protects the biological system from overload but also distorts signal amplitudes, reducing the ability to distinguish between strong stimuli. In communications terms, this resembles a receiver that is sensitive in the low signal-to-noise ratio (SNR) regime but compresses information at high SNR, forcing the system to rely more on detection of signal presence or timing rather than precise amplitude coding.

\subsubsection{ISI and Channel Memory}
Biologically, diffusion through tortuous pores, partitioning between air and water phases, and sorption–desorption on soil particles all contribute to long concentration tails. A single emission can therefore perturb the background concentration for hours or even days \cite{hillel1998environmental,Itakura2003}, making the soil environment a channel with slow clearance and strong memory effects. As discussed in the channel models section, from an ICT perspective, the soil channel exhibits very long memory and consequently severe ISI. As a result, successive signals inevitably overlap regardless of how long the symbol interval is. Realistic analyses of underground communication must therefore account for this persistent memory. Possible strategies include selecting molecules that degrade more rapidly, which shortens the impulse response at the expense of weaker signals, or applying ICT-inspired techniques such as baseline subtraction, error-control coding, or equalization methods designed for channels with extended delay spread.

\subsubsection{Noise and Multi-User Interference}
In belowground chemical communication, background organic compounds from microbes and detritus create a structured noise floor. Slow environmental variations in temperature and moisture induce baseline wander, while microbial populations may also consume or produce the same molecules used for signaling, confounding detection \cite{rivett2011unsaturated,Itakura2003}. Temporal averaging attenuates short-term fluctuations but cannot remove low-frequency drift. This is closely parallel to aboveground communication, where turbulence and overlapping plumes create non-white, non-stationary noise. In both cases, the effective noise is colored and time-varying rather than white and stationary, and multiple emitters compete in the same chemical “band,” creating multi-user interference. The difference lies in the dominant sources: in the air, turbulence drives variability, while in the soil, it is environmental cycles and microbial metabolism that control the baseline. From an ICT perspective, soil reception thus faces the same core challenges as aboveground communication but under slower dynamics. The received signal is corrupted by both stochastic variability and interference, analogous to fading channels with co-channel interference in RF. Robust reception requires ICT-inspired tools such as adaptive baselining to track drift, interference cancellation to suppress overlapping sources, or multi-user detection tailored to shared chemical pathways. The parallel with aboveground communication underscores that noise and interference are universal features of chemical channels, though their physical origins differ across environments.

\subsection{Mechanical Communication}

Mechanical communication encompasses information transfer mediated by physical forces and vibrations, including acoustic waves, touch-induced signals, and mechanically generated vibrations arising from interactions with the environment or other organisms. Unlike chemical signaling, where molecules serve as information carriers, here information is conveyed through disturbances propagating through air, soil, or plant tissues. From an ICT perspective, this pathway can be analyzed through transmitter, channel, and receiver blocks, where mechanical stimuli are generated, propagated through a physical medium, and detected by mechanosensitive structures in the receiver.

\subsubsection{Transmitter Models}
Transmitter models in mechanical communication describe how external or internal stimuli are converted into mechanically propagating signals. Information sources may include herbivore-induced vibrations, pollinator-generated sounds, touch stimuli, hydraulic fluctuations, or internally generated processes such as cavitation \cite{The_Behavioral_Ecology_of_Insect_Vibrational_Communication, Natural_and_LaserInduced_Cavitation_in_Corn_Stems_On_the_Mechanisms_of_Acoustic_Emissions, Flowers_Respond_to_Pollinator_Sound_within_Minutes_by_Increasing_Nectar_Sugar_Concentration, Caterpillar_Chewing_Vibrations_Cause_Changes_in_Plant_Hormones_and_Volatile_Emissions_in_Arabidopsis_Thaliana}. These inputs drive a mechanical emission process in which plant tissues, vascular structures, or contact interfaces act as sources of vibrations or force disturbances. From an ICT perspective, such transmitters may be abstracted as impulse sources, oscillatory sources, or stimulus-driven modulators, depending on whether signaling occurs through transient events, sustained vibrations, or structured mechanical patterns. The choice of transmitter model depends on the signaling submodality, the generating mechanism, and the degree to which the emitted mechanical waveform can encode information.

One of the main mechanical signals is the acoustic waves. However, it remains uncertain whether plants produce vibrations involuntarily or deliberately for communication purposes \cite{Demey2023}. Furthermore, the processes leading to sound production in plants are not fully understood. One known mechanism for sound emission is cavitation, which occurs when air bubbles form and collapse in the xylem during water transport, producing characteristic popping or clicking sounds \cite{Demey2023}, \cite{Khait2023,ponomarenko2014ultrasonic}. Although the exact mechanism for voluntary sound emission is still unknown, cavitation can explain the emission of involuntary sound waves \cite{Khait2023}. These emissions are impulsive and broadband, and can be treated as event-driven transmitters in which the presence of a pulse conveys information about plant stress. From an ICT perspective, this represents a stochastic encoding process where message content is limited to event timing rather than waveform modulation. Currently, several models for cavitation in plants and other systems have been explored and can be directly used for transmitter modeling \cite{Rockwell2014, Singhal2002, Peshkovsky2008, Vanhille2012, Roohi2019, Laborde1998, Zhu2024}. Other mechanical signals can be simply modeled as impulsive $F(t) = F_0 \delta(t-t_0)$ or oscillatory $F(t) = F_0 cos(\omega t + \phi)$.

\subsubsection{Channel Models}

After generation, mechanical signals propagate through a physical medium such as air, soil, water, or plant tissues. Signal transmission depends on material properties such as elasticity, density, damping, and boundary conditions, which together determine propagation speed, attenuation, reflection, and dispersion. From an ICT perspective, these effects define the channel impulse response and govern distortion, delay, and information transfer reliability.

In the literature, different modeling approaches have been used depending on the propagation medium and signaling scenario. For vibrations transmitted through plant structures, models have been developed to capture bending dynamics, structural resonances, and frequency-dependent transmission \cite{An_Analytical_Model_for_the_Propagation_of_Bending_Waves_on_a_Plant_Stem_Due_to_Vibration_of_an_Attached_Insect}. In addition, viscoelastic formulations have been used to represent attenuation and time-dependent dissipation in plant tissues \cite{Homogenization_of_a_Viscoelastic_Model_for_Plant_Cell_Wall_Biomechanics, Modeling_Mechanical_Behaviors_of_Plant_Stems_Undergoing_Microstructural_Changes}. However, for generality, a mass-spring-damper system can be used which is governed by \cite{Systems_Theory_An_Approach_to_MassDamperSpring_and_MassNondamperSpring_Systems, Calibration_of_MassSpringDamper_Equivalent_Systems_for_Real_Time_Assessment_of_the_Dynamics_of_Trees}
\begin{equation}
\label{eq:mass_spring_damper}
\begin{split}
m \ddot{x}(t) + c \dot{x}(t) + k x(t) = F(t),
\end{split}
\end{equation}
\noindent where $m$ is the mass, $c$ is the damping coefficient, $k$ is the spring constant, $F(t)$ is the externally impressed force, and $x(t)$ denotes the displacement of the mechanically vibrating or deforming plant structure from its equilibrium position. The mass–spring–damper model provides a general framework for representing a broad range of mechanical signal transmission processes. By varying the system parameters $m$, $c$, and $k$, different dynamic behaviors can be captured, including lightly damped oscillations, overdamped responses, transient impulses, and sustained vibrations. Likewise, different forms of the forcing term $F(t)$ can represent diverse mechanical stimuli, ranging from impulsive events to periodic or stochastic excitations. Through appropriate parameter selection, this framework can therefore describe many different mechanical signal transmission mechanisms within a unified model. While these approaches differ in formulation and assumptions, they collectively show that mechanical channels can exhibit attenuation, dispersion, reflections, and memory effects, all of which influence the reliability of information transfer. From an ICT perspective, attenuation limits communication range, dispersion broadens transmitted signals, and reflections may create multipath-like effects. Environmental vibrations and background mechanical disturbances may further act as noise sources, affecting detection at the receiver.

Sound wave propagation through a medium is also well-documented in the literature. The propagation of sound in the atmosphere, including effects like absorption, reflection, and scattering, is well-documented in \cite{Crocker1998}. Sound wave propagation through crops, shrubs, and trees is explored in \cite{Attenborough2014}, with methods for predicting propagation effects in these environments. For sound propagation through turbulent air near the ground, a numerical model is provided in \cite{Chevret1996}, while \cite{Daigle1986} addresses the impact of gradients and turbulence on sound propagation near the ground. The propagation of sound through trees, acting as scatterers, is discussed in \cite{Bullen1982}, and \cite{Chobeau2014} focuses on sound propagation within forests. Sound attenuation through trees is examined in \cite{Price1988}, while \cite{Peng1995} presents sound propagation models for agricultural environments. Additionally, \cite{Embleton1996} provides a tutorial on field measurements and simple physical interpretations of sound propagation in various settings. In soil, sound wave propagation is investigated in \cite{Albers2009, Wireless_Underground_Sensor_Communication_Using_Acoustic_Technology}. The attenuation and behavior of acoustic waves in farmland soil are modeled in \cite{Huang2022}.

\subsubsection{Receiver Models}

At the receiver, propagated mechanical disturbances must be detected and converted into a measurable signaling input. In plants, this reception is primarily mediated by mechanosensitive ion channels, which act as the receiving structures by responding to stress, strain, or membrane tension induced by incoming mechanical signals \cite{merdan2025acousticcommunicationmodelplants, Thigmomorphogenesis, Thigmomorphogenesis_a_complex_plant_response_to_mechano_stimulation}. When sufficiently activated, these channels undergo conformational changes that alter their permeability, enabling the incoming mechanical signal to be detected at the receiver. In this sense, mechanosensitive channels serve as the primary interface between the mechanical channel and signal reception \cite{Hamant2017}. After reception, the signal is transformed into secondary messengers like ion fluxes and reactive oxygen species (ROS) \cite{RodrigoMoreno2017}. Ion flux models from \cite{Shabala2006}, \cite{Watanabe1995}, and \cite{Gradmann2001} may be used to model mechanical signal absorption in plants.

In \cite{Feeling_Stressed_or_Strained_A_Biophysical_Model_for_Cell_Wall_Mechanosensing_in_Plants}, a mathematical framework is developed to describe how externally applied mechanical forces can be mapped into forces acting on membrane-associated mechanosensitive proteins. This is the first stage in the receiver for the mechanical communication paradigm. Secondly, mechanosensitive channels provide a tractable framework for formal modeling of mechanical detection at the cellular scale. Their gating can be represented by a standard two-state Boltzmann model, where the probability of being in the open state is given by 
\begin{equation}
P_{\mathrm{open}} = \frac{1}{1 + \exp\!\left(\frac{\Delta G_{\mathrm{tot}}}{k_B T}\right)},
\end{equation}
with $\Delta G_{\mathrm{tot}}$ the free energy difference between closed and open states, $k_B$ the Boltzmann constant, and $T$ the absolute temperature \cite{generalboltzmanref}. This formalism captures how mechanical perturbations shift channel energetics and thereby modulate ion fluxes. From an ICT perspective, this mechanism resembles a threshold-based detector: only when external mechanical input shifts the free energy sufficiently does the channel open, producing an ion flux that can be interpreted as a discrete signal event. In communication terms, mechanosensitive channels thus act as binary transducers, mapping continuous mechanical stimuli into digital-like electrical responses.

\subsubsection{ISI and Channel Memory}

Mechanical signals do not always vanish immediately after they are generated. Depending on the medium, vibrations may continue for some time due to residual oscillations, reflections, or slow relaxation. If a new symbol is transmitted while part of the previous mechanical disturbance is still present, the two signals may overlap and cause intersymbol interference. For this reason, the duration of these lingering effects influences channel memory and should be considered when modeling mechanical communication.

\subsubsection{Noise and Interference}
Mechanical communication is also affected by noise and interference arising from unwanted mechanical disturbances in the environment. Sources such as wind, rain, background substrate vibrations, and vibrations generated by other organisms may introduce fluctuations that can mask or distort the transmitted signal. In addition, interference may arise when multiple mechanical signals propagate simultaneously through the same medium and overlap at the receiver. These effects can reduce detection reliability and should be considered in evaluating communication performance.

\subsection{Electrical Communication}

A large portion of plant electrical signaling occurs in intra-organismal communication. Different cells / tissues / organs of the plant can send electrical signals to each other. Electrical communication can also be seen between plants and other plants, animals, insects, and fungi. This communication requires a close contact between the transmitter and the receiver. However, the range of the overall signal can be large due to the relaying behavior: the receiver can re-transmit the signal to other receivers.

\subsubsection{Transmitter Models}

The source of information for electrical communication in plants can be chemical, mechanical, electrical, or a combination of these \cite{Szechynska-Hebda2022}. From an ICT perspective, the transmitter encodes these into electrical signals. Physically, this is done via the ion channels of the cells whose activity is altered under external stimuli. For example, when herbivores attack the plant, specific chemicals are released at the attack site, and water pressure levels change. These chemical and mechanical signals activate the ion channels of the other cells, and an electrical signal is generated \cite{Diffusion_and_Bulk_Flow_of_Amino_Acids_Mediate_Calcium_Waves_in_Plants}.

The duration and intensity of the stimulus modulate the temporal and amplitude characteristics of the generated electrical signal, indicating that these features are inherently encoded in the signal itself \cite{Electrical_Signals_Plant_Tolerance_to_Actions_of_Stressors_and_Programmed_Cell_Death_Is_Interaction_Possible}, \cite{Exploring_Strategies_for_Classification_of_External_Stimuli_Using_Statistical_Features_of_the_Plant_Electrical_Response}. This encoding process is governed by both the instantaneous physiological state of the plant (e.g., ion channel availability, membrane potential, and hormonal levels) and its genetic configuration, which determines the expression and regulation of the underlying signaling machinery \cite{Ion_Channels_in_Electrical_Signaling_in_Higher_Plants}. A systematic characterization of how different stimuli are mapped onto electrical signal features, and how this mapping varies across physiological conditions and genetic backgrounds, remains an open research problem. From a communication-theoretic perspective, resolving these questions would directly enable a principled transmitter model for plant electrical signaling. Specifically, identifying how stimuli are mapped onto signal features would allow the definition of an input–output relationship analogous to a modulation scheme. This, in turn, would make it possible to formalize the encoding process, characterize the effective signal alphabet, and determine the controllable degrees of freedom at the transmitter. Moreover, understanding the dependence of this mapping on physiological state and genetic configuration would clarify the variability and constraints of the transmitter, enabling the development of adaptive or state-aware encoding strategies. Such a framework would be essential for quantifying information transfer, designing stimulation protocols, and ultimately interpreting plant electrical activity within a rigorous communication systems paradigm.

\subsubsection{Channel Models}

A plant can be represented as an electrical circuit. Circuit-level modeling has been applied at multiple scales: root hair networks have been represented as electrical networks \cite{Lew2009}; plant cells have been modeled using equivalent circuits between higher plant cells \cite{Spanswick1972}; and roots have been described as three-conductor transmission lines along which signals travel \cite{due1993}. Similar to the generation of electrical signals, the propagation is also based on ion exchange. Models of ion oscillations provide mechanistic insight into electrical communication, including a feedback-controlled oscillatory model for ion fluxes \cite{Shabala2006}, potential distribution curves that incorporate ion concentrations via Henderson’s equation for liquid junction (diffusion) potentials \cite{Watanabe1995}, and analyses of ionic relations in cells with free-running transmembrane voltage \cite{Gradmann2001}. For long-distance electrical signal transmission, plants use the phloem and xylem as a transport network within the plant \cite{nervous_system_plants}.

Mathematical descriptions of plant electrical activity are often adapted from Hodgkin–Huxley-type conductance-based models that capture ion-channel dynamics underlying spike generation and propagation \cite{A_Theory_Model_for_Description_of_the_Electrical_Signals_in_Plant_Part_I}, \cite{sukhova2017}. However, these models mainly rely on numerical techniques and do not provide analytical expressions that explicitly describe the channel mechanism. Such expressions would be useful for a rigorous ICT-based analysis. Nonetheless, the qualitative signal properties they capture — most notably the stereotyped, all-or-nothing character of action potentials versus the graded, summative character of variation potentials — are well established and form the basis for the channel-level characterizations developed below.

At the signal level, APs, VPs, and SPs have all been considered; AP and VP generation and propagation are relatively well studied, whereas SP remains less well understood \cite{sukhova2017}. Representative AP models include mathematical formulations for AP generation in vascular plant cells \cite{Sukhov2009}, \cite{Mathematical_Model_of_Action_Potential_in_Higher_Plants_with_Account_for_the_Involvement_of_Vacuole_in_the_Electrical_Signal_Generation}, propagation models \cite{Sukhov2011}, and stimulus-driven AP generation \cite{beilby1982}, with a detailed review provided in \cite{Beilby2007}. For VPs, mechanisms of generation and propagation are described in \cite{Sukhov2013}, \cite{Vodeneev2015}, and burning-induced propagation in wheat leaves is examined in \cite{Vodeneev2012}. In \cite{Volkov2017Aloe}, a model using cable theory is applied to electrostimulation in Aloe Vera leaves and to electrical signals between plants within the same pot. It also discusses the electrical stimulation of soil and of a single Aloe Vera plant through the soil.

It is important to address the propagation of electrical signals in both plants and soil. The conductivity of electric signals in soil is influenced by various environmental factors, including soil moisture content, texture, composition, and temperature \cite{Friedman2005,Ma2011}. Numerous studies have investigated the effects of these factors on soil conductivity. For example, \cite{Friedman2005} models the effect of soil solution on conductivity, while \cite{Ma2011} presents a temperature correction model for soil electrical conductivity measurements. The impact of soil salinity on electrical conductivity is examined in \cite{rhoades1989}, and \cite{Cai2017} presents a model for electrical conductivity in saturated porous media. Further research on estimating soil solution electrical conductivity from bulk soil conductivity in sandy soils can be found in \cite{amente2000}. Additionally, \cite{Sudduth2013} discusses modeling soil electrical conductivity–depth relationships using experimental data, and \cite{Rhoades1990} models soil conductivity, considering factors like volumetric water content, electrical conductivity of the soil water, and the average electrical conductivity of soil particles.

From an ICT perspective, plant conductive tissues function as biological transmission lines. In contrast to chemical channels, which are diffusion-limited and low-rate, electrical conduction provides a relatively high-rate backbone with reduced latency. However, the channel is spatially constrained to within the plant body, and inter-plant connectivity requires physical couplings such as grafts, shared tissues, or possibly fungal networks. This distinction highlights how electrical signaling can serve as a fast intra-plant bus that often triggers slower modalities, analogous to control signals that initiate secondary communication pathways in engineered networks.

\subsubsection{Receiver Models}
The receiver models for plant electrical signaling should focus on each type of electrical signal separately. APs are stereotyped, distinguishable from background fluctuations, and propagate robustly through excitable tissues. In contrast, VPs are graded changes in membrane potential, often triggered by wounding or hydraulic disturbances, and can vary in both amplitude and duration \cite{nervous_system_plants}. This distinction means that APs are relatively easy to identify, whereas VPs present greater uncertainty in decoding because their magnitudes reflect stimulus severity. From an ICT perspective, APs resemble digital symbols that can be detected by simple thresholding, much like binary signals in conventional receivers. Their uniform amplitude and stereotyped waveform minimize the chance of false positives. VPs, on the other hand, act more like analog waveforms, in which amplitude and temporal profile carry information about stimulus intensity. This introduces ambiguity in detection and requires receiver models that can handle continuous-valued signals, analogous to analog modulation schemes in communications.

\subsubsection{ISI and Channel Memory}
Biologically, APs reset after a refractory period, ensuring that individual spikes remain distinct and preventing overlap. This imposes an upper limit on firing frequency but also acts as a natural guard interval, reducing interference between successive spikes. VPs, in contrast, can summate and overlap when multiple stimuli occur close in time, creating lingering changes in potential that blur the boundaries between signals \cite{nervous_system_plants}.

From an ICT perspective, AP signaling approximates a near-memoryless binary channel: each spike is a clear, separable symbol, and the refractory period ensures temporal isolation. This makes APs highly reliable but rate-limited, with typical plants producing at most a few spikes per minute. VPs, however, correspond to analog fading channels with state dependence, since overlapping responses introduce intersymbol distortion and make decoding stimulus history more complex. In this sense, APs provide a high-fidelity but low-throughput digital link, while VPs represent a higher-variability analog channel where information is encoded in waveform shape and summation dynamics.

\subsubsection{Noise and Reliability}
Within a plant, thermal noise is negligible compared to the amplitude of APs, which typically span tens to over a hundred millivolts \cite{mudrilov2021electrical}. Instead, variability in signal propagation stems primarily from physiological and environmental conditions. For example, changes in temperature, light irradiation, and other abiotic stressors are known to influence both the propagation of electrical signals and their functional efficacy in plants \cite{mudrilov2021electrical}. Experimental studies also show that water stress and drought can reduce signal amplitude or speed—specifically, thermal stress has been reported to diminish the magnitude of electrical potentials, and conduction speed can vary under such conditions \cite{chaparro2021plant}. From an ICT perspective, this behavior resembles a channel whose gain and delay are environment-dependent. In controlled settings, signal transmission remains reliable; in natural or stressed environments, however, channel performance can degrade or become unpredictable.

%% ============================================================
%%  SECTION IV — EMPIRICAL AND EXPERIMENTAL APPROACHES
%%  Revised version 3
%% ============================================================

\section{Empirical and Experimental Approaches}
\label{sec:empexpapp}

Empirical and experimental studies are essential for validating theoretical models, identifying their limitations, and deepening understanding of plant communication in realistic conditions. This section reviews such studies across chemical, mechanical, and electrical modalities, together with the sensing methods used to capture each type of communication.

% -----------------------------------------------------------
\subsection{Chemical Communication}
% -----------------------------------------------------------

\subsubsection{Aboveground Chemical Communication}

\paragraph{Empirical and Experimental Studies}

Aboveground chemical communication has been investigated through two complementary lines of research: how signal identity and intensity shape VOC emission profiles, and how receiver plants respond to both naturally emitted and artificially applied signals.

Early field evidence for VOC-mediated interplant signaling comes from a study in which undamaged tobacco plants near clipped sagebrush showed reduced herbivore damage and elevated polyphenol oxidase activity relative to controls \cite{Karban2000}. This study provided a baseline for end-to-end signaling models by demonstrating that naturally emitted VOCs are sufficient to prime physiological responses under field conditions. Subsequent work established that receiver responses are dose-dependent: MeJA dosage governs the amplitude and kinetics of induced volatile responses \cite{Jiang2017}, while different MeJA epimers elicit similar overall activity but qualitatively distinct emission profiles \cite{Preston2004}. In maize, the herbivore-specific compound indole was released earlier than other induced volatiles, priming undamaged tissues for subsequent attack \cite{Erb2015}. From an ICT perspective, these findings constrain receiver models by specifying the temporal structure of the signal alphabet that arriving molecules must encode and that receivers must decode.

The influence of signal type and plant state on emission profiles has been characterized across herbivory, mechanical wounding, and abiotic stimuli. Herbivore feeding severity correlates directly with emission kinetics in \textit{Alnus glutinosa} \cite{copolovici2011volatile}, and species-specific responses emerge across five conifer species under increased herbivory \cite{Faiola2015}. Within a single species, damage type matters: silver birch responds rapidly to short-term foliage damage but more weakly to bark or chronic damage \cite{Maja2014}. Community-level effects are also evident, with emissions in red clover modulated by local species richness \cite{Kigathi2019}. Mechanical wounding produces emission changes that depend on cut type, severity, and interactions with ozone \cite{Portillo-Estrada2015, Kanagendran2018}. Under abiotic conditions, terpenoid synthesis exhibits a nonlinear, gene-regulated response to heat, with mild exposure yielding recoverable changes but severe heat producing irreversible declines in physiological activity \cite{Pazouki2016}. Collectively, these studies indicate that VOC emissions are not binary but form a graded, multidimensional signal space whose dimensions include compound identity, blend ratio, concentration, and temporal profile—a characterization directly relevant to understanding the modulation scheme and information capacity of the aboveground chemical channel.

\paragraph{Sensing Methods}

\begin{figure}[t]
    \centering
    \includegraphics[width=0.9\linewidth]{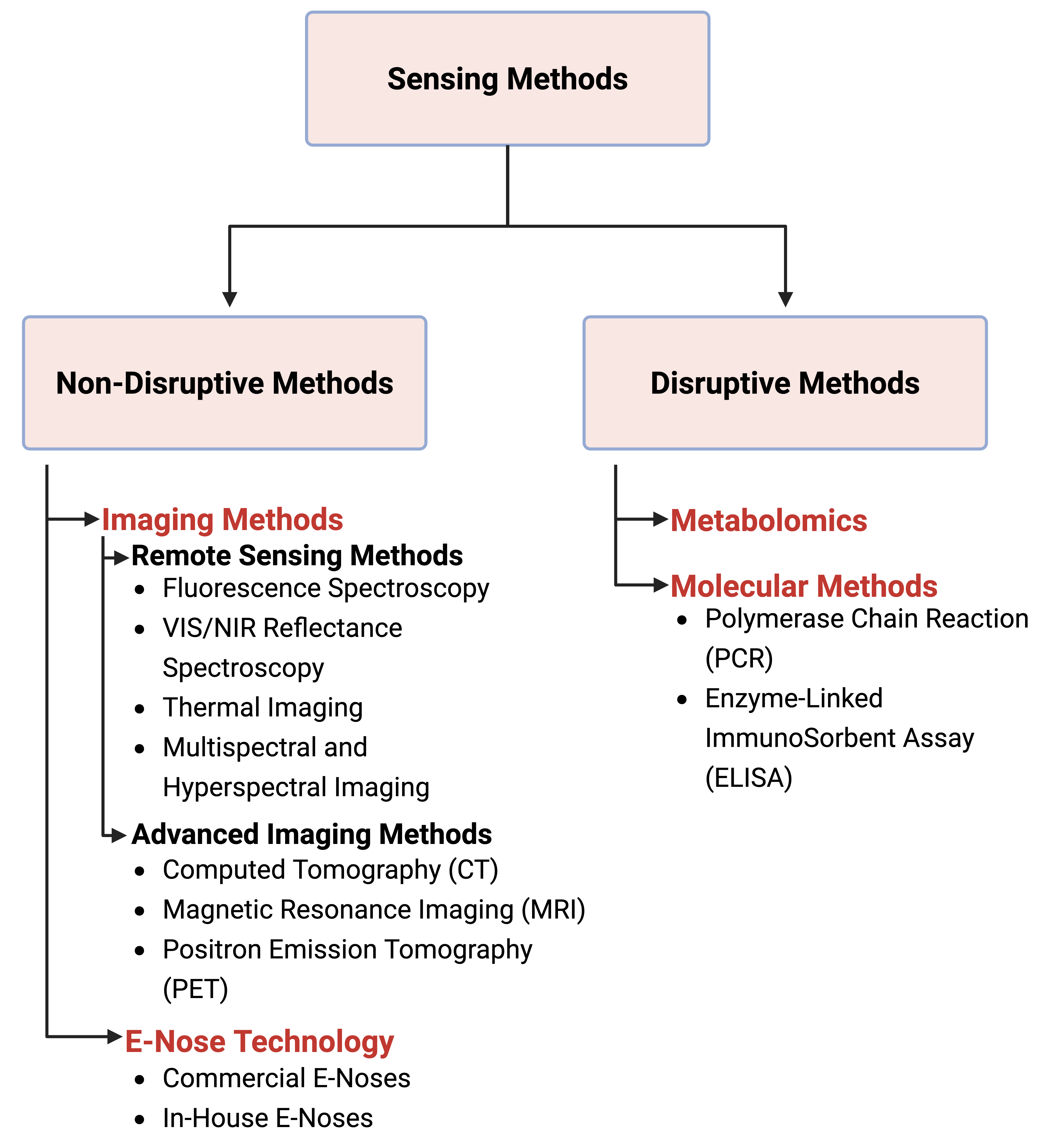}
    \caption{Overview of sensing methods for aboveground chemical communication \cite{Galieni2021}. Created with BioRender.com.}
    \label{fig:snsmethdsabv}
\end{figure}

VOC sensing methods are classified as disruptive or non-disruptive \cite{Galieni2021}, as summarized in Fig.~\ref{fig:snsmethdsabv}. Disruptive methods, though highly sensitive, hinder long-term studies. They include laboratory-based metabolomics for identifying biochemical changes, and molecular techniques such as PCR and ELISA for detecting gene expression changes and pathogens \cite{Galieni2021}. From a communication systems perspective, these methods provide ground-truth data for transmitter and receiver characterization but are incompatible with in-channel observation during active signaling.

Non-disruptive methods sacrifice some accuracy but enable field-scale monitoring \cite{Galieni2021}. Remote sensing approaches include fluorescence spectroscopy (chlorophyll fluorescence), VIS/NIR reflectance spectroscopy (structural traits, pigments, water content), thermal imaging (leaf temperature and water status), and multispectral or hyperspectral imaging. Advanced imaging adapted from medical physics—CT for 3D structural assessment, MRI for non-invasive visualization of water flow, and PET for tracking metabolic activity—offers higher resolution at greater cost \cite{Galieni2021}. These approaches capture receiver-side physiological changes but typically cannot resolve individual VOC species at the channel level.

Electronic nose (E-nose) systems directly sample the chemical channel by employing sensor arrays to absorb and classify ambient VOC profiles. Commercial devices span portable \cite{Sensigent_Cyranose320, Airsense_Pen3, NANOSENSORS_MSS8RM} and non-portable \cite{AlphaMOS_Heracles} form factors. Research-grade E-noses have been developed for specific applications including rapid aphid infestation diagnosis in greenhouse tomatoes \cite{Cui_FastENose}, pesticide detection \cite{Leccese_ElectronicNose}, salinity identification in citrus \cite{Sharma_GasSensorArray}, and noise-robust classification using olfactory bulb-inspired neural networks \cite{Martinelli_StableOdor}. The E-nose architecture—sensor array followed by pattern classifier—is a practical implementation of a matched filter receiver operating in the chemical domain, making it the sensing technology most directly aligned with the ICT receiver block.

\subsubsection{Belowground Chemical Communication}

\paragraph{Empirical and Experimental Studies}

Belowground chemical communication has been studied across three functional contexts: plant-to-plant signaling, plant-insect interactions, and allelopathic effects. The underground setting makes all of these more difficult to isolate experimentally than their aboveground counterparts, since soil acts simultaneously as channel, noise source, and biological processing medium.

Root-to-root communication was shown to increase resilience to osmotic challenge in neighboring plants, with connected and unconnected individuals both exhibiting stomatal closure \cite{Falik2012}. In mixed stands of \textit{Ambrosia dumosa} and \textit{Larrea tridentata}, species-specific root recognition behaviors were observed: \textit{Larrea} suppressed nearby root growth while \textit{Ambrosia} avoided conspecific roots \cite{Mahall1991}. Not all belowground signals are triggered by acute stimuli; bitou bush and sagebrush constitutively release VOCs that inhibit germination and growth of neighboring species \cite{ens2009}, and sagebrush roots emit phytotoxic volatiles suppressing competitors through oxidative disruption of root mitosis \cite{jassbi2010, romagni2000}. From an ICT perspective, these baseline emissions represent a persistent interference floor that any acute information-bearing signal must overcome, analogous to co-channel interference in RF networks.

Root-insect signaling represents a functionally distinct subset of belowground communication in which the receiver is an animal rather than a neighboring plant. Sesquiterpenes emitted by maize roots under herbivorous larval attack attract entomopathogenic nematodes that prey on the herbivores \cite{rasmann2005, rasmann2012}. Soil insects also locate plant hosts through herbivore-induced root VOCs \cite{robert2012}, and parasitic plants use root-emitted sesquiterpene lactones as germination cues \cite{bouwmeester2003}. Mycorrhizal networks extend the spatial reach of belowground chemical signals beyond what soil diffusion alone would permit: CMNs relay information between connected plants \cite{babikova2013underground}, transport allelochemicals at concentrations exceeding passive diffusion limits \cite{barto2011fungal}, and convey priming signals between healthy and pathogen-challenged individuals \cite{song2010interplant}. In ICT terms, CMNs function as a relay network that overcomes the severe attenuation of the soil channel, extending effective communication range at the cost of introducing additional delay and biological processing.

\paragraph{Sensing Methods}

Belowground sensing presents unique challenges because root extraction and soil disturbance both risk releasing VOCs that contaminate measurements \cite{delory2016root}. The primary analytical tools are GC-MS and PTR-MS. GC-MS separates compounds by mass and retention time for library-matched identification but requires prior sampling and VOC concentration; PTR-MS allows real-time monitoring with high temporal resolution, particularly for highly volatile low-molecular-weight compounds difficult to capture on packed adsorbents \cite{delory2016root, vanDam2012, Danner2015}. For mycorrhizal communication studies, methods shift from capturing emitted molecules to analyzing the receiver plant: gas chromatography of headspace gas from receiver tissue and PCR analysis of receiver leaves have been the principal approaches \cite{babikova2013underground, song2010interplant}. This distinction is significant from a systems perspective: aboveground sensing can probe the channel directly, whereas belowground mycorrhizal sensing is limited to observing receiver-side outputs, making end-to-end channel characterization substantially harder.

Sampling strategies include static, dynamic, and passive methods. Static approaches (SPME, HSME) capture headspace VOCs by adsorption or solvent extraction \cite{Tholl2006, delory2016root}. Dynamic methods use packed adsorbents for in situ collection from root systems grown in sandy soil or semi-vertical rhizotrons \cite{Abraham2015, Hiltpold2011}. Passive PDMS sorbents continuously capture non-polar compounds released by roots in mesocosms \cite{Mohney2009, Eilers2015}. Dynamic methods provide better in situ resolution while static methods remain faster for initial VOC identification.

% -----------------------------------------------------------
\subsection{Mechanical Communication}
% -----------------------------------------------------------

\subsubsection{Acoustic Communication}

\begin{figure*}[t]
    \centering
    \includegraphics[width = 0.75\linewidth]{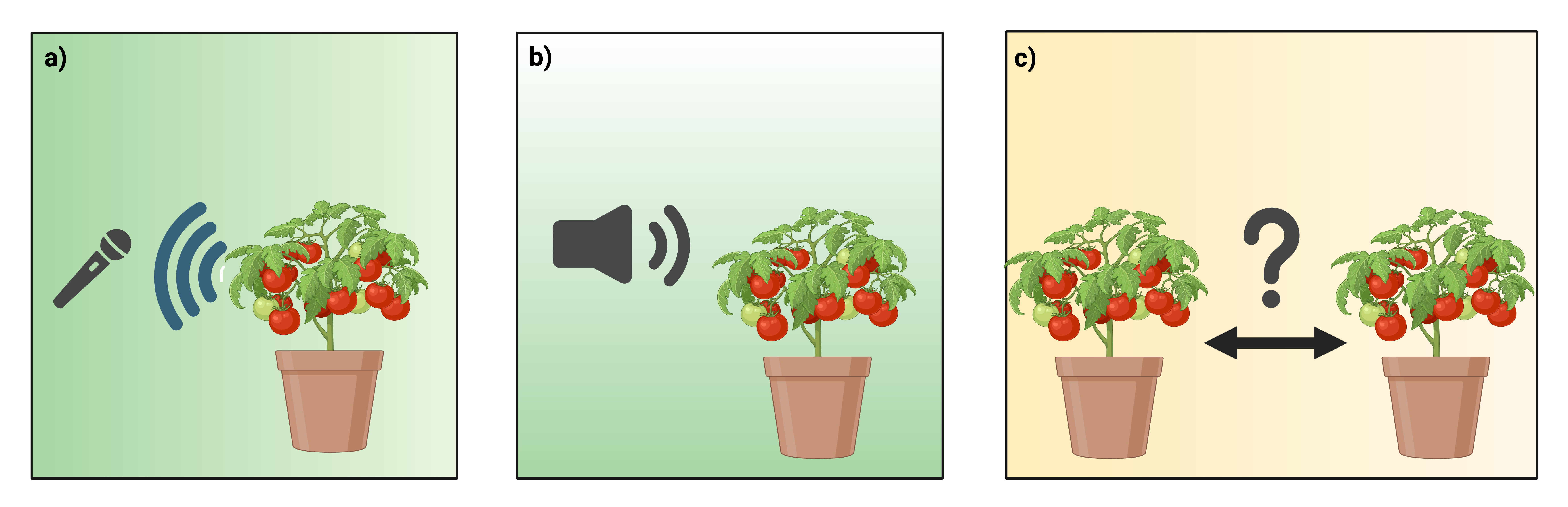}
    \caption{State of experimental studies on plant acoustic communication. (a) Detection of plant stress calls \cite{Khait2023}. (b) Plants' responses to stressor sounds \cite{Veits2019}. (c) Interplant acoustic communication remains undetected. Created with BioRender.com.}
    \label{fig:plntaccexp}
\end{figure*}

\paragraph{Empirical and Experimental Studies}

Empirical work on plant acoustic communication spans two directions: characterizing sounds emitted by plants and documenting physiological responses to external acoustic stimuli. An overview of the current state of experimental studies is illustrated in Fig.~\ref{fig:plntaccexp}.

Regarding emission, ultrasonic signals recorded from tomato and tobacco in both acoustic chambers and greenhouse environments demonstrated that machine learning classifiers could reliably distinguish plant conditions—including dehydration and mechanical damage—from the emitted sound profile alone \cite{Khait2023}. This result is significant from an ICT perspective: it provides empirical evidence that acoustic emissions carry sufficient mutual information about plant state to support reliable classification, suggesting that plants function as stochastic event-driven transmitters whose emission statistics encode physiological information. Earlier investigations established cavitation as a principal emission mechanism: clicks were detected in fern sporangia and Ricinus tissues \cite{Milburn1966}, and detached leaves produced ultrasound emissions during drying consistent with vein cavitation \cite{kikuta1997}. However, emission frequency spectra and waveforms do not fully conform to the classical cavitation-disruption model, suggesting additional mechanisms remain uncharacterized and that a complete transmitter model is not yet available \cite{Laschimke2006}.

Regarding perception, plants respond to acoustic stimuli in ways that are frequency-specific and ecologically relevant. \textit{Oenothera drummondii} flowers exposed to bee flight sounds rapidly increased nectar sugar concentration within three minutes, a response absent at higher frequencies, suggesting that floral structures function as acoustic sensory organs with a defined passband \cite{Veits2019}. Feeding vibrations provide cues sufficient for \textit{Arabidopsis} to distinguish herbivore identity and mount targeted responses \cite{Kollasch2020}. Sound vibrations also induce epigenetic changes and activate immune pathways against bacterial pathogens \cite{Jung2020}, and touch-regulated genes are upregulated by sound treatment, indicating overlapping mechanical transduction pathways \cite{Ghosh2016}. Evidence for acoustic sensing spans flowers, leaves, and roots \cite{Veits2019, Appel2014, RodrigoMoreno2017}, though direct plant-to-plant acoustic communication remains demonstrated in only a single study \cite{Gagliano2012}.

\paragraph{Sensing Methods}

Both remote and contact-based methods are used to capture plant-generated acoustic signals. Remote detection via microphones in controlled chambers has been demonstrated at practical distances \cite{Khait2023}. Contact-based methods include phonograph pick-up needles (simple and cost-effective, but susceptible to environmental noise and limited in frequency range), broadband and resonant ultrasonic sensors (effective for cavitation-frequency detection, requiring high-frequency analysis and periodic calibration), high-performance multi-channel data acquisition systems (high resolution but costly and storage-intensive), and X-ray micro-computed tomography for visual confirmation of cavitation (destructive, unsuitable for field use) \cite{DeRoo2016}. Remote methods remain less thoroughly characterized than contact methods, and no currently deployed system achieves reliable real-time field monitoring. From an ICT perspective, this sensing gap is significant: without a reliable non-invasive receiver at the channel output, empirical validation of acoustic channel models and quantification of acoustic information transfer remain infeasible.

\subsubsection{Touch Communication}

\paragraph{Empirical and Experimental Studies}

Physical touch is the earliest reliable cue for neighbor detection, preceding phytochrome-mediated shade avoidance signals. In dense \textit{Arabidopsis} stands, upward leaf movement (hyponasty) was initiated exclusively in leaves whose tips made physical contact with neighbors, before canopy density reached levels sufficient for a physiologically meaningful reduction in the red-to-far-red light ratio \cite{deWit2012}. Controlled experiments confirmed the causal role of touch: transparent tags maintaining leaf-tip contact alone were sufficient to induce hyponasty, while intermittent contact was not, indicating that continuous stimulation is required. Functional-structural plant modeling further revealed that touch-induced hyponasty is the physical prerequisite for subsequent phytochrome-based signaling, since horizontally growing rosette leaves produce insufficient far-red reflection until elevated by touch \cite{deWit2012}. From a communication-theoretic standpoint, touch here functions as an earlier, parallel signaling channel that gates access to a secondary channel—a form of channel multiplexing in which the mechanical and optical modalities operate in sequence rather than independently.

Physical contact also establishes community-wide signaling networks. In chains of three \textit{Arabidopsis} plants with overlapping leaf contact, signals propagated linearly through the contact points, activating over 2,000 genes in receiver plants within one hour \cite{PelaezVico2025}. Genetically disrupted mediator plants impaired in ROS synthesis or calcium-ROS accumulation severed transmission, confirming an active biological relay rather than passive diffusion. Receiver plants in functional chains showed significantly higher photosynthetic efficiency and lower cellular damage than isolated controls. This relay structure—transmitter, active repeater, receiver—maps directly onto a multi-hop communication network and reveals that the effective communication range of touch signaling extends beyond single contact interfaces through biological amplification at each node \cite{PelaezVico2025}.

Physical contact also reshapes chemical output through thigmomorphogenesis. Plants subjected to simulated neighbor contact exhibited a shifted VOC profile favoring defense-related sesquiterpenes over attractive monoterpenes, rendering touched plants less attractive to aphid pests in behavioral assays \cite{Hansen2025}. This mechanochemical coupling illustrates that touch and chemical channels are not independent: a mechanical input modulates the transmitter state of the chemical channel, representing cross-modality interference from an ICT perspective that must be accounted for in multimodal communication models.

\paragraph{Sensing Methods}

Flexible wearable sensors for monitoring thigmomorphogenic growth and touch-induced mechanical deformation employ piezoresistive, capacitive, or piezoelectric transduction mechanisms \cite{Yan2025wearable}. Substrates such as PDMS, Eco-flex silicone, and polyimide combined with conductive elements including PEDOT:PSS, carbon nanotubes, and graphene provide mechanical compliance and electrical sensitivity. Recent designs achieve detection limits of 10 Pa, sufficient to detect the landing of a small insect, while printable polymer films maintain structural integrity above 700\% strain. A key challenge is sensor breathability: intimate contact with leaf surfaces restricts stomatal gas exchange. Porous mesh and nanofiber architectures have partially addressed this, with leading designs achieving water vapor transmission rates compatible with sustained plant health during long-term monitoring \cite{Yan2025wearable}.

For monitoring internal ionic fluxes associated with touch-induced action potentials, microneedle (MN) sensor arrays penetrate the epidermis and cortex to directly sample vascular sap. Solid-contact potentiometric arrays functionalized with ion-selective membranes profile real-time potassium and sodium fluxes within vascular bundles, providing high-resolution data on ionic signal propagation following mechanical stimulation \cite{Wang2025microneedle}. These sensors effectively read out the electrical channel at the transmitter side, enabling direct measurement of the ionic signals that constitute the information-bearing waveform in touch communication.

Genetically encoded FRET biosensors provide subcellular spatial and temporal resolution of the calcium and hormone dynamics underlying touch responses. Calcium-sensitive constructs such as Yellow Cameleon variants allow real-time visualization of calcium bursts within milliseconds of mechanical stimulation. Downstream hormonal dynamics are traceable through sensors including ABACUS (abscisic acid) and Jas9-VENUS (jasmonic acid) \cite{WaliaJones2018}. For ROS profiling, SWCNT nanosensors functionalized with specific DNA sequences or peptides enter mesophyll cells through stomata and undergo quantifiable fluorescence shifts upon contact with hydrogen peroxide, nitric oxide, or salicylic acid, enabling non-invasive continuous monitoring of the ROS cascade associated with touch signaling \cite{Shaw2022nanosensor}. For reading bioelectric signals in field settings, conformable surface electrodes constructed from PEDOT:PSS adhere to the plant epidermis via van der Waals forces without penetrating tissue. Machine learning pipelines applied to the resulting time-series data via CNN architectures processing mel-spectrograms have achieved classification accuracies above 85\% for distinguishing environmental states from short signal segments \cite{Gloor2025bioelec}, demonstrating that plant bioelectric signals carry recoverable information about external stimuli and that ICT-style detection algorithms can serve as practical plant signal decoders.

% -----------------------------------------------------------
\subsection{Electrical Communication}
% -----------------------------------------------------------

\subsubsection{Empirical and Experimental Studies}

Research on plant electrical communication spans four areas: inter-plant potential transfer, soil-mediated propagation, plant-as-circuit modeling, and signal classification.

Inter-plant electrical transfer has been demonstrated across multiple experimental configurations. Sinusoidal electrostimulation of Aloe vera leaves induced responses with a 90-degree phase shift in connected tissues, confirming electrical network activity in the absence of biotic or abiotic perturbation \cite{Volkov2016}. When soils in separate pots were connected by metallic wires, stimulation of Aloe vera, tomato, or cabbage transmitted electropotentials across species boundaries \cite{Volkov2019}. Foliar signals in dandelion propagated to neighboring plants through direct physical contact, inducing systemic physiological changes across species \cite{Szechynska-Hebda2022}. These results collectively establish that electrical coupling enables rapid interplant coordination that operates independently of chemical diffusion—from an ICT perspective, electrical signaling provides a low-latency channel that complements the higher-capacity but slower chemical channel, analogous to the role of control signaling alongside data channels in layered communication architectures.

Soil-mediated electrical propagation has been characterized by varying stimulus waveforms and measuring spatial decay. Pulse trains and sinusoidal or triangular voltage stimuli applied to soil generated electrical waves that attenuated with distance \cite{Volkov2017Aloe}. The same study found that mycorrhizal disruption via antifungal treatment did not eliminate electrical responses, suggesting that soil conductivity rather than hyphal networks is the primary coupling pathway in these experiments. Square-pulse stimulation induced passive potentials in tomato without biological amplification \cite{Volkov2018}, while soil electrical conductivity is governed by moisture, texture, salinity, and temperature \cite{Friedman2005, Ma2011, rhoades1989}. The distance-dependent attenuation and environmental sensitivity of soil conductivity establish the soil electrical channel as strongly time-varying, with channel gain fluctuating on timescales governed by weather and irrigation.

Modeling plants as electrical circuits has provided mechanistic insight into specific behaviors. Venus flytrap closure was replicated using a discrete circuit model \cite{Volkov2013}, and electrode placement topology confirmed that action potentials in the trap can reach speeds up to 10 m/s \cite{Volkov2019fly}. Direct current application at small increments altered growth and ion uptake in tomato, linking external electrical stimulation to physiological state change \cite{black1971}. These circuit models establish a direct mapping between plant electrical behavior and lumped-element network theory, enabling the use of standard circuit analysis tools for characterizing channel impedance, signal attenuation, and propagation delay.

Signal classification studies address the challenge of extracting information from plant electrical records, which is equivalent to the receiver design problem in an ICT framework. Automated classification of APs using template matching achieved 96\% accuracy, outperforming backpropagation neural networks, support vector machines, and deep learning methods in a controlled comparison \cite{Chen2016}. A multi-channel monitoring system simultaneously tracking environmental variables and electrical signals in cucumber demonstrated feasibility for both laboratory and greenhouse deployment \cite{Wang2007}. Aphid bioelectrodes for phloem potential measurement identified GLR3.3 and GLR3.6 glutamate receptor-like genes as contributors to systemic AP transmission between leaves, while also establishing that phloem APs can propagate independently of these receptors \cite{Hedrich2016}. The high classification accuracy achieved by template matching—equivalent to a matched filter receiver—suggests that plant APs maintain sufficient waveform consistency to support reliable symbol detection, consistent with the binary digital channel interpretation of AP signaling described in Section III.

\subsubsection{Sensing Methods}

Four main sensing methods are used in plant electrical communication research \cite{Yan2009}. Extracellular measurement detects surface potentials using contact or metal electrodes; it is non-invasive and supports whole-plant long-term monitoring but is susceptible to drying and minor mechanical damage at the contact site. Intracellular measurement via microelectrode insertion provides precise membrane potential data at single-cell resolution but is invasive and limited to short-term experiments. The patch-clamp technique isolates and records individual ion channel activity with high precision, yielding mechanistic insights into ionic conductances, at the cost of requiring cell wall removal and substantial technical expertise. The non-invasive microelectrode vibrating probe measures ion fluxes around cells without contact, enabling real-time monitoring across tissue types, though with lower temporal resolution than patch-clamp. Each method occupies a distinct position in the trade-off between spatial resolution, invasiveness, and experimental duration, and the appropriate choice depends on whether the goal is whole-plant signal monitoring or mechanistic channel-level characterization.

%% ============================================================
%%  SECTION V — LESSONS LEARNED
%% ============================================================

\section{Lessons Learned}

The preceding sections establish that plant communication can be meaningfully analyzed through an ICT framework, but they also expose a set of recurring tensions between biological reality and engineering abstraction. This section distills the cross-cutting insights that a communications researcher should carry away from this survey, organized around five themes that span modeling, channel behavior, multimodal interaction, receiver design, and observability.

% -----------------------------------------------------------
\subsection{Biology-to-ICT Mappings}
% -----------------------------------------------------------

The transmitter-channel-receiver abstraction applies to plant communication, but the fidelity of the mapping varies substantially across modalities and across the transmitter, channel, and receiver blocks within each modality. On the transmitter side, the aboveground chemical transmitter is the best-characterized: the compartmental model partitions production into aqueous, lipid, and gas pools and yields a well-defined emitted flux, and information encoding via CSK and RSK has direct analogs in molecular communication. The belowground chemical transmitter and the electrical transmitter are modeled with reasonable completeness, though the former draws on soil science literature adapted to an emission context rather than purpose-built MC models, and the latter lacks a unified treatment of how stimulus identity maps to signal waveform. The mechanical transmitter is the least developed: cavitation explains involuntary acoustic emission, but no model exists for voluntary or information-directed sound production.

On the channel side, the aboveground diffusion-advection equation and the soil porous-media balance are well-established, and their ICT interpretation as dispersive, memory-laden channels with colored noise is productive and accurate. The electrical channel through plant tissue and soil is characterized by cable theory and circuit models, but the time-varying nature of soil conductivity means no static channel model applies reliably in field conditions. The acoustic channel through air, soil, and plant tissue has multiple candidate propagation models, but none has been validated end-to-end for plant-to-plant acoustic communication. On the receiver side, the Robin boundary condition and Michaelis-Menten kinetics for chemical uptake map cleanly onto partial-absorption and saturation-nonlinearity models familiar from MC. Electrical signal reception via MS ion channels maps onto threshold detection. Touch signal reception through calcium-ROS cascades maps onto a relay with biological amplification. Where the mapping is weakest, the lesson is the same: the biological mechanism is known qualitatively, but a quantitative input-output model connecting the channel output to a receiver response function does not yet exist.

% -----------------------------------------------------------
\subsection{Distinctive Channel Properties}
% -----------------------------------------------------------

Across all modalities, plant communication channels share a set of properties that distinguish them from engineered channels and that any ICT analysis must account for.

The most consequential is extreme memory. The soil chemical channel retains signal energy for hours or days due to dual-phase partitioning and sorption, creating intersymbol interference that cannot be cleared by lengthening the symbol interval. The aboveground chemical channel has shorter but still significant memory governed by the Péclet number: diffusion-dominated conditions produce long tails, while advection shortens memory at the cost of directional dependence. The electrical channel through plant tissue has the shortest memory — AP refractory periods enforce a natural guard interval — but VP summation can accumulate over timescales of seconds to minutes, introducing state-dependent ISI. The mechanical channel through plant tissue is intermediate, with residual oscillations and multipath-like reflections that decay faster than chemical signals but slower than digital electrical ones.

A second shared property is that channel parameters are not constants but functions of biological and environmental state. Soil moisture, temperature, microbial activity, and plant physiological condition all modulate effective diffusivity, conductivity, and mechanical compliance simultaneously. This means the channel impulse response is not just time-varying but coupled to the transmitter and receiver states — a class of channel not typically treated in standard communication theory. Standard approaches such as estimating channel parameters from pilot signals and then treating the channel as fixed will fail here; adaptive methods that track slow parameter drift are required.

A third property is that multiple emitters sharing the same medium create interference that is chemically structured rather than spectrally structured. In the aboveground chemical channel, the interfering signal from a neighboring plant is not white noise but a correlated process whose statistics depend on the interferer's physiological state. This makes interference cancellation harder than in RF co-channel scenarios and means that standard SINR-based performance metrics may not translate directly.

% -----------------------------------------------------------
\subsection{Multimodal Coupling}
% -----------------------------------------------------------

A critical lesson from this survey is that the modalities are not independent channels operating in parallel — they are coupled both at the transmitter and at the receiver, and this coupling is functionally significant rather than a secondary effect.

At the transmitter, touch-induced thigmomorphogenesis reshapes VOC emission profiles: a mechanical input modulates the state of the chemical transmitter, producing a systematic cross-channel effect that would appear as interference if the two channels were modeled independently. Similarly, electrical signals generated by herbivory trigger jasmonic acid biosynthesis, which in turn drives VOC emission — so the electrical and chemical channels share a common upstream source that introduces correlation between their outputs. Treating these as independent channels therefore overestimates the degrees of freedom available for information transfer and may lead to incorrect conclusions about channel capacity that future information-theoretic treatments of multimodal plant signaling should consider.

At the receiver, plants integrate signals across modalities: VOC priming followed by electrical signaling or touch-induced calcium waves enhances the response to subsequent chemical signals. This integrative behavior resembles diversity combining at a multi-antenna receiver, but the combining rule is nonlinear and dependent on timing. Modeling receiver behavior as a linear superposition of modality-specific responses will underestimate the system's actual response under combined stimuli.

The practical implication for ICT modeling is that end-to-end plant communication should be approached as a multimodal system from the outset, rather than as a collection of single-channel systems. Joint transmitter-channel-receiver models that account for cross-modality coupling will be necessary for accurate information-theoretic analysis, and the design of any sensing or actuation interface that interacts with plant communication will need to account for these cross-modal effects.

% -----------------------------------------------------------
\subsection{Receiver Design Principles}
% -----------------------------------------------------------

The empirical evidence surveyed in Section~\ref{sec:empexpapp} contains concrete information about how plant receivers function as detectors, and this information directly constrains what ICT receiver designs are biologically plausible.

APs are the most tractable signal class: their stereotyped waveform, fixed amplitude, and all-or-nothing generation make them well-suited to threshold detection, and the 96\% classification accuracy achieved by template matching \cite{Chen2016} is consistent with a matched filter operating at high signal-to-noise ratio. This suggests that AP-based electrical signaling supports reliable symbol detection with relatively simple receiver architectures, and that detection performance will degrade primarily when the channel introduces amplitude distortion or timing jitter rather than thermal noise. VPs, by contrast, carry graded information in their amplitude and temporal profile, making threshold detection insufficient; an analog receiver that tracks waveform shape is needed, and detection becomes harder because plant-to-plant VP transmission through soil introduces additional amplitude scaling.

For chemical signals, the Michaelis-Menten receiver saturates at high concentrations, effectively clipping the received amplitude. This means the receiver operates linearly only in a limited concentration range, and information encoded in high-concentration signals is partially lost. The lesson for ICT is that concentration-based modulation schemes such as CSK face a fundamental dynamic range constraint at the receiver that is not present in RF systems, and that ratio-based schemes such as RSK are preferable when signal amplitude is uncertain — because blend ratios are preserved under amplitude scaling. The empirical finding that plants respond to blend composition even at concentrations too low to elicit single-compound responses is consistent with this analysis.

For touch and mechanical signals, the two-stimulus integration mechanism of the Venus flytrap and the requirement for continuous rather than intermittent contact in neighbor detection both indicate that plant mechanical receivers implement temporal integration rather than instantaneous detection. This resembles a matched filter with a finite integration window, and it provides natural noise immunity against brief spurious contacts. The calcium-ROS relay cascade then amplifies and propagates the received signal systemically, functioning as a distributed regenerative amplifier that extends the effective range of touch communication beyond the initial contact interface.

% -----------------------------------------------------------
\subsection{Observability and Sensing Constraints}
% -----------------------------------------------------------

A lesson that emerges consistently from Section~\ref{sec:empexpapp} is that the observability of plant communication differs fundamentally between modalities, and that many channel properties that are straightforward to measure in engineered systems cannot be directly accessed in plant systems.

Aboveground chemical communication is the most observable: GC-MS and PTR-MS can resolve the chemical channel output at the transmitter side, E-nose systems sample the channel mid-propagation, and receiver-side uptake can be inferred from foliar VOC measurements. This chain of observability has enabled end-to-end empirical validation of transmitter, channel, and receiver models \cite{kilic2024endtoendmathematicalmodelingstress, unluturk2016end}. Aboveground electrical communication is similarly well-instrumented: surface electrodes, intracellular microelectrodes, and patch-clamp techniques span from macroscopic to single-channel resolution, and signal classification from recorded waveforms has been demonstrated with high accuracy.

Belowground chemical communication faces a fundamental observability problem: sampling the channel requires disturbing it, since soil extraction releases VOCs that contaminate the measurement. As a result, most empirical studies of belowground communication can only observe receiver-side outputs — changes in gene expression, stomatal aperture, or root morphology in the receiver plant — and must infer channel and transmitter behavior indirectly. This is equivalent to designing a receiver without access to channel pilots or transmitted reference signals, and it makes model validation substantially harder. The mycorrhizal chemical sub-channel is even less directly observable; its contribution is typically inferred by comparing outcomes with and without fungal network disruption.

Mechanical communication through soil and plant tissue is accessible only at the ends: emission can be detected remotely or by contact sensors, and reception can be inferred from plant physiological responses, but mid-channel propagation through the vascular system or soil matrix cannot currently be observed non-invasively. The most significant sensing gap overall is therefore in the mechanical and belowground chemical channels, where the absence of in-channel observability prevents the kind of end-to-end characterization that ICT channel modeling requires.

The development of non-invasive in-channel sensing tools — such as implantable nanosensors that can report chemical concentrations within the vascular system, or distributed MEMS arrays for substrate vibration profiling — would be the single most enabling technical advance for validating the ICT models reviewed in this paper and for closing the gap between theoretical analysis and empirical reality.

%% ============================================================
%%  SECTION VI — CONCLUSION
%%  Revised
%% ============================================================

\section{Conclusion}
To conclude, plants communicate through diverse modalities, including chemical, mechanical, and electrical signaling, engaging not only neighboring plants but also insects, animals, fungi, and soil microorganisms across a broad range of ecological functions. Understanding the principles behind these modalities is crucial not only for comprehending plant communication but also for developing methods to interface with and learn from these natural systems. While numerous studies have examined these mechanisms individually, a deeper and more systematic perspective is required — particularly from the lens of ICT.
 
This paper has reviewed the fundamental mechanisms of plant communication across chemical, mechanical, and electrical modalities, summarizing biological background, empirical and experimental findings, and the sensing techniques used to study them. It has further provided an ICT-based framework for modeling plant communication, offering transmitter, channel, and receiver models for each modality and distilling cross-cutting lessons on channel properties, multimodal coupling, receiver design, and observability constraints.
 
Looking forward, the lessons learned from this survey point to several concrete directions. Uneven mapping fidelity across modalities calls for purpose-built models, particularly for mechanical communication where no validated end-to-end model exists. The distinctive channel properties identified here motivate new analytical tools beyond standard frameworks. Multimodal coupling establishes joint modeling as a necessary next step for accurate information-theoretic analysis. Saturation-bounded and temporally integrating receiver mechanisms invite designs that depart from linear assumptions. Finally, the observability gap in belowground and mechanical channels highlights non-invasive in-channel sensing as the most enabling technical advance the field could make. Overall, this paper establishes a foundation for advancing plant communication research and highlights the opportunities for interdisciplinary exploration at the interface of biology and ICT.
 
\bibliographystyle{IEEEtran}
\bibliography{references.bib}

@article{bais2004underground,
  title = {How plants communicate using the underground information superhighway},
  author = {Bais, Harsh P. and Park, Sang Wook and Weir, Tiffany L. and Callaway, Ragan M. and Vivanco, Jorge M.},
  journal = {Trends in Plant Science},
  volume = {9},
  number = {1},
  pages = {26--32},
  year = {2004},
  doi = {10.1016/j.tplants.2003.11.008},
}

@article{plantxcom,
author = {Rodríguez, Ana and Alquézar, Berta and Peña, Leandro},
title = {Fruit aromas in mature fleshy fruits as signals of readiness for predation and seed dispersal},
journal = {New Phytologist},
volume = {197},
number = {1},
pages = {36-48},
doi = {https://doi.org/10.1111/j.1469-8137.2012.04382.x},
eprint = {https://nph.onlinelibrary.wiley.com/doi/pdf/10.1111/j.1469-8137.2012.04382.x},
year = {2013}
}

@article{Clarke2017,
  author    = {Dominic Clarke and Erica Morley and Daniel Robert},
  title     = {The bee, the flower, and the electric field: electric ecology and aerial electroreception},
  journal   = {Journal of Comparative Physiology A},
  year      = {2017},
  volume    = {203},
  number    = {9},
  pages     = {737--748},
  month     = {Sep},
  doi       = {10.1007/s00359-017-1176-6},
}

@Article{Farag2013,
  author          = {Farag, Mohamed A. and Zhang, Huiming and Ryu, Choong-Min},
  title           = {Dynamic Chemical Communication between Plants and Bacteria through Airborne Signals: Induced Resistance by Bacterial Volatiles},
  journal         = {Journal of Chemical Ecology},
  year            = {2013},
  volume          = {39},
  number          = {8},
  pages           = {1007--1018},
  doi             = {10.1007/s10886-013-0317-9},
  issn            = {1573-1561}
}

@article{detectthyfamily,
author = {Anten, Niels P. R. and Chen, Bin J. W.},
title = {Detect thy family: Mechanisms, ecology and agricultural aspects of kin recognition in plants},
journal = {Plant, Cell \& Environment},
volume = {44},
number = {4},
pages = {1059-1071},
doi = {https://doi.org/10.1111/pce.14011},
year = {2021}
}

@Article{insects10100366,
AUTHOR = {Wang, Zhengwei and Tan, Ken},
TITLE = {Honey Bee Alarm Pheromone Mediates Communication in Plant–Pollinator–Predator Interactions},
JOURNAL = {Insects},
VOLUME = {10},
YEAR = {2019},
NUMBER = {10},
ARTICLE-NUMBER = {366},
ISSN = {2075-4450},
DOI = {10.3390/insects10100366}
}

@article{ninkovic2021plant,
  title = {Plant volatiles as cues and signals in plant communication},
  author = {Ninkovic, Velemir and Markovic, Dijana and Rensing, Stefan A.},
  journal = {Plant, Cell \& Environment},
  volume = {44},
  number = {4},
  pages = {1030--1043},
  year = {2021},
  doi = {10.1111/pce.13910},
}

@article{lin2007volatile,
  author = {Chun Lin and Susan M. Owen and Josep Peñuelas},
  title = {Volatile organic compounds in the roots and rhizosphere of Pinus spp.},
  journal = {Soil Biology and Biochemistry},
  volume = {39},
  number = {4},
  pages = {951--960},
  year = {2007},
  issn = {0038-0717},
  doi = {10.1016/j.soilbio.2006.11.007},
}

@article{midzi2022stress,
  title = {Stress-Induced Volatile Emissions and Signalling in Inter-Plant Communication},
  author = {Midzi, Junior and Jeffery, David W. and Baumann, Ute and Rogiers, Suzy and Tyerman, Stephen D. and Pagay, Vinay},
  journal = {Plants},
  volume = {11},
  number = {19},
  pages = {2566},
  year = {2022},
  doi = {10.3390/plants11192566},
}

@article{preston2001methyl,
  title = {Methyl jasmonate is blowing in the wind, but can it act as a plant–plant airborne signal?},
  author = {Preston, Catherine A. and Laue, Gerhard and Baldwin, Ian T.},
  journal = {Biochemical Systematics and Ecology},
  volume = {29},
  number = {10},
  pages = {1007--1023},
  year = {2001},
}

@article{Jiang2017,
  author    = {Y. Jiang and J. Ye and S. Li and U. Niinemets},
  title     = {Methyl jasmonate-induced emission of biogenic volatiles is biphasic in cucumber: a high-resolution analysis of dose dependence},
  journal   = {Journal of Experimental Botany},
  volume    = {68},
  number    = {16},
  pages     = {4679--4694},
  year      = {2017},
  doi       = {10.1093/jxb/erx273}
}

@article{Maja2014,
  author    = {M. M. Maja and A. Kasurinen and P. Yli-Pirilä and J. Joutsensaari and T. Klemola and T. Holopainen and J. K. Holopainen},
  title     = {Contrasting responses of silver birch VOC emissions to short-and long-term herbivory},
  journal   = {Tree Physiology},
  volume    = {34},
  number    = {3},
  pages     = {241--252},
  year      = {2014},
  doi       = {10.1093/treephys/tpu014}
}

@article{Faiola2015,
  author = {Faiola, C. L. and Jobson, B. T. and VanReken, T. M.},
  title = {Impacts of simulated herbivory on volatile organic compound emission profiles from coniferous plants},
  journal = {Biogeosciences},
  year = {2015},
  volume = {12},
  number = {2},
  pages = {527-547},
  doi = {10.5194/bg-12-527-2015}
}

@article{faiola2020impact,
  author = {Faiola, C. and Taipale, D.},
  title = {Impact of insect herbivory on plant stress volatile emissions from trees: A synthesis of quantitative measurements and recommendations for future research},
  journal = {Atmospheric Environment: X},
  volume = {5},
  pages = {100060},
  year = {2020}
}

@article{Kigathi2019,
  author    = {R. N. Kigathi and W. W. Weisser and M. Reichelt and J. Gershenzon and S. B. Unsicker},
  title     = {Plant volatile emission depends on the species composition of the neighboring plant community},
  journal   = {BMC Plant Biology},
  volume    = {19},
  pages     = {1--17},
  year      = {2019},
  doi       = {10.1186/s12870-019-1777-1}
}

@article{Karban2000,
  author    = {Richard Karban and Ian T. Baldwin and Kathryn J. Baxter and Gerald Laue and Gary W. Felton},
  title     = {Communication between plants: induced resistance in wild tobacco plants following clipping of neighboring sagebrush},
  journal   = {Oecologia},
  volume    = {125},
  pages     = {66--71},
  year      = {2000},
  doi       = {10.1007/PL00008892}
}

@ARTICLE{kilic2024endtoendmathematicalmodelingstress,
  author={Kilic, Ahmet Burak and Akan, Ozgur B.},
  journal={IEEE Transactions on Molecular, Biological, and Multi-Scale Communications}, 
  title={End-to-End Mathematical Modeling of Stress Communication Between Plants}, 
  year={2026},
  volume={12},
  number={},
  pages={69-78},
  doi={10.1109/TMBMC.2025.3626218}}

@article{Guenneau2015,
  author    = {S. Guenneau and D. Petiteau and M. Zerrad and C. Amra and T. Puvirajesinghe},
  title     = {Transformed Fourier and Fick equations for the control of heat and mass diffusion},
  journal   = {AIP Advances},
  volume    = {5},
  number    = {5},
  pages     = {053404},
  year      = {2015},
  month     = {May},
  doi       = {10.1063/1.4917492},
}

@article{Silva2013,
  author    = {E.J.G. Silva and T. Tirabassi and M.T. Vilhena and D. Buske},
  title     = {A puff model using a three-dimensional analytical solution for the pollutant diffusion process},
  journal   = {Atmospheric Research},
  volume    = {134},
  pages     = {131--136},
  year      = {2013},
  issn      = {0169-8095},
  doi       = {10.1016/j.atmosres.2013.07.009},
}

@article{unluturk2016end,
  title = {An end-to-end model of plant pheromone channel for long range molecular communication},
  author = {Unluturk, B. D. and Akyildiz, I. F.},
  journal = {IEEE Transactions on Nanobioscience},
  volume = {16},
  number = {1},
  pages = {11--20},
  year = {2016},
  doi = {10.1109/TNB.2016.2518520}
}

@incollection{harley2013roles,
  author = {Harley, Peter},
  title = {The Roles of Stomatal Conductance and Compound Volatility in Controlling the Emission of Volatile Organic Compounds from Leaves},
  booktitle = {Biology, Controls and Models of Tree Volatile Organic Compound Emissions},
  pages = {181--208},
  year = {2013},
  doi = {10.1007/978-94-007-6606-8_7}
}

@article{Portillo-Estrada2015,
  author = {Portillo-Estrada, M. and Kazantsev, T. and Talts, E. and Tosens, T. and Niinemets, U.},
  title = {Emission timetable and quantitative patterns of wound-induced volatiles across different leaf damage treatments in aspen (Populus tremula)},
  journal = {Journal of Chemical Ecology},
  year = {2015},
  volume = {41},
  pages = {1105-1117},
  doi = {10.1007/s10886-015-0654-1}
}

@article{Kanagendran2018,
  author = {Kanagendran, A. and Pazouki, L. and Niinemets, U.},
  title = {Differential regulation of volatile emission from Eucalyptus globulus leaves upon single and combined ozone and wounding treatments through recovery and relationships with ozone uptake},
  journal = {Environmental and Experimental Botany},
  year = {2018},
  volume = {145},
  pages = {21-38},
  doi = {10.1016/j.envexpbot.2017.11.015}
}

@article{Erb2015,
  author = {Erb, M. and Veyrat, N. and Robert, C. A. and Xu, H. and Frey, M. and Ton, J. and Turlings, T. C.},
  title = {Indole is an essential herbivore-induced volatile priming signal in maize},
  journal = {Nature Communications},
  year = {2015},
  volume = {6},
  number = {1},
  pages = {6273},
  doi = {10.1038/ncomms7273}
}

@article{Pazouki2016,
  author = {Pazouki, L. and Kanagendran, A. and Li, S. and Kännaste, A. and Memari, H. R. and Bichele, R. and Niinemets, U.},
  title = {Mono-and sesquiterpene release from tomato (Solanum lycopersicum) leaves upon mild and severe heat stress and through recovery: From gene expression to emission responses},
  journal = {Environmental and Experimental Botany},
  year = {2016},
  volume = {132},
  pages = {1-15},
  doi = {10.1016/j.envexpbot.2016.08.005}
}

@incollection{grote2013leaf,
  author = {Grote, RUdiger and Monson, Russell and Niinemets, Ulo},
  title = {Leaf-Level Models of Constitutive and Stress-Driven Volatile Organic Compound Emissions},
  booktitle = {Biology, Controls and Models of Tree Volatile Organic Compound Emissions},
  year = {2013},
  doi = {10.1007/978-94-007-6606-8_12}
}

@article{guenther2012model,
  author = {Guenther, Alex and Jiang, Xiaoyan and Heald, Colette and Sakulyanontvittaya, Tanarit and Duhl, Tiffany and Emmons, Louisa and Wang, Xuemei},
  title = {The Model of Emissions of Gases and Aerosols from Nature version 2.1 (MEGAN2.1): an extended and updated framework for modeling biogenic emissions},
  journal = {Geoscientific Model Development Discussions},
  volume = {5},
  year = {2012},
  doi = {10.5194/gmdd-5-1503-2012}
}

@article{grote2019new,
  author = {Grote, RUdiger and Sharma, Monica and Ghirardo, Andrea and Schnitzler, JOrg-Peter},
  title = {A New Modeling Approach for Estimating Abiotic and Biotic Stress-Induced de novo Emissions of Biogenic Volatile Organic Compounds From Plants},
  journal = {Frontiers in Forests and Global Change},
  volume = {2},
  pages = {26},
  year = {2019},
  doi = {10.3389/ffgc.2019.00026}
}

@article{guenther1995global,
  author = {Guenther, A. and Hewitt, C. and Erickson, D. and Fall, R. and Geron, C. and Graedel, T. and et al.},
  title = {A global model of natural volatile organic compound emissions},
  journal = {Journal of Geophysical Research},
  volume = {100},
  number = {D5},
  pages = {8873--8892},
  year = {1995},
  report_number = {D5},
}

@article{lerdau1997plant,
  author = {Lerdau, Manuel and Guenther, Alex and Monson, Russ},
  title = {Plant Production and Emission of Volatile Organic Compounds},
  journal = {BioScience},
  volume = {47},
  number = {6},
  pages = {373--383},
  year = {1997},
  doi = {10.2307/1313152}
}

@article{trapp2007fruit,
  author = {Trapp, S.},
  title = {Fruit tree model for uptake of organic compounds from soil and air},
  journal = {SAR and QSAR in Environmental Research},
  volume = {18},
  number = {3-4},
  pages = {367--387},
  year = {2007}
}

@article{rein2011new,
  author = {Rein, A. and Legind, C. N. and Trapp, S.},
  title = {New concepts for dynamic plant uptake models},
  journal = {SAR and QSAR in Environmental Research},
  volume = {22},
  number = {1-2},
  pages = {191--215},
  year = {2011}
}

@article{trapp2022generic,
  author = {Trapp, S. and Matthies, M.},
  title = {Generic one-compartment model for uptake of organic chemicals by foliar vegetation},
  journal = {Environmental Science and Pollution Research},
  volume = {29},
  number = {1},
  pages = {123--134},
  year = {2022}
}

@article{vu2007nonlinear,
  author       = {Vu, Tra Thi and Vohradsky, Jiri},
  title        = {Nonlinear Differential Equation Model for Quantification of Transcriptional Regulation Applied to Microarray Data of Saccharomyces cerevisiae},
  journal      = {Nucleic Acids Research},
  volume       = {35},
  issue        = {1},
  pages        = {279--287},
  year         = {2007},
  doi          = {10.1093/nar/gkl1001}
}

@article{hung1997uptake,
  author    = {Hung, Hayley and Mackay, Donald},
  title     = {A novel and simple model of the uptake of organic chemicals by vegetation from air and soil},
  journal   = {Chemosphere},
  volume    = {35},
  number    = {5},
  pages     = {959--977},
  year      = {1997},
  doi       = {10.1016/S0045-6535(97)00182-3}
}

@ARTICLE{farsad2016,
  author={Farsad, Nariman and Yilmaz, H. Birkan and Eckford, Andrew and Chae, Chan-Byoung and Guo, Weisi},
  journal={IEEE Communications Surveys \& Tutorials},
  title={A Comprehensive Survey of Recent Advancements in Molecular Communication}, 
  year={2016},
  volume={18},
  number={3},
  pages={1887-1919},
  doi={10.1109/COMST.2016.2527741}}

@article{paterson1994model,
  author    = {Paterson, Stephen and Mackay, Donald and McFarlane, Charles},
  title     = {A model of organic chemical uptake by plants from soil and the atmosphere},
  journal   = {Environmental Science \& Technology},
  volume    = {28},
  number    = {13},
  pages     = {2259--2266},
  year      = {1994},
  doi       = {10.1021/es00062a009}
}

@article{paterson1991fugacity,
  author    = {Paterson, Sally and Mackay, Donald and Gladman, Aviv},
  title     = {A fugacity model of chemical uptake by plants from soil and air},
  journal   = {Chemosphere},
  volume    = {23},
  number    = {4},
  pages     = {539--565},
  year      = {1991},
  issn      = {0045-6535},
  doi       = {10.1016/0045-6535(91)90203-P}
}

@article{kumar2010analytical,
  author    = {Kumar, Atul and Jaiswal, Dilip Kumar and Kumar, Naveen},
  title     = {Analytical solutions to one-dimensional advection–diffusion equation with variable coefficients in semi-infinite media},
  journal   = {Journal of Hydrology},
  volume    = {380},
  number    = {3--4},
  pages     = {330--337},
  year      = {2010},
  issn      = {0022-1694},
  doi       = {10.1016/j.jhydrol.2009.11.008},
}

@article{zoppouspatial,
  author    = {Zoppou, C. and Knight, J. H.},
  title     = {Analytical Solutions for Advection and Advection-Diffusion Equations with Spatially Variable Coefficients},
  journal   = {Journal of Hydraulic Engineering},
  volume    = {123},
  number    = {2},
  pages     = {144--148},
  year      = {1997},
  doi       = {10.1061/(ASCE)0733-9429(1997)123:2(144)},
  eprint    = {https://ascelibrary.org/doi/pdf/10.1061/%28ASCE%290733-9429%281997%29123%3A2%28144%29}
}

@article{sun2022approximate,
  author    = {Sun, Yubiao and Jayaraman, Amitesh S. and Chirikjian, Gregory S.},
  title     = {Approximate solutions of the advection–diffusion equation for spatially variable flows},
  journal   = {Physics of Fluids},
  volume    = {34},
  number    = {3},
  pages     = {033318},
  year      = {2022},
  doi       = {10.1063/5.0084789},
}

@article{sanskrityayn2021generalized,
  author    = {Sanskrityayn, A. and Suk, H. and Chen, J.-S. and Park, E.},
  title     = {Generalized Analytical Solutions of The Advection-Dispersion Equation with Variable Flow and Transport Coefficients},
  journal   = {Sustainability},
  volume    = {13},
  number    = {14},
  pages     = {7796},
  year      = {2021},
  doi       = {10.3390/su13147796},
}

@article{delory2016root,
  author    = {Delory, B. M. and Delaplace, P. and Fauconnier, M. L. and Du Jardin, P.},
  title     = {Root-emitted volatile organic compounds: Can they mediate belowground plant-plant interactions?},
  journal   = {Plant and Soil},
  volume    = {402},
  pages     = {1--26},
  year      = {2016}
}

@article{Kollasch2020,
  author    = {Kollasch, A. M. and Abdul-Kafi, A. R. and Body, M. J. and Pinto, C. F. and Appel, H. M. and Cocroft, R. B.},
  title     = {Leaf vibrations produced by chewing provide a consistent acoustic target for plant recognition of herbivores},
  journal   = {Oecologia},
  volume    = {194},
  pages     = {1--13},
  year      = {2020}
}

@article{hamamoto2012organic,
  author    = {Hamamoto, Shoichiro and Moldrup, Per and Kawamoto, Ken and Komatsu, Toshiko},
  title     = {Organic Matter Fraction Dependent Model for Predicting the Gas Diffusion Coefficient in Variably Saturated Soils},
  journal   = {Vadose Zone Journal},
  volume    = {11},
  year      = {2012},
  doi       = {10.2136/vzj2011.0065},
}

@article{uteau2013root,
  author    = {Uteau, Daniel and Pagenkemper, Sebastian and Peth, Stephan},
  title     = {Root and Time Dependent Soil Structure Formation and Its Influence on Gas Transport in the Subsoil},
  journal   = {Soil and Tillage Research},
  volume    = {132},
  pages     = {69--76},
  year      = {2013},
  doi       = {10.1016/j.still.2013.05.001},
}

@article{moldrup2000predicting,
  author    = {Moldrup, Per and Olesen, Torben and Schjønning, Per and Yamaguchi, Toshiko and Rolston, Dennis E.},
  title     = {Predicting the Gas Diffusion Coefficient in Undisturbed Soil from Soil Water Characteristics},
  journal   = {Soil Science Society of America Journal},
  volume    = {64},
  pages     = {94--100},
  year      = {2000},
  doi       = {10.2136/sssaj2000.64194x},
}

@article{mendozadiffmodel,
author = {Mendoza, Carl A. and Frind, Emil O.},
title = {Advective-dispersive transport of dense organic vapors in the unsaturated zone: 1. Model development},
journal = {Water Resources Research},
volume = {26},
number = {3},
pages = {379-387},
doi = {https://doi.org/10.1029/WR026i003p00379},
eprint = {https://agupubs.onlinelibrary.wiley.com/doi/pdf/10.1029/WR026i003p00379},
year = {1990}
}

@article{sleepdiffmodel,
author = {Sleep, B. E. and Sykes, J. F.},
title = {Modeling the transport of volatile organics in variably saturated media},
journal = {Water Resources Research},
volume = {25},
number = {1},
pages = {81-92},
doi = {https://doi.org/10.1029/WR025i001p00081},
eprint = {https://agupubs.onlinelibrary.wiley.com/doi/pdf/10.1029/WR025i001p00081},
year = {1989}
}

@article{soilmicrobialdiff,
author = {Schjønning, Per and Thomsen, Ingrid K. and Moldrup, Per and Christensen, Bent T.},
title = {Linking Soil Microbial Activity to Water- and Air-Phase Contents and Diffusivities},
journal = {Soil Science Society of America Journal},
volume = {67},
number = {1},
pages = {156-165},
doi = {https://doi.org/10.2136/sssaj2003.1560},
eprint = {https://acsess.onlinelibrary.wiley.com/doi/pdf/10.2136/sssaj2003.1560},
year = {2003}
}

@article{soilarands,
author = {Arands, R. and Lam, T. and Massry, I. and Berler, D. H. and Muzzio, F. J. and Kosson, D. S.},
title = {Modeling and experimental validation of volatile organic contaminant diffusion through an unsaturated soil},
journal = {Water Resources Research},
volume = {33},
number = {4},
pages = {599-609},
doi = {https://doi.org/10.1029/96WR03976},
eprint = {https://agupubs.onlinelibrary.wiley.com/doi/pdf/10.1029/96WR03976},
year = {1997}
}

@article{moldrup2004three,
  author    = {Moldrup, Per and Olesen, Torben and Yoshikawa, Shinsuke and Komatsu, Toshiko and Rolston, Dennis E.},
  title     = {Three-Porosity Model for Predicting the Gas Diffusion Coefficient in Undisturbed Soil},
  journal   = {Soil Science Society of America Journal},
  volume    = {68},
  pages     = {750--759},
  year      = {2004},
  doi       = {10.2136/sssaj2004.7500},
}

@article{Wang2024,
  author = {Wang, S. and Song, L. and He, H. and others},
  title = {A Two-dimensional Analytical Model for Volatile Organic Compound Diffusion Through the Unsaturated Soil and Horizontal Permeable Reactive Barriers},
  journal = {Water, Air, \& Soil Pollution},
  volume = {235},
  pages = {414},
  year = {2024},
  doi = {10.1007/s11270-024-07211-4}
}

@article{moldrup2007predictive,
  author    = {Moldrup, Per and Olesen, Torben and Blendstrup, Helle and Komatsu, Toshiko and de Jonge, Lis W. and Rolston, Dennis E.},
  title     = {Predictive-Descriptive Models for Gas and Solute Diffusion Coefficients in Variably Saturated Porous Media Coupled to Pore-Size Distribution: IV. Solute Diffusivity and the Liquid Phase Impedance Factor},
  journal   = {Soil Science},
  volume    = {172},
  number    = {10},
  pages     = {741--750},
  month     = {October},
  year      = {2007},
  doi       = {10.1097/SS.0b013e3180d0a423},
}

@article{fantke2011plant,
  author    = {Fantke, P. and Charles, R. and de Alencastro, L. F. and Friedrich, R. and Jolliet, O.},
  title     = {Plant Uptake of Pesticides and Human Health: Dynamic Modeling of Residues in Wheat and Ingestion Intake},
  journal   = {Chemosphere},
  volume    = {85},
  number    = {10},
  pages     = {1639--1647},
  year      = {2011},
  doi       = {10.1016/j.chemosphere.2011.08.030},
}

@article{fantke2013dynamics,
  author    = {Fantke, Peter and Wieland, Peter and Wannaz, Cédric and Friedrich, Rainer and Jolliet, Olivier},
  title     = {Dynamics of Pesticide Uptake into Plants: From System Functioning to Parsimonious Modeling},
  journal   = {Environmental Modelling \& Software},
  volume    = {40},
  pages     = {316--324},
  year      = {2013},
  issn      = {1364-8152},
  doi       = {10.1016/j.envsoft.2012.09.016},
}

@article{trapp2003fruit,
  author    = {Trapp, Stefan and Rasmussen, D. and Samsøe-Petersen, L.},
  title     = {Fruit Tree Model for Uptake of Organic Compounds from Soil},
  journal   = {SAR and QSAR in Environmental Research},
  volume    = {14},
  number    = {1},
  pages     = {17--26},
  year      = {2003},
  doi       = {10.1080/1062936021000058755},
}

@article{niinemets2004physiological,
  author = {Niinemets, U. and Loreto, F. and Reichstein, M.},
  title = {Physiological and physicochemical controls on foliar volatile organic compound emissions},
  journal = {Trends in Plant Science},
  volume = {9},
  number = {4},
  pages = {180--186},
  year = {2004}
}

@article{niinemets2010mild,
  author = {Niinemets, U.},
  title = {Mild versus severe stress and BVOCs: Thresholds, priming and consequences},
  journal = {Trends in Plant Science},
  volume = {15},
  number = {3},
  pages = {145--153},
  year = {2010},
}

@article{ueda2012plant,
  author = {Ueda, H. and Kikuta, Y. and Matsuda, K.},
  title = {Plant communication: Mediated by individual or blended {VOCs}?},
  journal = {Plant Signaling \& Behavior},
  volume = {7},
  number = {2},
  pages = {222--226},
  year = {2012},
  doi = {10.4161/psb.19216}
}

@article{copolovici2011volatile,
  author = {Copolovici, L. and Kännaste, A. and Remmel, T. and Vislap, V. and Niinemets, U.},
  title = {Volatile emissions from \textit{Alnus glutinosa} induced by herbivory are quantitatively related to the extent of damage},
  journal = {Journal of Chemical Ecology},
  volume = {37},
  number = {1},
  pages = {18--28},
  year = {2011},
  doi = {10.1007/s10886-010-9894-8}
}

@article{wang2021root,
  author = {Wang, N. Q. and Kong, C. H. and Wang, P. and Meiners, S. J.},
  title = {Root exudate signals in plant--plant interactions},
  journal = {Plant, Cell \& Environment},
  volume = {44},
  number = {4},
  pages = {1044--1058},
  year = {2021},
  doi = {10.1111/pce.13966}
}

@article{dudareva2013,
author = {Dudareva, Natalia and Klempien, Antje and Muhlemann, Joëlle K. and Kaplan, Ian},
title = {Biosynthesis, function and metabolic engineering of plant volatile organic compounds},
journal = {New Phytologist},
volume = {198},
number = {1},
pages = {16-32},
doi = {https://doi.org/10.1111/nph.12145},
eprint = {https://nph.onlinelibrary.wiley.com/doi/pdf/10.1111/nph.12145},
year = {2013}
}

@ARTICLE{aktas2023odorbasedmolecularcommunicationsstateoftheart,
  author={Aktas, Dilara and Ortlek, Beyza E. and Civas, Meltem and Baradari, Elham and Kilic, Ahmet B. and Bilgen, Fatih E. and Okcu, Ayse S. and Whitfield, Melanie and Cetinkaya, Oktay and Akan, Ozgur B.},
  journal={IEEE Communications Surveys \& Tutorials},
  title={Odor-Based Molecular Communications: State-of-the-Art, Vision, Challenges, and Frontier Directions}, 
  year={2025},
  volume={27},
  number={4},
  pages={2658-2692}}

@article{horn1994,
  author    = {R. Horn and H. Taubner and M. Wuttke and T. Baumgartl},
  title     = {Soil physical properties related to soil structure},
  journal   = {Soil and Tillage Research},
  volume    = {30},
  number    = {2-4},
  pages     = {187--216},
  year      = {1994},
  publisher = {Elsevier}
}

@article{soildiff,
author = {Laemmel, T. and Maier, M. and Schack-Kirchner, H. and Lang, F.},
title = {An in situ method for real-time measurement of gas transport in soil},
journal = {European Journal of Soil Science},
volume = {68},
number = {2},
pages = {156-166},
doi = {https://doi.org/10.1111/ejss.12412},
eprint = {https://bsssjournals.onlinelibrary.wiley.com/doi/pdf/10.1111/ejss.12412},
year = {2017}
}

@article{soilpore,
author = {Arthur, Emmanuel and Moldrup, Per and Schjønning, Per and de Jonge, Lis W.},
title = {Linking Particle and Pore Size Distribution Parameters to Soil Gas Transport Properties},
journal = {Soil Science Society of America Journal},
volume = {76},
number = {1},
pages = {18-27},
doi = {https://doi.org/10.2136/sssaj2011.0125},
eprint = {https://acsess.onlinelibrary.wiley.com/doi/pdf/10.2136/sssaj2011.0125},
year = {2012}
}

@article{eden2012,
  author    = {Marie Eden and Per Moldrup and Per Schjønning and Kate M. Scow and Lis W. de Jonge},
  title     = {Soil-Gas Phase Transport and Structure Parameters for a Soil Under Different Management Regimes and at Two Moisture Levels},
  journal   = {Soil Science},
  volume    = {177},
  number    = {9},
  pages     = {527--534},
  year      = {2012},
  month     = {Sep},
  doi       = {10.1097/SS.0b013e318267ec85}
}

@article{thomsen1999,
  author    = {Anne Belinda Thomsen and Kaj Henriksen and Christian Grøn and Per Møldrup},
  title     = {Sorption, Transport, and Degradation of Quinoline in Unsaturated Soil},
  journal   = {Environmental Science \& Technology},
  volume    = {33},
  number    = {17},
  pages     = {2891--2898},
  year      = {1999},
  doi       = {10.1021/es981065t}
}

@article{nye1966,
  author    = {P.H. Nye},
  title     = {The effect of the nutrient intensity and buffering power of a soil, and the absorbing power, size and root hairs of a root, on nutrient absorption by diffusion},
  journal   = {Plant and Soil},
  volume    = {25},
  pages     = {81--105},
  year      = {1966},
  doi       = {10.1007/BF01347964}
}

@article{boyno2022plant,
  author    = {Boyno, G. and Demir, S.},
  title     = {Plant-mycorrhiza communication and mycorrhizae in inter-plant communication},
  journal   = {Symbiosis},
  volume    = {86},
  number    = {2},
  pages     = {155--168},
  year      = {2022}
}

@article{babikova2013underground,
  author    = {Babikova, Z. and Gilbert, L. and Bruce, T. J. A. and Birkett, M. and Caulfield, J. C. and Woodcock, C. and Johnson, D.},
  title     = {Underground signals carried through common mycelial networks warn neighbouring plants of aphid attack},
  journal   = {Ecology Letters},
  volume    = {16},
  pages     = {835--843},
  year      = {2013},
  doi       = {10.1111/ele.12115},
}

@article{Mahall1991,
  author    = {B. E. Mahall and R. M. Callaway},
  title     = {Root communication among desert shrubs},
  journal   = {Proceedings of the National Academy of Sciences of the United States of America},
  volume    = {88},
  number    = {3},
  pages     = {874--876},
  year      = {1991},
  month     = {Feb},
  doi       = {10.1073/pnas.88.3.874},
  pmid      = {11607151},
  pmcid     = {PMC50916}
}

@article{song2010interplant,
  author    = {Song, Y. Y. and Zeng, R. S. and Xu, J. A. F. and Li, J. and Shen, X. A. and Yihdego, W. G.},
  title     = {Inter-plant communication of tomato plants through underground common mycorrhizal networks},
  journal   = {PLoS One},
  volume    = {5},
  pages     = {Art. no. e13324},
  year      = {2010},
  doi       = {10.1371/journal.pone.0013324}
}

@article{oelmuller2019interplant,
  author    = {Oelm{\"u}ller, R.},
  title     = {Interplant communication via hyphal networks},
  journal   = {Plant Physiology Reports},
  volume    = {24},
  number    = {4},
  pages     = {463--473},
  year      = {2019}
}

@article{barto2011fungal,
  author    = {Barto, E. K. and others},
  title     = {The fungal fast lane: Common mycorrhizal networks extend bioactive zones of allelochemicals in soils},
  journal   = {PLoS ONE},
  volume    = {6},
  pages     = {Art. no. e27195},
  year      = {2011},
  doi       = {10.1371/journal.pone.0027195}
}

@article{Volkov2016,
  author    = {Volkov, A. G. and Shtessel, Y. B.},
  title     = {Propagation of electrotonic potentials in plants: experimental study and mathematical modeling},
  journal   = {AIMS Biophysics},
  volume    = {3},
  number    = {3},
  pages     = {358-379},
  year      = {2016},
  doi       = {10.3934/biophy.2016.3.358}
}

@article{Volkov2019fly,
  author    = {Volkov, Alexander G.},
  title     = {Signaling in electrical networks of the Venus flytrap (Dionaea muscipula Ellis)},
  journal   = {Bioelectrochemistry},
  volume    = {125},
  pages     = {25-32},
  year      = {2019},
  issn      = {1567-5394},
  doi       = {10.1016/j.bioelechem.2018.09.001},
}

@article{Volkov2018,
  author    = {Alexander G. Volkov and Yuri B. Shtessel},
  title     = {Electrical signal propagation within and between tomato plants},
  journal   = {Bioelectrochemistry},
  volume    = {124},
  pages     = {195--205},
  year      = {2018},
  issn      = {1567-5394},
  doi       = {10.1016/j.bioelechem.2018.08.001},
}

@article{Chen2016,
  author    = {Chen, Y. and Zhao, D.-J. and Wang, Z.-Y. and Wang, Z.-Y. and Tang, G. and Huang, L.},
  title     = {Plant Electrical Signal Classification Based on Waveform Similarity},
  journal   = {Algorithms},
  volume    = {9},
  number    = {4},
  pages     = {70},
  year      = {2016},
  doi       = {10.3390/a9040070}
}

@article{Wang2007,
  author    = {Wang, C. and Huang, L. and Wang, Z. and Qiao, X.},
  title     = {Monitoring and analysis of electrical signals in water-stressed plants},
  journal   = {New Zealand Journal of Agricultural Research},
  volume    = {50},
  number    = {5},
  pages     = {823--829},
  year      = {2007},
  doi       = {10.1080/00288230709510356}
}

@article{Volkov2013,
  author    = {Volkov, A. G. and Vilfranc, C. L. and Murphy, V. A. and Mitchell, C. M. and Volkova, M. I. and O'Neal, L. and Markin, V. S.},
  title     = {Electrotonic and action potentials in the Venus flytrap},
  journal   = {Journal of Plant Physiology},
  volume    = {170},
  number    = {9},
  pages     = {838-846},
  year      = {2013},
  month     = {Jun 15},
  doi       = {10.1016/j.jplph.2013.01.009},
  pmid      = {23422156}
}

@article{black1971,
author = {Black, J. D. and Forsyth, F. R. and Fensom, D. S. and Ross, R. B.},
title = {Electrical stimulation and its effects on growth and ion accumulation in tomato plants},
journal = {Canadian Journal of Botany},
volume = {49},
number = {10},
pages = {1809-1815},
year = {1971},
doi = {10.1139/b71-255},
eprint = {https://doi.org/10.1139/b71-255}}

@article{Volkov2017Aloe,
  author    = {Alexander G. Volkov and Yuri B. Shtessel},
  title     = {Electrotonic signal transduction between Aloe vera plants using underground pathways in soil: Experimental and analytical study},
  journal   = {AIMS Biophysics},
  volume    = {4},
  number    = {4},
  year      = {2017},
  pages     = {576--593},
  doi       = {10.3934/biophy.2017.4.576},
}

@article{Volkov2019,
  author    = {Alexander G. Volkov and Shannon Toole and Mwangi WaMaina},
  title     = {Electrical signal transmission in the plant-wide web},
  journal   = {Bioelectrochemistry},
  volume    = {129},
  pages     = {70--78},
  year      = {2019},
  issn      = {1567-5394},
  doi       = {10.1016/j.bioelechem.2019.05.003},
}

@incollection{Lew2009,
  author    = {Lew, Roger R.},
  title     = {Root Hair Electrophysiology},
  booktitle = {Root Hairs},
  editor    = {Emons, Andrea M. C. and Ketelaar, Tijs},
  series    = {Plant Cell Monographs},
  volume    = {12},
  publisher = {Springer, Berlin, Heidelberg},
  year      = {2009},
  pages     = {149--165},
  doi       = {10.1007/978-3-540-79405-9_9},
}

@article{Spanswick1972,
  author    = {Spanswick, R. M.},
  title     = {Electrical coupling between cells of higher plants: A direct demonstration of intercellular communication},
  journal   = {Planta},
  volume    = {102},
  year      = {1972},
  pages     = {215--227},
  doi       = {10.1007/BF00386892},
}

@article{Watanabe1995,
  author    = {Yumi Watanabe and Satoshi Takeuchi and Mitsuo Ashisada and Yasunari Ikezawa and Tsutomu Takamura},
  title     = {Potential Distribution and Ionic Concentration at the Bean Root Surface of the Growing Tip and Lateral Root Emerging Points},
  journal   = {Plant and Cell Physiology},
  volume    = {36},
  number    = {4},
  pages     = {691--698},
  year      = {1995},
  month     = {June},
  doi       = {10.1093/oxfordjournals.pcp.a078810},
}

@book{Crocker1998,
  editor    = {Malcolm J. Crocker},
  title     = {Handbook of Acoustics},
  year      = {1998},
  publisher = {Wiley},
  address   = {New York},
  isbn      = {978-0-471-25293-1},
  pages     = {1488}
}

@incollection{Attenborough2014,
  author    = {Keith Attenborough},
  title     = {Sound Propagation in the Atmosphere},
  booktitle = {Springer Handbook of Acoustics},
  editor    = {Thomas D. Rossing},
  publisher = {Springer, New York, NY},
  series    = {Springer Handbooks},
  year      = {2014},
  doi       = {10.1007/978-1-4939-0755-7_4}
}

@article{Chevret1996,
  author    = {P. Chevret and Ph. Blanc‐Benon and D. Juvé},
  title     = {A numerical model for sound propagation through a turbulent atmosphere near the ground},
  journal   = {Journal of the Acoustical Society of America},
  volume    = {100},
  number    = {6},
  pages     = {3587--3599},
  year      = {1996},
  doi       = {10.1121/1.417224}
}

@article{Daigle1986,
  author    = {G. A. Daigle and T. F. W. Embleton and J. E. Piercy},
  title     = {Propagation of sound in the presence of gradients and turbulence near the ground},
  journal   = {Journal of the Acoustical Society of America},
  volume    = {79},
  number    = {3},
  pages     = {613--627},
  year      = {1986},
  doi       = {10.1121/1.393451}
}

@article{Bullen1982,
  author    = {R. Bullen and F. Fricke},
  title     = {Sound propagation through vegetation},
  journal   = {Journal of Sound and Vibration},
  volume    = {80},
  number    = {1},
  pages     = {11--23},
  year      = {1982},
  issn      = {0022-460X},
  doi       = {10.1016/0022-460X(82)90387-X},
}

@phdthesis{Chobeau2014,
  author    = {Pierre Chobeau},
  title     = {Modeling of sound propagation in forests using the Transmission Line Matrix method},
  school    = {Université du Maine},
  year      = {2014},
  type      = {Ph.D. Thesis},
  language  = {English},
}

@article{Huang2022,
  author    = {Shenghai Huang and Caiyun Lu and Hongwen Li and Jin He and Qingjie Wang and Zhen Gao and Panpan Yuan and Yunxiang Li},
  title     = {The attenuation mechanism and regular of the acoustic wave on propagation path in farmland soil},
  journal   = {Computers and Electronics in Agriculture},
  volume    = {199},
  year      = {2022},
  pages     = {107138},
  issn      = {0168-1699},
  doi       = {10.1016/j.compag.2022.107138}
}

@article{Peng1995,
  author    = {C. Peng and J.A. Lines},
  title     = {Noise Propagation in the Agricultural Environment},
  journal   = {Journal of Agricultural Engineering Research},
  volume    = {60},
  number    = {3},
  year      = {1995},
  pages     = {155-165},
  issn      = {0021-8634},
  doi       = {10.1006/jaer.1995.1009}
}

@article{Price1988,
  author    = {Margaret A. Price and Keith Attenborough and Nicholas W. Heap},
  title     = {Sound attenuation through trees: Measurements and models},
  journal   = {Journal of the Acoustical Society of America},
  volume    = {84},
  number    = {5},
  year      = {1988},
  pages     = {1836-1844},
  doi       = {10.1121/1.397150}
}

@article{Albers2009,
  author    = {B. Albers},
  title     = {Analysis of the Propagation of Sound Waves in Partially Saturated Soils by Means of a Macroscopic Linear Poroelastic Model},
  journal   = {Transport in Porous Media},
  volume    = {80},
  year      = {2009},
  pages     = {173--192},
  doi       = {10.1007/s11242-009-9360-y}
}

@article{Embleton1996,
  author    = {Tony F. W. Embleton},
  title     = {Tutorial on sound propagation outdoors},
  journal   = {Journal of the Acoustical Society of America},
  volume    = {100},
  number    = {1},
  pages     = {31--48},
  year      = {1996},
  doi       = {10.1121/1.415879}
}

@article{due1993,
author = {DUE, G.},
title = {Interpretation of the electrical potential on the surface of plant roots},
journal = {Plant, Cell \& Environment},
volume = {16},
number = {5},
pages = {501-510},
doi = {https://doi.org/10.1111/j.1365-3040.1993.tb00897.x},
eprint = {https://onlinelibrary.wiley.com/doi/pdf/10.1111/j.1365-3040.1993.tb00897.x},
year = {1993}
}

@article{Szechynska-Hebda2022,
  author    = {Magdalena Szechyńska-Hebda and Maria Lewandowska and Damian Witoń and Yosef Fichman and Ron Mittler and Stanisław M. Karpiński},
  title     = {Aboveground plant-to-plant electrical signaling mediates network acquired acclimation},
  journal   = {The Plant Cell},
  volume    = {34},
  number    = {8},
  pages     = {3047--3065},
  year      = {2022},
  doi       = {10.1093/plcell/koac150},
}

@article{Vodeneev2016,
  author    = {Vodeneev, V. A. and Katicheva, L. A. and Sukhov, V. S.},
  title     = {Electrical signals in higher plants: Mechanisms of generation and propagation},
  journal   = {BIOPHYSICS},
  volume    = {61},
  pages     = {505--512},
  year      = {2016},
  doi       = {10.1134/S0006350916030209}
}

@article{Kilic2024MRSK,
  author    = {Kilic, B. A. and Akan, O. B.},
  title     = {Multi Ratio Shift Keying (MRSK) for Molecular Communication},
  journal   = {arXiv preprint},
  year      = {2024},
  eprint    = {2412.20161},
  archivePrefix = {arXiv},
  primaryClass = {cs.IT}
}

@article{ahmed2022molecular,
  author = {Ahmed, S. and Hu, J. and Naqvi, S. M. and Zhang, Y. and Linze, L. and Iderawumi, A. M.},
  title = {Molecular communication network and its applications in crop sciences},
  journal = {Planta},
  volume = {255},
  number = {6},
  pages = {128},
  year = {2022},
}

@article{Sukhov2009,
  author    = {Sukhov, V. and Vodeneev, V.},
  title     = {A mathematical model of action potential in cells of vascular plants},
  journal   = {Journal of Membrane Biology},
  volume    = {232},
  number    = {1-3},
  pages     = {59--67},
  year      = {2009},
  month     = {December},
  doi       = {10.1007/s00232-009-9218-9},
  pmid      = {19921324}
}

@article{Beilby2007,
  author    = {Mary Jane Beilby},
  title     = {Action Potential in Charophytes},
  journal   = {International Review of Cytology},
  publisher = {Academic Press},
  volume    = {257},
  pages     = {43-82},
  year      = {2007},
  issn      = {0074-7696},
  isbn      = {9780123737014},
  doi       = {10.1016/S0074-7696(07)57002-6}
}

@article{Shabala2006,
  author    = {Sergey Shabala and Lana Shabala and Dietrich Gradmann and Zhonghua Chen and Ian Newman and Stefano Mancuso},
  title     = {Oscillations in plant membrane transport: model predictions, experimental validation, and physiological implications},
  journal   = {Journal of Experimental Botany},
  volume    = {57},
  number    = {1},
  pages     = {171--184},
  year      = {2006},
  doi       = {10.1093/jxb/erj022}
}

@article{Cai2017,
  author    = {Jianchao Cai and Wei Wei and Xiangyun Hu and David A. Wood},
  title     = {Electrical conductivity models in saturated porous media: A review},
  journal   = {Earth-Science Reviews},
  volume    = {171},
  pages     = {419--433},
  year      = {2017},
  issn      = {0012-8252},
  doi       = {10.1016/j.earscirev.2017.06.013}
}

@article{rhoades1989,
author = {Rhoades, J. D. and Manteghi, N. A. and Shouse, P. J. and Alves, W. J.},
title = {Soil Electrical Conductivity and Soil Salinity: New Formulations and Calibrations},
journal = {Soil Science Society of America Journal},
volume = {53},
number = {2},
pages = {433-439},
doi = {https://doi.org/10.2136/sssaj1989.03615995005300020020x},
eprint = {https://acsess.onlinelibrary.wiley.com/doi/pdf/10.2136/sssaj1989.03615995005300020020x},
year = {1989}
}

@article{Sudduth2013,
  author    = {Kenneth A. Sudduth and D. Brenton Myers and Newell R. Kitchen and Scott T. Drummond},
  title     = {Modeling soil electrical conductivity–depth relationships with data from proximal and penetrating ECa sensors},
  journal   = {Geoderma},
  volume    = {199},
  pages     = {12--21},
  year      = {2013},
  issn      = {0016-7061},
  doi       = {10.1016/j.geoderma.2012.10.006}
}

@article{Rhoades1990,
  author    = {J. D. Rhoades and D. L. Corwin},
  title     = {Soil electrical conductivity: Effects of soil properties and application to soil salinity appraisal},
  journal   = {Communications in Soil Science and Plant Analysis},
  volume    = {21},
  number    = {11--12},
  pages     = {837--860},
  year      = {1990},
  doi       = {10.1080/00103629009368274}
}

@article{amente2000,
author = {Amente, G. and Baker, John M. and Reece, Clive F.},
title = {Estimation of Soil Solution Electrical Conductivity from Bulk Soil Electrical Conductivity in Sandy Soils},
journal = {Soil Science Society of America Journal},
volume = {64},
number = {6},
pages = {1931-1939},
doi = {https://doi.org/10.2136/sssaj2000.6461931x},
eprint = {https://acsess.onlinelibrary.wiley.com/doi/pdf/10.2136/sssaj2000.6461931x},
year = {2000}
}

@article{Ma2011,
  author    = {R. Ma and A. McBratney and B. Whelan and et al.},
  title     = {Comparing temperature correction models for soil electrical conductivity measurement},
  journal   = {Precision Agriculture},
  volume    = {12},
  pages     = {55--66},
  year      = {2011},
  doi       = {10.1007/s11119-009-9156-7}
}

@article{Friedman2005,
  author    = {Shmulik P. Friedman},
  title     = {Soil properties influencing apparent electrical conductivity: a review},
  journal   = {Computers and Electronics in Agriculture},
  volume    = {46},
  number    = {1--3},
  pages     = {45--70},
  year      = {2005},
  issn      = {0168-1699},
  doi       = {10.1016/j.compag.2004.11.001}
}

@article{Gradmann2001,
  author    = {D. Gradmann},
  title     = {Impact of Apoplast Volume on Ionic Relations in Plant Cells},
  journal   = {J. Membrane Biol.},
  volume    = {184},
  pages     = {61--69},
  year      = {2001},
  doi       = {10.1007/s00232-001-0074-5}
}

@article{beilby1982,
author = {Beilby, M. J.  and Keynes, Richard Darwin  and Walker, N. A.  and Keynes, Richard Darwin  and Ellory, J. C. },
title = {Cl<sup>-</sup> channels in <i>Chara</i>},
journal = {Philosophical Transactions of the Royal Society of London. B, Biological Sciences},
volume = {299},
number = {1097},
pages = {435-445},
year = {1982},
doi = {10.1098/rstb.1982.0142},
eprint = {https://royalsocietypublishing.org/doi/pdf/10.1098/rstb.1982.0142}}

@article{Sukhov2011,
  author    = {Vladimir Sukhov and Vladimir Nerush and Lyubov Orlova and Vladimir Vodeneev},
  title     = {Simulation of action potential propagation in plants},
  journal   = {Journal of Theoretical Biology},
  volume    = {291},
  pages     = {47-55},
  year      = {2011},
  issn      = {0022-5193},
  doi       = {10.1016/j.jtbi.2011.09.019}
}

@article{Sukhov2013,
  author    = {V. Sukhov and E. Akinchits and L. Katicheva and et al.},
  title     = {Simulation of Variation Potential in Higher Plant Cells},
  journal   = {J Membrane Biol},
  volume    = {246},
  pages     = {287--296},
  year      = {2013},
  doi       = {10.1007/s00232-013-9529-8}
}

@article{Vodeneev2012,
  author    = {Vladimir Vodeneev and Anna Orlova and Ekaterina Morozova and Lyubov Orlova and Elena Akinchits and Olga Orlova and Vladimir Sukhov},
  title     = {The mechanism of propagation of variation potentials in wheat leaves},
  journal   = {Journal of Plant Physiology},
  volume    = {169},
  number    = {10},
  pages     = {949-954},
  year      = {2012},
  issn      = {0176-1617},
  doi       = {10.1016/j.jplph.2012.02.013}
}

@article{cskmsk,
    author={Kuran, M. S. and Yilmaz, H. B. and Tugcu, T. and Akyildiz, I. F.},
    booktitle={2011 IEEE International Conference on Communications (ICC)}, 
    title={Modulation Techniques for Communication via Diffusion in Nanonetworks}, 
    year={2011},
    volume={},
    number={},
    pages={1-5},
    doi={10.1109/icc.2011.5962989}}

@article{Vodeneev2015,
  author    = {V. Vodeneev and E. Akinchits and V. Sukhov},
  title     = {Variation potential in higher plants: Mechanisms of generation and propagation},
  journal   = {Plant Signaling \& Behavior},
  volume    = {10},
  number    = {9},
  year      = {2015},
  doi       = {10.1080/15592324.2015.1057365},
}

@article{Zimmermann2009,
  author    = {Matthias R. Zimmermann and Heiko Maischak and Axel MithOfer and Wilhelm Boland and Hubert H. Felle},
  title     = {System Potentials, a Novel Electrical Long-Distance Apoplastic Signal in Plants, Induced by Wounding},
  journal   = {Plant Physiology},
  volume    = {149},
  number    = {3},
  pages     = {1593--1600},
  year      = {2009},
  doi       = {10.1104/pp.108.133884},
}

@article{Khait2023,
  author       = {Khait, Itzhak and Lewin-Epstein, Ohad and Sharon, Roni and Saban, Keren and Goldstein, Raaya and Anikster, Yonatan and Zeron, Yair and Agassy, Clara and Nizan, Sarit and Sharabi, Gal and Perelman, Rami and Boonman, Arjan and Sade, Naomi and Yovel, Yossi and Hadany, Lilach},
  title        = {Sounds emitted by plants under stress are airborne and informative},
  journal      = {Cell},
  volume       = {186},
  number       = {7},
  pages        = {1328--1336.e10},
  year         = {2023},
  doi          = {10.1016/j.cell.2023.03.009},
}

@article{karban2021plant,
  author = {Karban, R.},
  title = {Plant communication},
  journal = {Annual Review of Ecology, Evolution, and Systematics},
  volume = {52},
  number = {1},
  pages = {1--24},
  year = {2021}
}

@article{Mishra2016,
  author = {R.C. Mishra and R. Ghosh and H. Bae},
  title = {Plant acoustics: In the search of a sound mechanism for sound signaling in plants},
  journal = {Journal of Experimental Botany},
  volume = {67},
  number = {15},
  pages = {4483--4494},
  year = {2016},
  month = {August},
  doi = {10.1093/jxb/erw235},
  pmid = {27342223},
}

@article{Demey2023,
  author       = {Demey, Michelle L. and Mishra, Rishav C. and Van Der Straeten, Dominique},
  title        = {Sound perception in plants: from ecological significance to molecular understanding},
  journal      = {Trends in Plant Science},
  volume       = {28},
  number       = {7},
  pages        = {825--840},
  year         = {2023},
  doi          = {10.1016/j.tplants.2023.03.003},
}

@article{Rockwell2014,
  author = {Fulton E. Rockwell and James K. Wheeler and N. Michele Holbrook},
  title = {Cavitation and Its Discontents: Opportunities for Resolving Current Controversies},
  journal = {Plant Physiology},
  volume = {164},
  number = {4},
  pages = {1649--1660},
  year = {2014},
  month = {April},
  doi = {10.1104/pp.113.233817},
}

@article{Singhal2002,
  author = {A. K. Singhal and M. M. Athavale and H. Li and Y. Jiang},
  title = {Mathematical Basis and Validation of the Full Cavitation Model},
  journal = {Journal of Fluids Engineering},
  volume = {124},
  number = {3},
  pages = {617--624},
  year = {2002},
  month = {September},
  doi = {10.1115/1.1486223},
}

@article{Peshkovsky2008,
  author = {Sergei L. Peshkovsky and Alexey S. Peshkovsky},
  title = {Shock-wave model of acoustic cavitation},
  journal = {Ultrasonics Sonochemistry},
  volume = {15},
  number = {4},
  pages = {618--628},
  year = {2008},
  issn = {1350-4177},
  doi = {10.1016/j.ultsonch.2007.07.006},
}

@article{Roohi2019,
  author = {Reza Roohi and Elahe Abedi and Seyed Mohammad Bagher Hashemi and Krystian Marszałek and Jose Manuel Lorenzo and Francisco J. Barba},
  title = {Ultrasound-assisted bleaching: Mathematical and 3D computational fluid dynamics simulation of ultrasound parameters on microbubble formation and cavitation structures},
  journal = {Innovative Food Science \& Emerging Technologies},
  volume = {55},
  pages = {66--79},
  year = {2019},
  issn = {1466-8564},
  doi = {10.1016/j.ifset.2019.05.014},
}

@article{Laborde1998,
  author = {J.-L. Laborde and C. Bouyer and J.-P. Caltagirone and A. Gérard},
  title = {Acoustic cavitation field prediction at low and high frequency ultrasounds},
  journal = {Ultrasonics},
  volume = {36},
  number = {1-5},
  pages = {581--587},
  year = {1998},
  issn = {0041-624X},
  doi = {10.1016/S0041-624X(97)00106-6},
}

@article{Zhu2024,
  author = {Xianglu Zhu and Rahel Suchintita Das and Mysore Lokesh Bhavya and Marco Garcia-Vaquero and Brijesh K. Tiwari},
  title = {Acoustic cavitation for agri-food applications: Mechanism of action, design of new systems, challenges and strategies for scale-up},
  journal = {Ultrasonics Sonochemistry},
  volume = {105},
  pages = {106850},
  year = {2024},
  issn = {1350-4177},
  doi = {10.1016/j.ultsonch.2024.106850},
}

@article{sukhova2017,
  author    = {Sukhova, E. and Akinchits, E. and Sukhov, V.},
  title     = {Mathematical Models of Electrical Activity in Plants},
  journal   = {Journal of Membrane Biology},
  volume    = {250},
  pages     = {407--423},
  year      = {2017},
  doi       = {10.1007/s00232-017-9969-7}
}

@article{Vanhille2012,
  author = {Christian Vanhille and Cleofé Campos-Pozuelo},
  title = {Acoustic cavitation mechanism: A nonlinear model},
  journal = {Ultrasonics Sonochemistry},
  volume = {19},
  number = {2},
  pages = {217--220},
  year = {2012},
  issn = {1350-4177},
  doi = {10.1016/j.ultsonch.2011.06.019},
}

@article{Gagliano2012,
  author    = {Monica Gagliano and Martin Renton and Ohad Duvdevani and Mandy Timmins and Stefano Mancuso},
  title     = {Out of Sight but Not out of Mind: Alternative Means of Communication in Plants},
  journal   = {PLOS ONE},
  volume    = {7},
  number    = {5},
  pages     = {Art. no. e37382},
  year      = {2012},
  month     = {May},
  doi       = {10.1371/journal.pone.0037382},
}

@article{Ghosh2016,
  author       = {Ghosh, Ritesh and Mishra, Ratnesh Chandra and Choi, Byung-Ho and Kwon, Yong Seok and Bae, Do Won and Park, Sang Chul and Bae, Hanhong},
  title        = {Exposure to sound vibrations lead to transcriptomic, proteomic and hormonal changes in Arabidopsis},
  journal      = {Scientific Reports},
  volume       = {6},
  number       = {1},
  pages        = {33370},
  year         = {2016},
  doi          = {10.1038/srep33370},
}

@article{Milburn1966,
  author    = {Milburn, J. A. and Johnson, R. P. C.},
  title     = {The conduction of sap},
  journal   = {Planta},
  volume    = {69},
  pages     = {43--52},
  year      = {1966},
  doi       = {10.1007/BF00380209}
}

@article{Laschimke2006,
  author    = {Laschimke, Ralf and Burger, Maria and Vallen, Hartmut},
  title     = {Acoustic emission analysis and experiments with physical model systems reveal a peculiar nature of the xylem tension},
  journal   = {Journal of Plant Physiology},
  volume    = {163},
  number    = {10},
  pages     = {996--1007},
  year      = {2006},
  issn      = {0176-1617},
  doi       = {10.1016/j.jplph.2006.05.004},
}

@article{kikuta1997,
author = {KIKUTA, S. B. and LO GULLO, M. A. and NARDINI, A. and RICHTER, H. and SALLEO, S.},
title = {Ultrasound acoustic emissions from dehydrating leaves of deciduous and evergreen trees},
journal = {Plant, Cell \& Environment},
volume = {20},
number = {11},
pages = {1381-1390},
doi = {https://doi.org/10.1046/j.1365-3040.1997.d01-34.x},
eprint = {https://onlinelibrary.wiley.com/doi/pdf/10.1046/j.1365-3040.1997.d01-34.x},
year = {1997}
}

@article{Jung2020,
  author       = {Jung, Jae and Kim, Seok Ki and Jung, Seung Ho and Jeong, Mi Jung and Ryu, Choong-Min},
  title        = {Sound vibration-triggered epigenetic modulation induces plant root immunity against *Ralstonia solanacearum*},
  journal      = {Frontiers in Microbiology},
  volume       = {11},
  pages        = {1978},
  year         = {2020},
  doi          = {10.3389/fmicb.2020.01978},
}

@article{Appel2014,
  author       = {Appel, H. M. and Cocroft, R. B.},
  title        = {Plants respond to leaf vibrations caused by insect herbivore chewing},
  journal      = {Oecologia},
  volume       = {175},
  number       = {4},
  pages        = {1257--1266},
  year         = {2014}
}

@article{RodrigoMoreno2017,
  author       = {Rodrigo-Moreno, A. and Bazihizina, N. and Azzarello, E. and Masi, E. and Tran, D. and Bouteau, F. and Mancuso, S.},
  title        = {Root phonotropism: early signalling events following sound perception in *Arabidopsis* roots},
  journal      = {Plant Science},
  volume       = {264},
  pages        = {9--15},
  year         = {2017}
}

@article{Veits2019,
  author       = {Veits, M. and Khait, I. and Obolski, U. and Zinger, E. and Boonman, A. and Goldshtein, A. and Hadany, L.},
  title        = {Flowers respond to pollinator sound within minutes by increasing nectar sugar concentration},
  journal      = {Ecology Letters},
  volume       = {22},
  number       = {9},
  pages        = {1483--1492},
  year         = {2019}
}

@article{gagliano2013green,
  author = {Gagliano, M.},
  title = {Green symphonies: A call for studies on acoustic communication in plants},
  journal = {Behavioral Ecology},
  volume = {24},
  number = {4},
  pages = {789--796},
  year = {2013}
}

@article{Gagliano2017,
  author       = {Gagliano, M. and Grimonprez, M. and Depczynski, M. and Renton, M.},
  title        = {Tuned in: Plant roots use sound to locate water},
  journal      = {Oecologia},
  volume       = {184},
  number       = {1},
  pages        = {151--160},
  year         = {2017}
}

@article{Hamant2017,
  author       = {Hamant, O. and Haswell, E. S.},
  title        = {Life behind the wall: sensing mechanical cues in plants},
  journal      = {BMC Biology},
  volume       = {15},
  pages        = {1--9},
  year         = {2017}
}

@article{DeRoo2016,
  author = {De Roo, L. and Vergeynst, L. L. and De Baerdemaeker, N. J. F. and Steppe, K.},
  title = {Acoustic Emissions to Measure Drought-Induced Cavitation in Plants},
  journal = {Applied Sciences},
  year = {2016},
  volume = {6},
  number = {3},
  pages = {71},
  doi = {10.3390/app6030071},
}

@article{Yan2009,
  author = {Xiaofei Yan and Zhongyi Wang and Lan Huang and Cheng Wang and Ruifeng Hou and Zhilong Xu and Xiaojun Qiao},
  title = {Research progress on electrical signals in higher plants},
  journal = {Progress in Natural Science},
  volume = {19},
  number = {5},
  pages = {531--541},
  year = {2009},
  issn = {1002-0071},
  doi = {10.1016/j.pnsc.2008.08.009},
}

@article{Simard2012,
  author    = {Suzanne W. Simard and Kevin J. Beiler and Marcus A. Bingham and Julie R. Deslippe and Leanne J. Philip and François P. Teste},
  title     = {Mycorrhizal networks: Mechanisms, ecology and modelling},
  journal   = {Fungal Biology Reviews},
  volume    = {26},
  number    = {1},
  pages     = {39-60},
  year      = {2012},
  issn      = {1749-4613},
  doi       = {10.1016/j.fbr.2012.01.001},
}

@article{simard2004mycorrhizal,
  author = {Simard, S. W. and Durall, D. M.},
  title = {Mycorrhizal networks: A review of their extent, function, and importance},
  journal = {Canadian Journal of Botany},
  volume = {82},
  number = {8},
  pages = {1140--1165},
  year = {2004}
}

@article{Preston2004,
  author    = {C. A. Preston and G. Laue and I. T. Baldwin},
  title     = {Plant–plant signaling: Application of trans- or cis-methyl jasmonate equivalent to sagebrush releases does not elicit direct defenses in native tobacco},
  journal   = {Journal of Chemical Ecology},
  volume    = {30},
  pages     = {2193--2214},
  year      = {2004},
  doi       = {10.1023/B:JOEC.0000048785.39814.ef}
}

@article{Falik2012,
  author    = {Omer Falik and Yonat Mordoch and Daniel Ben-Natan and Miriam Vanunu and Oron Goldstein and Ariel Novoplansky},
  title     = {Plant responsiveness to root–root communication of stress cues},
  journal   = {Annals of Botany},
  volume    = {110},
  number    = {2},
  pages     = {271--280},
  year      = {2012},
  month     = {July},
  doi       = {10.1093/aob/mcs045},
}

@misc{Sensigent_Cyranose320,
  author = {Sensigent},
  title = {Cyranose 320},
  howpublished = {Online},
  note = {Accessed: 2025-02-11},
  url = {https://www.sensigent.com/cyranose-320.html}
}

@misc{Airsense_Pen3,
  author = {Airsense Analytics},
  title = {Pen3 portable electronic nose},
  howpublished = {Flyer, Online},
  note = {Accessed: 2025-02-11},
  url = {https://airsense.com/sites/default/files/flyer_pen.pdf}
}

@misc{AlphaMOS_Heracles,
  author = {Alpha MOS},
  title = {Smell analysis - Heracles electronic nose},
  howpublished = {Online},
  note = {Accessed: 2025-02-11},
  url = {https://www.alpha-mos.com/smell-analysis-heracles-electronic-nose}
}

@misc{NANOSENSORS_MSS8RM,
  author = {NANOSENSORS},
  title = {MSS-8RM – A readout module for membrane-type surface stress sensors (MSS)},
  howpublished = {Online},
  month = {October},
  day = {30},
  year = {2017},
  note = {Accessed: 2025-02-11},
  url = {https://mss-sensor.com/NANOSENSORS_MSS-8RM.pdf}
}

@article{Martinelli_StableOdor,
  author = {E. Martinelli and G. Magna and D. Polese et al.},
  title = {Stable Odor Recognition by a neuro-adaptive Electronic Nose},
  journal = {Scientific Reports},
  volume = {5},
  pages = {10960},
  year = {2015},
  doi = {10.1038/srep10960},
}

@inproceedings{Leccese_ElectronicNose,
  author = {Fabio Leccese et al.},
  title = {Electronic Nose: A First Sensors Array Optimization for Pesticides Detection Based on Wilks' A-Statistic},
  booktitle = {2018 5th IEEE International Workshop on Metrology for AeroSpace (MetroAeroSpace)},
  year = {2018},
  pages = {440-445}
}

@article{Cui_FastENose,
  author = {S. Cui and E. A. A. Inocente and N. Acosta and H. M. Keener and H. Zhu and P. P. Ling},
  title = {Development of fast e-nose system for early-stage diagnosis of aphid-stressed tomato plants},
  journal = {Sensors},
  volume = {19},
  number = {16},
  pages = {3480},
  year = {2019}
}

@article{Sharma_GasSensorArray,
  author = {C. Sharma and A. Dey and H. Khatun and J. Das and U. Sarma},
  title = {Design and development of a gas sensor array to detect salinity stress in Khasi Mandarin Orange plants},
  journal = {IEEE Transactions on Instrumentation and Measurement},
  year = {2023}
}

@article{Galieni2021,
    author = {Galieni, Angelica and D'Ascenzo, Nicola and Stagnari, Fabio and Pagnani, Giancarlo and Xie, Qingguo and Pisante, Michele},
    title = {Past and Future of Plant Stress Detection: An Overview From Remote Sensing to Positron Emission Tomography},
    journal = {Frontiers in Plant Science},
    volume = {11},
    year = {2021},
    doi = {10.3389/fpls.2020.609155},
    issn = {1664-462X}
}

@article{ens2009,
  author = {Ens, E. J. and Bremner, J. B. and French, K. and Korth, J.},
  title = {Identification of volatile compounds released by roots of an invasive plant, bitou bush (*Chrysanthemoides monilifera* spp. *rotundata*), and their inhibition of native seedling growth},
  journal = {Biological Invasions},
  volume = {11},
  pages = {275-287},
  year = {2009},
  doi = {10.1007/s10530-008-9232-3}
}

@article{jassbi2010,
  author = {Jassbi, A. R. and Zamanizadehnajari, S. and Baldwin, I. T.},
  title = {Phytotoxic volatiles in the roots and shoots of *Artemisia tridentata* as detected by headspace solid-phase microextraction and gas chromatographic-mass spectrometry analysis},
  journal = {Journal of Chemical Ecology},
  volume = {36},
  pages = {1398-1407},
  year = {2010},
  doi = {10.1007/s10886-010-9885-0}
}

@article{bouwmeester2003,
  author = {Bouwmeester, H. J. and Matusova, R. and Zhongkui, S. and Beale, M. H.},
  title = {Secondary metabolite signalling in host–parasitic plant interactions},
  journal = {Current Opinion in Plant Biology},
  volume = {6},
  pages = {358-364},
  year = {2003},
  doi = {10.1016/S1369-5266(03)00065-7}
}

@incollection{rasmann2012,
  author = {Rasmann, S. and Hiltpold, I. and Ali, J.},
  title = {The role of root-produced volatile secondary metabolites in mediating soil interactions},
  booktitle = {Advances in Selected Plant Physiology Aspects},
  editor = {Montanaro, G. and Bartolomeo, D.},
  publisher = {InTech},
  address = {Rijeka},
  pages = {269-290},
  year = {2012}
}

@article{robert2012,
  author    = {Robert CAM and Erb M and Duployer M and et al.},
  title     = {Herbivore-induced plant volatiles mediate host selection by a root herbivore},
  journal   = {New Phytologist},
  volume    = {194},
  pages     = {1061--1069},
  year      = {2012},
  doi       = {10.1111/j.1469-8137.2012.04127.x}
}

@article{Rasmann2005,
  author    = {Rasmann, S. and KOllner, T. G. and Degenhardt, J. and et al.},
  title     = {Recruitment of entomopathogenic nematodes by insect-damaged maize roots},
  journal   = {Nature},
  volume    = {434},
  pages     = {732--737},
  year      = {2005},
  doi       = {10.1038/nature03451}
}

@article{Romagni2000,
  author    = {Romagni, J. G. and Allen, S. N. and Dayan, F. E.},
  title     = {Allelopathic effects of volatile cineoles on two weedy plant species},
  journal   = {Journal of Chemical Ecology},
  volume    = {26},
  pages     = {303--313},
  year      = {2000}
}

@article{Tholl2006,
  author    = {Tholl, D. and Boland, W. and Hansel, A. and et al.},
  title     = {Practical approaches to plant volatile analysis},
  journal   = {Plant Journal},
  volume    = {45},
  pages     = {540--560},
  year      = {2006},
  doi       = {10.1111/j.1365-313X.2005.02612.x}
}

@article{Eilers2015,
  author    = {Eilers, E. J. and Pauls, G. and Rillig, M. C. and et al.},
  title     = {Novel set-up for low-disturbance sampling of volatile and non-volatile compounds from plant roots},
  journal   = {Journal of Chemical Ecology},
  volume    = {41},
  pages     = {253--266},
  year      = {2015},
  doi       = {10.1007/s10886-015-0559-9}
}

@article{Mohney2009,
  author    = {Mohney, B. K. and Matz, T. and LaMoreaux, J. and et al.},
  title     = {In situ silicone tube microextraction: a new method for undisturbed sampling of root-exuded thiophenes from Marigold (\textit{Tagetes erecta} L.) in soil},
  journal   = {Journal of Chemical Ecology},
  volume    = {35},
  pages     = {1279--1287},
  year      = {2009},
  doi       = {10.1007/s10886-009-9711-8}
}

@article{Hiltpold2011,
  author    = {Hiltpold, I. and Erb, M. and Robert, C. A. M. and Turlings, T. C. J.},
  title     = {Systemic root signalling in a belowground, volatile-mediated tritrophic interaction},
  journal   = {Plant, Cell \& Environment},
  volume    = {34},
  pages     = {1267--1275},
  year      = {2011}
}

@article{Abraham2015,
  author    = {Abraham, J. and Giacomuzzi, V. and Angeli, S.},
  title     = {Root damage to apple plants by cockchafer larvae induces a change in volatile signals below- and above-ground},
  journal   = {Entomologia Experimentalis et Applicata},
  volume    = {156},
  pages     = {279--289},
  year      = {2015},
  doi       = {10.1111/eea.12330}
}

@article{vanDam2012,
  author    = {van Dam, N. M. and Samudrala, D. and Harren, F. J. M. and Cristescu, S. M.},
  title     = {Real-time analysis of sulfur-containing volatiles in \textit{Brassica} plants infested with root-feeding \textit{Delia radicum} larvae using proton-transfer reaction mass spectrometry},
  journal   = {AoB Plants},
  year      = {2012},
  volume    = {2012},
  pages     = {pls021},
  doi       = {10.1093/aobpla/pls021}
}

@article{Danner2015,
  author    = {Danner, H. and Brown, P. and Cator, E. A. and et al.},
  title     = {Aboveground and belowground herbivores synergistically induce volatile organic sulfur compound emissions from shoots but not from roots},
  journal   = {Journal of Chemical Ecology},
  year      = {2015},
  doi       = {10.1007/s10886-015-0601-y}
}

@article{Hedrich2016,
  author    = {Hedrich, R. and Salvador-Recatalà, V. and Dreyer, I.},
  title     = {Electrical Wiring and Long-Distance Plant Communication},
  journal   = {Trends in Plant Science},
  volume    = {21},
  number    = {5},
  pages     = {376-387},
  year      = {2016},
  month     = {May},
  doi       = {10.1016/j.tplants.2016.01.016},
  pmid      = {26880317}
}

@book{nye1977solute,
  author    = {Nye, P.H. and Tinker, P.B.},
  title     = {Solute Movement in the Soil--Root System},
  publisher = {University of California Press},
  year      = {1977},
  address   = {Berkeley, CA},
  isbn      = {9780520032338}
}

@book{barber1984bioavailability,
  author    = {Barber, Stanley A.},
  title     = {Soil Nutrient Bioavailability: A Mechanistic Approach},
  publisher = {John Wiley \& Sons},
  year      = {1995},
  address   = {New York, NY},
  isbn      = {9780471080794}
}

@article{darrah1991models,
  author    = {Darrah, P.R.},
  title     = {Models of the rhizosphere: I. Microbial population dynamics around a root releasing soluble exudates},
  journal   = {Plant and Soil},
  volume    = {133},
  number    = {2},
  pages     = {187--199},
  year      = {1991},
  doi       = {0.1007/BF00009191}
}

@article{roose2016rhizosphere,
  author    = {Roose, T. and Keyes, S.D. and Daly, K.R.},
  title     = {Challenges in modelling the rhizosphere at different scales},
  journal   = {Plant and Soil},
  volume    = {407},
  number    = {1--2},
  pages     = {9--38},
  year      = {2016},
  doi       = {10.1007/s11104-016-2878-2}
}

@book{jury2004soil,
  author    = {Jury, William A. and Horton, Robert},
  title     = {Soil Physics},
  publisher = {Wiley},
  edition   = {6},
  year      = {2004},
  address   = {Hoboken, NJ},
  isbn      = {9780471059653}
}

@book{leij1991breakthrough,
  author    = {Leij, Feike J. and Toride, Nobuo and van Genuchten, M. Th.},
  title     = {The CXTFIT code for estimating transport parameters from laboratory or field tracer experiments},
  publisher = {Salinity Laboratory},
  year      = {1995},
}

@book{hillel1998environmental,
  author    = {Hillel, Daniel},
  title     = {Introduction to Environmental Soil Physics},
  publisher = {Academic Press},
  year      = {2003},
  isbn      = {9781493300631}
}

@article{Itakura2003,
  author  = {Itakura, T. and Airey, D. W. and Leo, C. J.},
  title   = {The diffusion and sorption of volatile organic compounds through kaolinitic clayey soils},
  journal = {Journal of Contaminant Hydrology},
  year    = {2003},
  volume  = {65},
  number  = {3-4},
  pages   = {219--243},
  doi     = {10.1016/S0169-7722(03)00002-0}
}

@misc{vakilipoor2025mcagricultureframeworknatureinspired,
      title={{MC} for Agriculture: A Framework for Nature-inspired Sustainable Pest Control},
      author={Fardad Vakilipoor and Nora Hirschmann and Julian Schladt and Stefan Schwab and Annette Reineke and Robert Schober and Kathrin Castiglione and Maximilian Schaefer},
      year={2025},
      eprint={2506.20637},
      archivePrefix={arXiv},
      primaryClass={eess.SP},
      url={https://arxiv.org/abs/2506.20637}, 
}

@misc{gulec2025decodingengineeringphytobiomecommunication,
      title={Decoding and Engineering the Phytobiome Communication for Smart Agriculture}, 
      author={Fatih Gulec and Hamdan Awan and Nigel Wallbridge and Andrew W. Eckford},
      year={2025},
      eprint={2508.03584},
      archivePrefix={arXiv},
      primaryClass={eess.SP},
      url={https://arxiv.org/abs/2508.03584}, 
}

@article{rivett2011unsaturated,
  author  = {Rivett, M.O. and Wealthall, G.P. and Dearden, R.A. and McAlary, T.A.},
  title   = {Review of unsaturated-zone transport and attenuation of volatile organic compound (VOC) plumes leached from shallow source zones},
  journal = {Journal of Contaminant Hydrology},
  year    = {2011},
  volume  = {123},
  number  = {3-4},
  pages   = {130--156},
  doi     = {10.1016/j.jconhyd.2010.12.013}
}

@article{nervous_system_plants,
  author    = {Miguel-Tom{\'e}, Susana and Llin{\'a}s, Rodolfo R.},
  title     = {Broadening the definition of a nervous system to better understand the evolution of plants and animals},
  journal   = {Plant Signaling \& Behavior},
  volume    = {16},
  number    = {10},
  pages     = {1927562},
  year      = {2021},
  publisher = {Taylor \& Francis},
  doi       = {10.1080/15592324.2021.1927562},
}

@article{mudrilov2021electrical,
  title        = {Electrical Signaling of Plants under Abiotic Stressors: Transmission of Stimulus-Specific Information},
  author       = {Mudrilov, Mikhail and Ladeynova, Margarita and Grinberg, Mikhail and Balalaeva, Irina and Vodeneev, Vladimir},
  journal      = {International Journal of Molecular Sciences},
  volume       = {22},
  number       = {19},
  pages        = {10715},
  year         = {2021},
  publisher    = {MDPI},
  doi          = {10.3390/ijms221910715},
}

@article{chaparro2021plant,
  title        = {Plant electrophysiology: bibliometric analysis, methods and applications in the monitoring of plant-environment interactions},
  author       = {Chaparro-C{\'a}rdenas, Silvia and Ramirez, Julian and Gamboa, Wilson and Moreno-Chac{\'o}n, Andr{\'e}s and Vargas-Tangua, Frank},
  journal      = {DYNA},
  volume       = {88},
  number       = {218},
  pages        = {112--123},
  year         = {2021},
  doi          = {10.15446/dyna.v88n218.92405},
}

@article{ponomarenko2014ultrasonic,
  title={Ultrasonic emissions reveal individual cavitation bubbles in water-stressed wood},
  author={Ponomarenko, A. and Vincent, O. and Pietriga, A. and Cochard, H. and Badel, {\'E}. and Marmottant, P.},
  journal={Journal of the Royal Society Interface},
  volume={11},
  number={99},
  pages={20140480},
  year={2014},
  publisher={The Royal Society},
  doi={10.1098/rsif.2014.0480},
}

@article{generalboltzmanref,
  title = {Membrane mechanics as a probe of ion-channel gating mechanisms},
  author = {Reeves, Daniel and Ursell, Tristan and Sens, Pierre and Kondev, Jane and Phillips, Rob},
  journal = {Phys. Rev. E},
  volume = {78},
  issue = {4},
  pages = {041901},
  numpages = {11},
  year = {2008},
  month = {Oct},
  publisher = {American Physical Society},
  doi = {10.1103/PhysRevE.78.041901},
}

@article{An_Absorption_Model_of_Volatile_Organic_Compound_by_Plant_Leaf_The_Most_Influential_Site_in_the_Absorption_Pathway,
  author = {Yamane, Mizuki and Tani, Akira},
  doi = {10.1016/j.aeaoa.2024.100274},
  issn = {2590-1621},
  journal = {Atmospheric Environment: X},
  month = {August},
  pages = {100274},
  shorttitle = {An Absorption Model of Volatile Organic Compound by Plant Leaf},
  title = {An Absorption Model of Volatile Organic Compound by Plant Leaf: {{The}} Most Influential Site in the Absorption Pathway},
  volume = {23},
  year = {2024}
}

@article{Basu2017,
  author    = {Basu, Debarati and Haswell, Elizabeth S.},
  title     = {Plant mechanosensitive ion channels: an ocean of possibilities},
  journal   = {Current Opinion in Plant Biology},
  volume    = {40},
  pages     = {43--48},
  year      = {2017},
  doi       = {10.1016/j.pbi.2017.07.002}
}

@article{Hamilton2015,
  author    = {Hamilton, Eric S. and Schlegel, Angela M. and Haswell, Elizabeth S.},
  title     = {United in diversity: mechanosensitive ion channels in plants},
  journal   = {Annual Review of Plant Biology},
  volume    = {66},
  pages     = {113--137},
  year      = {2015},
  doi       = {10.1146/annurev-arplant-043014-114700}
}

@article{Hansen2025,
  author    = {Hansen, Annalene and Gladala-Kostarz, Agnieszka and Hindhaugh, Rebecca
               and Doonan, John H. and Bosch, Maurice},
  title     = {Mechanical stimulation in plants: molecular insights, morphological
               adaptations, and agricultural applications in monocots},
  journal   = {BMC Biology},
  volume    = {23},
  number    = {1},
  pages     = {58},
  year      = {2025},
  doi       = {10.1186/s12915-025-02157-3}
}

@article{Pouliquen2012,
  author  = {Scorza, L. C. T. and Dornelas, M. C.},
  title   = {Plants on the move: Towards common mechanisms governing mechanically-induced plant movements},
  journal = {Plant Signaling \& Behavior},
  volume  = {6},
  number  = {12},
  pages   = {1979--1986},
  year    = {2011},
  month   = {December}
}

@article{deWit2012,
  author    = {{de Wit}, Mieke and Kegge, Wouter and Evers, Jochem B.
               and {Vergeer-van Eijk}, Marleen H. and Gankema, Paulien
               and Voesenek, Laurentius A. C. J. and Pierik, Ronald},
  title     = {Plant neighbor detection through touching leaf tips precedes
               phytochrome signals},
  journal   = {Proceedings of the National Academy of Sciences},
  volume    = {109},
  number    = {36},
  pages     = {14705--14710},
  year      = {2012},
  doi       = {10.1073/pnas.1205437109}
}

@article{vanderZee2021,
  author    = {{van der Zee}, Jens and Lau, Alvaro and Shenkin, Alexander},
  title     = {Understanding crown shyness from a {3-D} perspective},
  journal   = {Annals of Botany},
  volume    = {128},
  number    = {6},
  pages     = {725--736},
  year      = {2021},
  doi       = {10.1093/aob/mcab035}
}

@article{PelaezVico2025,
  author    = {Pel\'{a}ez-Vico, Mar\'{i}a \'{A}ngeles and Fichman, Yosef
               and Ghani, Abdul and Dhakal, Sameep and Karpi\'{n}ski, Stanis{\l}aw M.
               and Joshi, Trupti and Mittler, Ron},
  title     = {United we stand: plants that physically touch each other are more
               resilient to excess light stress},
  journal   = {bioRxiv},
  year      = {2025},
  doi       = {10.1101/2025.09.02.673745}
}

@article{Diffusion_and_Bulk_Flow_of_Amino_Acids_Mediate_Calcium_Waves_in_Plants,
  author = {Bellandi, Annalisa and Papp, Diana and Breakspear, Andrew and Joyce, Joshua and Johnston, Matthew G. and {de Keijzer}, Jeroen and Raven, Emma C. and Ohtsu, Mina and Vincent, Thomas R. and Miller, Anthony J. and Sanders, Dale and Hogenhout, Saskia A. and Morris, Richard J. and Faulkner, Christine},
  doi = {10.1126/sciadv.abo6693},
  journal = {Science Advances},
  month = {October},
  number = {42},
  pages = {Art. no. eabo6693},
  publisher = {American Association for the Advancement of Science},
  title = {Diffusion and Bulk Flow of Amino Acids Mediate Calcium Waves in Plants},
  volume = {8},
  year = {2022}
}

@article{Electrical_Signals_Plant_Tolerance_to_Actions_of_Stressors_and_Programmed_Cell_Death_Is_Interaction_Possible,
  author = {Sukhova, Ekaterina and Sukhov, Vladimir},
  copyright = {http://creativecommons.org/licenses/by/3.0/},
  doi = {10.3390/plants10081704},
  issn = {2223-7747},
  journal = {Plants},
  langid = {english},
  month = {August},
  number = {8},
  pages = {1704},
  publisher = {Multidisciplinary Digital Publishing Institute},
  shorttitle = {Electrical {{Signals}}, {{Plant Tolerance}} to {{Actions}} of {{Stressors}}, and {{Programmed Cell Death}}},
  title = {Electrical {{Signals}}, {{Plant Tolerance}} to {{Actions}} of {{Stressors}}, and {{Programmed Cell Death}}: {{Is Interaction Possible}}?},
  volume = {10},
  year = {2021}
}

@article{Exploring_Strategies_for_Classification_of_External_Stimuli_Using_Statistical_Features_of_the_Plant_Electrical_Response,
  author = {Chatterjee, Shre Kumar and Das, Saptarshi and Maharatna, Koushik and Masi, Elisa and Santopolo, Luisa and Mancuso, Stefano and Vitaletti, Andrea},
  doi = {10.1098/rsif.2014.1225},
  issn = {1742-5689},
  journal = {Journal of The Royal Society Interface},
  month = {March},
  number = {104},
  pages = {20141225},
  title = {Exploring Strategies for Classification of External Stimuli Using Statistical Features of the Plant Electrical Response},
  volume = {12},
  year = {2015}
}

@article{Ion_Channels_in_Electrical_Signaling_in_Higher_Plants,
  author = {Mudrilov, Maxim A. and Ladeynova, Maria M. and Kuznetsova, Darya V. and Vodeneev, Vladimir A.},
  doi = {10.1134/S000629792310005X},
  issn = {1608-3040},
  journal = {Biochemistry (Moscow)},
  langid = {english},
  month = {October},
  number = {10},
  pages = {1467--1487},
  title = {Ion {{Channels}} in {{Electrical Signaling}} in {{Higher Plants}}},
  volume = {88},
  year = {2023}
}

@article{Mathematical_Model_of_Action_Potential_in_Higher_Plants_with_Account_for_the_Involvement_of_Vacuole_in_the_Electrical_Signal_Generation,
  author = {Novikova, E. M. and Vodeneev, V. A. and Sukhov, V. S.},
  doi = {10.1134/S1990747817010068},
  issn = {1990-7494},
  journal = {Biochemistry (Moscow), Supplement Series A: Membrane and Cell Biology},
  langid = {english},
  month = {April},
  number = {2},
  pages = {151--167},
  title = {Mathematical Model of Action Potential in Higher Plants with Account for the Involvement of Vacuole in the Electrical Signal Generation},
  volume = {11},
  year = {2017}
}

@inproceedings{A_Theory_Model_for_Description_of_the_Electrical_Signals_in_Plant_Part_I,
  address = {Boston, MA},
  author = {Wang, Zhongyi and Huang, Lan and Yan, Xiaofei and Wang, Cheng and Xu, Zhilong and Hou, Ruifeng and Qiao, Xiaojun},
  booktitle = {Computer {{And Computing Technologies In Agriculture}}, {{Volume I}}},
  doi = {10.1007/978-0-387-77251-6_69},
  editor = {Li, Daoliang},
  isbn = {978-0-387-77251-6},
  langid = {english},
  pages = {637--643},
  publisher = {Springer US},
  title = {A {{Theory Model}} for {{Description}} of the {{Electrical Signals}} in {{Plant Part I}}},
  year = {2008}
}

@misc{maitra2026modelinganalysisvocbasedinterplant,
      title={Modeling and Analysis of {VOC}-based Interplant Molecular Communication Channel},
      author={Bitop Maitra and Ozgur B. Akan},
      year={2026},
      eprint={2512.12035},
      archivePrefix={arXiv},
      primaryClass={eess.SP},
      url={https://arxiv.org/abs/2512.12035}, 
}

@misc{merdan2026odorcommunicationgreenleaf,
      title={Odor Communication with Green Leaf Volatiles for Stress Signalling in the Internet of Plants}, 
      author={Fatih Merdan and Ozgur B. Akan},
      year={2026},
      eprint={2602.16850},
      archivePrefix={arXiv},
      primaryClass={eess.SP},
      url={https://arxiv.org/abs/2602.16850}, 
}

@misc{merdan2026airborneparticlecommunicationtimevarying,
      title={Airborne Particle Communication Through Time-varying Diffusion-Advection Channels}, 
      author={Fatih Merdan and Ozgur B. Akan},
      year={2026},
      eprint={2601.08534},
      archivePrefix={arXiv},
      primaryClass={eess.SP},
      url={https://arxiv.org/abs/2601.08534}, 
}

@article{THE_ROLE_OF_ROOT_EXUDATES_IN_RHIZOSPHERE_INTERACTIONS_WITH_PLANTS_AND_OTHER_ORGANISMS,
  author = {Bais, Harsh P. and Weir, Tiffany L. and Perry, Laura G. and Gilroy, Simon and Vivanco, Jorge M.},
  doi = {10.1146/annurev.arplant.57.032905.105159},
  issn = {1543-5008, 1545-2123},
  journal = {Annual Review of Plant Biology},
  langid = {english},
  month = {June},
  number = {1},
  pages = {233--266},
  title = {{{THE ROLE OF ROOT EXUDATES IN RHIZOSPHERE INTERACTIONS WITH PLANTS AND OTHER ORGANISMS}}},
  volume = {57},
  year = {2006}
}

@article{Root_Exudates_Mediate_Kin_Recognition_in_Plants,
  author = {Biedrzycki, Meredith L. and Jilany, Tafari A. and Dudley, Susan A. and Bais, Harsh P.},
  doi = {10.4161/cib.3.1.10118},
  issn = {null},
  journal = {Communicative \& Integrative Biology},
  month = {January},
  number = {1},
  pages = {28--35},
  pmid = {20539778},
  publisher = {Taylor \& Francis},
  title = {Root Exudates Mediate Kin Recognition in Plants},
  volume = {3},
  year = {2010}
}

@article{The_Behavioral_Ecology_of_Insect_Vibrational_Communication,
    author = {Cocroft, Reginald B. and Rodríguez, Rafael L.},
    title = {The Behavioral Ecology of Insect Vibrational Communication},
    journal = {BioScience},
    volume = {55},
    number = {4},
    pages = {323-334},
    year = {2005},
    month = {04},
    issn = {0006-3568},
    doi = {10.1641/0006-3568(2005)055[0323:TBEOIV]2.0.CO;2},
    eprint = {https://academic.oup.com/bioscience/article-pdf/55/4/323/26896785/55-4-323.pdf},
}

@article{Natural_and_LaserInduced_Cavitation_in_Corn_Stems_On_the_Mechanisms_of_Acoustic_Emissions,
  author = {Fern{\'a}ndez, E. and Fern{\'a}ndez, R. J. and Bilmes, Gabriel Mario},
  copyright = {Copyright (c) 2012 E. Fern\'andez, R. J. Fern\'andez, Gabriel Mario Bilmes},
  doi = {10.4279/pip.040003},
  issn = {1852-4249},
  journal = {Papers in Physics},
  langid = {english},
  month = {May},
  pages = {040003--040003},
  shorttitle = {Natural and Laser-Induced Cavitation in Corn Stems},
  title = {Natural and Laser-Induced Cavitation in Corn Stems: {{On}} the Mechanisms of Acoustic Emissions},
  volume = {4},
  year = {2012}
}

@article{Flowers_Respond_to_Pollinator_Sound_within_Minutes_by_Increasing_Nectar_Sugar_Concentration,
  author = {Veits, Marine and Khait, Itzhak and Obolski, Uri and Zinger, Eyal and Boonman, Arjan and Goldshtein, Aya and Saban, Kfir and Seltzer, Rya and {Ben-Dor}, Udi and Estlein, Paz and Kabat, Areej and Peretz, Dor and Ratzersdorfer, Ittai and Krylov, Slava and Chamovitz, Daniel and Sapir, Yuval and Yovel, Yossi and Hadany, Lilach},
  copyright = {\copyright{} 2019 The Authors. Ecology Letters published by CNRS and John Wiley \& Sons Ltd.},
  doi = {10.1111/ele.13331},
  issn = {1461-0248},
  journal = {Ecology Letters},
  langid = {english},
  number = {9},
  pages = {1483--1492},
  title = {Flowers Respond to Pollinator Sound within Minutes by Increasing Nectar Sugar Concentration},
  volume = {22},
  year = {2019}
}

@article{Caterpillar_Chewing_Vibrations_Cause_Changes_in_Plant_Hormones_and_Volatile_Emissions_in_Arabidopsis_Thaliana,
  author = {Body, M{\'e}lanie J. A. and Neer, William C. and Vore, Caitlin and Lin, Chung-Ho and Vu, Danh C. and Schultz, Jack C. and Cocroft, Reginald B. and Appel, Heidi M.},
  doi = {10.3389/fpls.2019.00810},
  issn = {1664-462X},
  journal = {Frontiers in Plant Science},
  langid = {english},
  month = {June},
  publisher = {Frontiers},
  title = {Caterpillar {{Chewing Vibrations Cause Changes}} in {{Plant Hormones}} and {{Volatile Emissions}} in {{Arabidopsis}} Thaliana},
  volume = {10},
  year = {2019}
}

@article{Systems_Theory_An_Approach_to_MassDamperSpring_and_MassNondamperSpring_Systems,
  author = {Sunday, J.},
  doi = {10.31559/GLM2016.3.3.4},
  issn = {25199269, 25199277},
  journal = {General Letters in Mathematics},
  langid = {english},
  month = {December},
  number = {3},
  shorttitle = {Systems {{Theory}}},
  title = {Systems {{Theory}}: {{An Approach}} to {{Mass-Damper-Spring}} and {{Mass-Nondamper-Spring Systems}}},
  volume = {3},
  year = {2017}
}

@article{Calibration_of_MassSpringDamper_Equivalent_Systems_for_Real_Time_Assessment_of_the_Dynamics_of_Trees,
  author = {Grande, Ernesto and Franceschini, Raffaella},
  doi = {10.1016/j.compag.2024.109610},
  issn = {0168-1699},
  journal = {Computers and Electronics in Agriculture},
  month = {December},
  pages = {109610},
  title = {Calibration of Mass-Spring-Damper Equivalent Systems for Real Time Assessment of the Dynamics of Trees},
  volume = {227},
  year = {2024}
}

@article{An_Analytical_Model_for_the_Propagation_of_Bending_Waves_on_a_Plant_Stem_Due_to_Vibration_of_an_Attached_Insect,
  author = {Miles, R. N.},
  doi = {10.1016/j.heliyon.2016.e00086},
  issn = {2405-8440},
  journal = {Heliyon},
  month = {March},
  number = {3},
  pages = {Art. no. e00086},
  title = {An Analytical Model for the Propagation of Bending Waves on a Plant Stem Due to Vibration of an Attached Insect},
  volume = {2},
  year = {2016}
}

@article{Homogenization_of_a_Viscoelastic_Model_for_Plant_Cell_Wall_Biomechanics,
  author = {Ptashnyk, Mariya and Seguin, Brian},
  doi = {10.1051/cocv/2016060},
  issn = {1292-8119, 1262-3377},
  journal = {ESAIM: Control, Optimisation and Calculus of Variations},
  langid = {english},
  month = {October},
  number = {4},
  pages = {1447--1471},
  title = {Homogenization of a Viscoelastic Model for Plant Cell Wall Biomechanics},
  volume = {23},
  year = {2017}
}

@article{Modeling_Mechanical_Behaviors_of_Plant_Stems_Undergoing_Microstructural_Changes,
  author = {Song, Ruyue and Muliana, Anastasia},
  doi = {10.1016/j.mechmat.2019.103175},
  issn = {0167-6636},
  journal = {Mechanics of Materials},
  month = {December},
  pages = {103175},
  title = {Modeling Mechanical Behaviors of Plant Stems Undergoing Microstructural Changes},
  volume = {139},
  year = {2019}
}

@article{Wireless_Underground_Sensor_Communication_Using_Acoustic_Technology,
  author = {Al Moshi, Md Adnan and Hardie, Marcus and Choudhury, Tanveer and Kamruzzaman, Joarder},
  copyright = {http://creativecommons.org/licenses/by/3.0/},
  doi = {10.3390/s24103113},
  issn = {1424-8220},
  journal = {Sensors},
  langid = {english},
  month = {January},
  number = {10},
  pages = {3113},
  publisher = {Multidisciplinary Digital Publishing Institute},
  title = {Wireless {{Underground Sensor Communication Using Acoustic Technology}}},
  volume = {24},
  year = {2024}
}

@article{Feeling_Stressed_or_Strained_A_Biophysical_Model_for_Cell_Wall_Mechanosensing_in_Plants,
  author = {Fruleux, Antoine and Verger, St{\'e}phane and Boudaoud, Arezki},
  doi = {10.3389/fpls.2019.00757},
  issn = {1664-462X},
  journal = {Frontiers in Plant Science},
  langid = {english},
  month = {June},
  publisher = {Frontiers},
  shorttitle = {Feeling {{Stressed}} or {{Strained}}?},
  title = {Feeling {{Stressed}} or {{Strained}}? {{A Biophysical Model}} for {{Cell Wall Mechanosensing}} in {{Plants}}},
  volume = {10},
  year = {2019}
}

@misc{merdan2025acousticcommunicationmodelplants,
      title={An Acoustic Communication Model in Plants}, 
      author={Fatih Merdan and Ozgur B. Akan},
      year={2025},
      eprint={2512.01096},
      archivePrefix={arXiv},
      primaryClass={eess.SP},
      url={https://arxiv.org/abs/2512.01096}, 
}

@article{Thigmomorphogenesis,
  author = {Braam, Janet and Chehab, E. Wassim},
  doi = {10.1016/j.cub.2017.07.008},
  issn = {0960-9822},
  journal = {Current Biology},
  langid = {english},
  month = {September},
  number = {17},
  pages = {R863-R864},
  pmid = {28898652},
  publisher = {Elsevier},
  title = {Thigmomorphogenesis},
  volume = {27},
  year = {2017}
}

@article{Thigmomorphogenesis_a_complex_plant_response_to_mechano_stimulation,
    author = {Chehab, E. Wassim and Eich, Elizabeth and Braam, Janet},
    title = {Thigmomorphogenesis: a complex plant response to mechano-stimulation},
    journal = {Journal of Experimental Botany},
    volume = {60},
    number = {1},
    pages = {43-56},
    year = {2009},
    month = {01},
    issn = {0022-0957},
    doi = {10.1093/jxb/ern315},
    eprint = {https://academic.oup.com/jxb/article-pdf/60/1/43/1535073/ern315.pdf},
}

@article{Yan2025wearable,
  author    = {Yan, Xuexin and Pang, Yawen and Niu, Kaiwen and Hu, Bowen
               and Zhu, Zhengbo and Tan, Zuojun and Lei, Hongwei},
  title     = {Wearable Sensors for Plants: Status and Prospects},
  journal   = {Biosensors},
  volume    = {15},
  number    = {1},
  pages     = {53},
  year      = {2025},
  doi       = {10.3390/bios15010053}
}

@article{Wang2025microneedle,
  author    = {Wang, Qianyu and Molinero-Fern{\'a}ndez, {\'A}gueda
               and Acosta-Motos, Jos{\'e} Ram{\'o}n and Crespo, Gast{\'o}n A.},
  title     = {Microneedle Sensors for Ion Monitoring in Plants.
               One Step Closer to Smart Agriculture},
  journal   = {ACS Sensors},
  volume    = {10},
  number    = {7},
  pages     = {4771--4784},
  year      = {2025},
  doi       = {10.1021/acssensors.5c01215}
}

@article{Gloor2025bioelec,
  author    = {Gloor, Peter A.},
  title     = {Plant Bioelectrical Signals for Environmental and
               Emotional State Classification},
  journal   = {Biosensors},
  volume    = {15},
  number    = {11},
  pages     = {744},
  year      = {2025},
  doi       = {10.3390/bios15110744}
}

@article{WaliaJones2018,
  author    = {Walia, Ankit and Waadt, Rainer and Jones, Alexander M.},
  title     = {Genetically Encoded Biosensors in Plants: Pathways to Discovery},
  journal   = {Annual Review of Plant Biology},
  volume    = {69},
  pages     = {497--524},
  year      = {2018},
  doi       = {10.1146/annurev-arplant-042817-040104}
}

@article{Shaw2022nanosensor,
  author    = {Shaw, Daniel S. and Honeychurch, Kevin C.},
  title     = {Nanosensor Applications in Plant Science},
  journal   = {Biosensors},
  volume    = {12},
  number    = {9},
  pages     = {675},
  year      = {2022},
  doi       = {10.3390/bios12090675}
}

\vspace{-1cm}
\begin{IEEEbiography}
[{\includegraphics[width=1in, height=1.25in, clip, keepaspectratio]{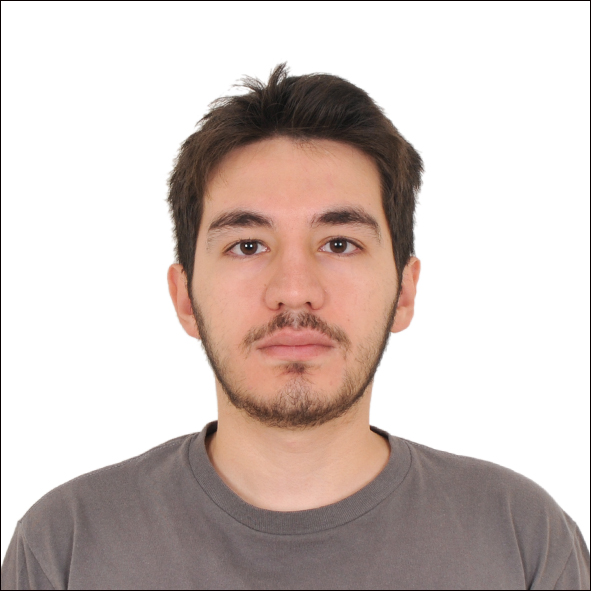}}]{Ahmet Burak Kilic} received the B.Sc. degree in Electrical and Electronics Engineering and the B.A. degree in Business Administration from Koç University, Istanbul, Turkey. He is currently pursuing the M.Sc. degree in Electrical and Electronics Engineering at Koç University under the supervision of Prof. Akan.
\end{IEEEbiography}
\begin{IEEEbiography}[{\includegraphics[width=1in,height=1.25in,clip,keepaspectratio]{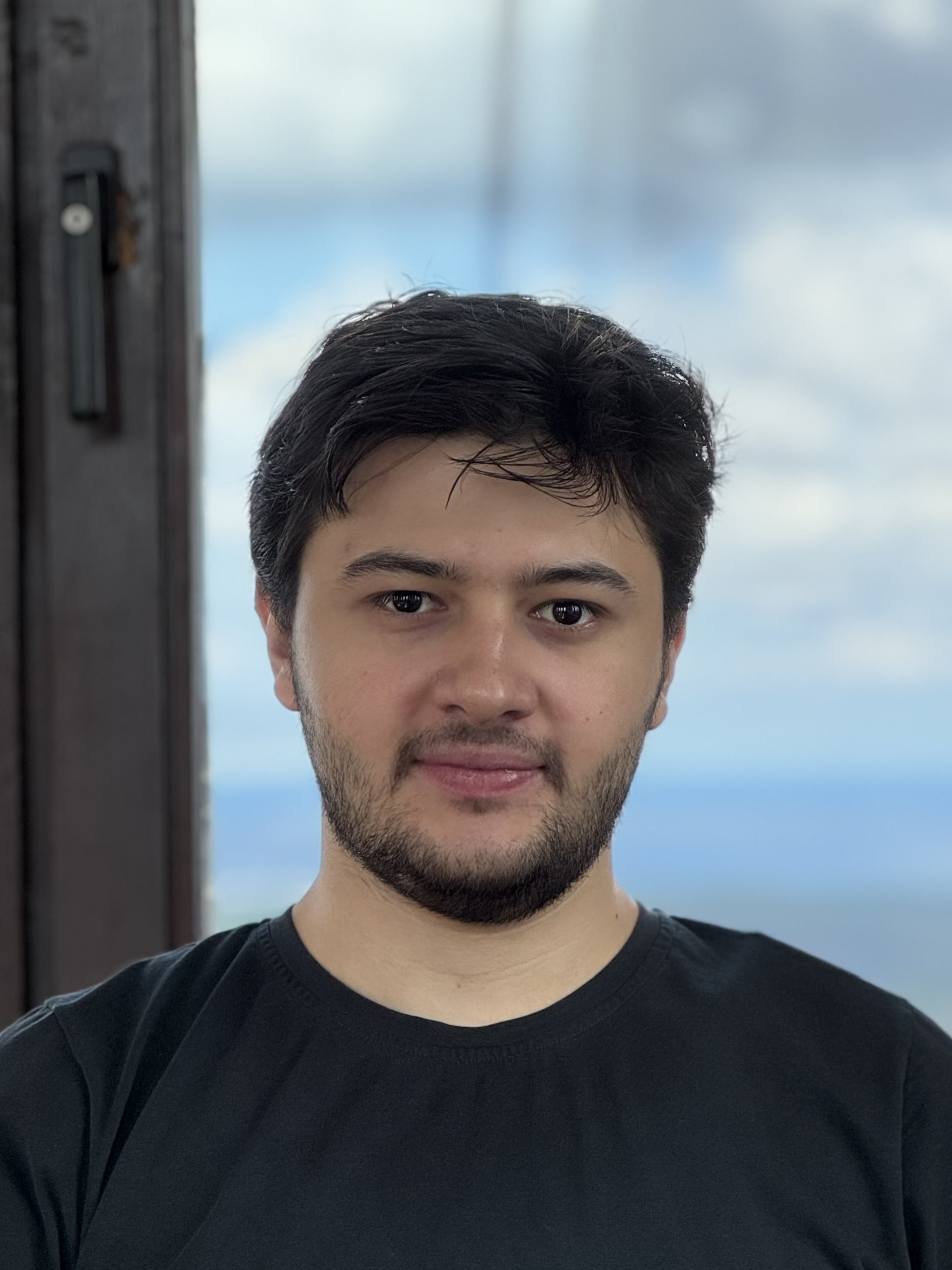}}]{Fatih Merdan}
completed his high school education at Kırıkkale Science High School. He received his B.Sc. degree in Electrical and Electronics Engineering from Middle East Technical University. He is currently pursuing his M.Sc. degree in Electrical and Electronics Engineering under the supervision of Prof. Akan at Koç University, Istanbul, Turkey.
\end{IEEEbiography}
\begin{IEEEbiography}
[{\includegraphics[width=1in,height=1.25in,clip,keepaspectratio]{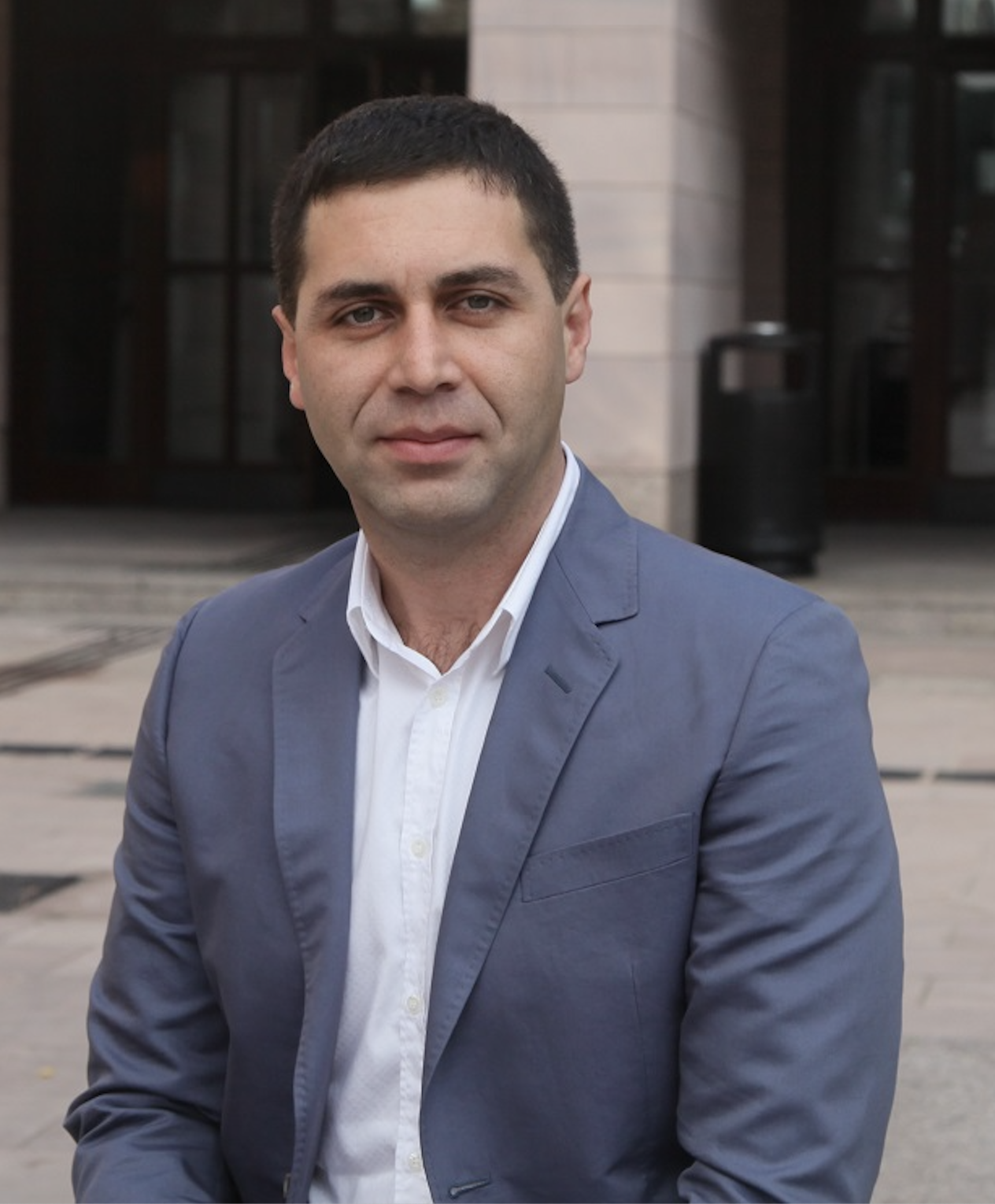}}]{Ozgur B. Akan (Fellow, IEEE)}
received the Ph.D. degree from the School of Electrical and Computer Engineering, Georgia Institute of Technology, Atlanta, in 2004. He is currently the Head of Centre for neXt Communications (CXC) at the Department of Engineering, University of Cambridge, UK and the Director of Centre for neXt-generation Communications (CXC), Koç University, Turkey. His research interests include wireless, nano, and molecular communications and Internet of Everything.
\end{IEEEbiography}
\end{document}